\documentclass[aps,twocolumn,nofootinbib,groupedaddress,superscriptaddress,longbibliography]{revtex4-1}

\usepackage{amsmath,amssymb}
\usepackage[normalem]{ulem}
\usepackage{graphicx}
\usepackage{multirow,makecell}
\usepackage{bm}
\usepackage{mathtools}
\usepackage{dsfont}
\usepackage{amsfonts}
\usepackage{url}
\usepackage{rotating}
\usepackage{wasysym}
\usepackage{diagbox}
\usepackage[dvipsnames]{xcolor}

\usepackage[colorlinks=true,linkcolor=RoyalBlue,citecolor=RoyalBlue,urlcolor=RoyalBlue,hypertexnames=false]{hyperref}

\newcommand{\nocontentsline}[3]{}
\newcommand{\tocless}[2]{\bgroup\let\addcontentsline=\nocontentsline#1{#2}\egroup}
\def\ba#1\ea{\begin{align}#1\end{align}}
\def\bg#1\eg{\begin{gather}#1\end{gather}}
\def\bpm{\begin{pmatrix}}
\def\epm{\end{pmatrix}}

\newcommand{\nn}{\nonumber \\ }
\newcommand{\td}[1]{\widetilde{#1}}

\newcommand{\hf}{\frac{1}{2}}

\newcommand{\bb}[1]{{\mathbf #1}}
\newcommand{\mbb}[1]{\mathbb{#1}}
\newcommand{\bs}[1]{\boldsymbol{#1}}
\newcommand{\bk}{\bb k}
\newcommand{\bR}{\bb R}

\newcommand{\bba}{\bb a}
\newcommand{\br}{\bb r}

\newcommand{\mc}[1]{\mathcal{#1}}
\newcommand{\mf}[1]{\mathfrak{#1}}
\newcommand{\der}{\partial}
\newcommand{\dg}{\dagger}
\newcommand{\nc}{N_c}
\newcommand{\ua}{\uparrow}
\newcommand{\da}{\downarrow}
\newcommand{\dmu}{\der_\mu}
\newcommand{\dnu}{\der_\nu}

\newcommand{\al}{\alpha}
\newcommand{\be}{\beta}
\newcommand{\gm}{\gamma}
\newcommand{\om}{\omega}
\newcommand{\sg}{\sigma}
\newcommand{\ld}{\lambda}
\newcommand{\vph}{\varphi}
\newcommand{\vep}{\varepsilon}
\newcommand{\ep}{\epsilon}

\newcommand{\ket}[1]{|#1\rangle}
\newcommand{\bra}[1]{\langle#1|}
\newcommand{\brk}[2]{\langle#1|#2\rangle}
\newcommand{\kbr}[2]{|#1\rangle\langle#2|}
\newcommand{\expv}[1]{\langle#1\rangle}

\newcommand{\vac}{\ket{0}}
\newcommand{\id}{\mathds{1}}

\newcommand{\Sec}[1]{Sec.~\ref{#1}}
\newcommand{\eq}[1]{Eq.~\eqref{#1}}

\newcommand{\fig}[1]{Fig.~\ref{#1}}

\newcommand{\dera}{\der_a}
\newcommand{\derb}{\der_b}

\newcommand{\chat}{\hat{c}}
\newcommand{\cdag}{\chat^{\dg}}
\newcommand{\phat}{\hat{\psi}}
\newcommand{\pdag}{\phat^{\dg}}

\newcommand{\ns}{N_s}
\newcommand{\bz}{\bs{\zeta}}
\newcommand{\ca}{\mc{A}}
\newcommand{\bx}{\hat{\bb x}}
\newcommand{\bpi}{\hat{\bs{\pi}}}
\newcommand{\bu}{\hat{\bb{u}}}
\newcommand{\bS}{\hat{\bb S}}
\newcommand{\shat}{\hat{S}}
\newcommand{\kxy}{\kappa_{xy}^{ph}}
\newcommand{\fem}{F^{em}}
\newcommand{\bn}{\bb{n}}
\newcommand{\Hel}{H_{el}}
\newcommand{\rion}{\br_{nu}}

\newcommand{\bv}{\bb{v}}
\newcommand{\vc}{\mc{V}}
\newcommand{\aem}{A^{em}}
\newcommand{\vd}{V_d}
\newcommand{\vo}{V_o}
\newcommand{\ein}{\vep_{\parallel}}
\newcommand{\eout}{\vep_{\perp}}
\newcommand{\vin}{V_{\parallel}}
\newcommand{\vout}{V_{\perp}}
\newcommand{\wdag}{W^{\dg}}
\newcommand{\Htd}{\td{\mc{H}}_{\ld}}
\newcommand{\Hld}{\mc{H}_{\ld}}
\newcommand{\Ho}{\mc{H}_0}
\newcommand{\vth}{\vartheta}
\newcommand{\fab}{\fem_{ab}}
\newcommand{\Zeta}{\mc{Z}}
\newcommand{\bxi}{\bs{\xi}}
\newcommand{\cf}{\mc{F}}
\newcommand{\imtr}[1]{Im[Tr[#1]]}
\newcommand{\cp}{\mc{P}}
\newcommand{\ktot}{\kappa_{\mu \nu}^{tot}}
\newcommand{\kmag}{\kappa_{\mu \nu}^{rest}}
\newcommand{\mr}{M_r}

\newcommand{\nion}{N_{nu}}
\newcommand{\Vb}{\hat{V}_{BO}}

\newcommand{\hH}{\hat{H}}
\newcommand{\bP}{\hat{\bb{P}}}
\newcommand{\kion}{\hat{K}_{nu}}
\newcommand{\Htot}{\hH_{tot}}
\newcommand{\hHel}{\hH_{el}}
\newcommand{\Vdag}{\vc^\dg}

\newcommand{\Sld}{S(\ld)}
\newcommand{\Tld}{T(\ld)}
\newcommand{\bt}{\boldsymbol{\tau}}

\newcommand{\Hp}{\hat{H}_p}
\newcommand{\uhat}{\hat{u}}

\newcommand{\fc}{\mc{F}}

\newcommand{\pihat}{\hat{\pi}}

\newcommand{\hthi}{\frac{\vth_i}{2}}
\newcommand{\hthj}{\frac{\vth_j}{2}}
\newcommand{\ijp}{e^{\frac{i}{2}(\vph_i-\vph_j)}}
\newcommand{\ijm}{e^{-\frac{i}{2}(\vph_i-\vph_j)}}
\newcommand{\brkf}[4]{\brk{#1}{#2}\brk{#2}{#3}\brk{#3}{#4}\brk{#4}{#1}}
\newcommand{\chif}[4]{(#1\chi_{ijk} #2\chi_{jkl} #3 \chi_{kli} #4 \chi_{lij})}
\newcommand{\bmu}{\bs{\mu}}
\newcommand{\bep}{\bs{\vep}}
\newcommand{\heff}{\mc{M}}
\newcommand{\bby}{\bb{y}}
\newcommand{\bhat}{\hat{b}}
\newcommand{\xhat}{\hat{x}}
\newcommand{\bd}{\bs{\delta}}

\newcommand{\Pg}{P_g}
\newcommand{\Og}{O_g}
\newcommand{\Gg}{\Gamma_g}

\newcommand{\ndeg}{[n]}
\newcommand{\nk}{\ndeg \bk}
\newcommand{\ehat}{\hat{\bb{e}}}
\newcommand{\trs}{\mc{T}}
\newcommand{\phg}{\phi_g}
\newcommand{\gbar}[1]{\overline{#1}^g}
\newcommand{\gel}{\mc{G}_{el}}
\newcommand{\gph}{\mc{G}_{p}}

\newcommand{\uion}{\bb{u}_{ion}}
\newcommand{\Ug}{U_g}
\newcommand{\dtdt}[4]{\Delta_{#1 #2}^a t_{#2 #3} \Delta_{#3 #4}^b t_{#4 #1}}
\newcommand{\gtd}{\td{\Gamma}_g}
\newcommand{\cb}{\mc{B}}
\newcommand{\kmn}{\kappa_{\mu\nu}^{ph}}
\newcommand{\omn}{\Omega_{[n]}(\bk)_{\mu\nu}}
\newcommand{\bka}{\bs{\kappa}^{ph}}
\newcommand{\kl}{k_L}
\newcommand{\kt}{k_T}
\newcommand{\cug}{\mc{U}_g}

\newcommand{\ourtitle}{
Microscopic Theory of the Phonon Thermal Hall Effect in Chiral Mott Insulators
}
\allowdisplaybreaks

\begin{document}
\title{\textbf{\ourtitle}}

\author{Junha \surname{Kang}}
%\thanks {These authors contributed equally to this work.}
\affiliation{Department of Physics and Astronomy, Seoul National University, Seoul 08826, Korea}

\author{Taekoo \surname{Oh}}
\email{taekoo.oh@ssu.ac.kr}
\affiliation{Department of Physics, Soongsil University, Seoul 06978, Korea}
\affiliation{Origin of Matter and Evolution of Galaxies (OMEG) institute, Soongsil University, Seoul 06978, Korea}
%%%%%%%%%%%%%%%%%%%%%%

%%%%%%%%%%%%%%%%%%%%%%
\begin{abstract}
The thermal Hall effect (THE) probes charge-neutral excitations in insulators, where the charge gap blocks electronic transport. Recently, phonons have been shown to induce a THE comparable in magnitude to the spin contribution, underscoring their critical role in thermal transport. Here, we develop a microscopic theory of the phonon thermal Hall effect (PTHE) in chiral Mott insulators. First, we derive the analytic form of the effective Raman interaction in half-filled Mott insulators, showing that its strength is directly proportional to the scalar spin chirality. Next, we demonstrate the intrinsic PTHE explicitly on the kagome lattice. Crucially, our formulation reveals a temperature-dependent crossover in the transport behavior under isotopic substitution. Using this result, we establish a scaling law that quantitatively separates the phonon contribution to the THE from other background signals. Our results not only provide the first fully microscopic derivation of the PTHE, but also establish a definitive experimental standard for isolating microscopic heat carriers in chiral Mott insulators.
\end{abstract}
%%%%%%%%%%%%%%%%%%%%%%

\maketitle

\let\oldaddcontentsline\addcontentsline
\renewcommand{\addcontentsline}[3]{}

%%%%%%%%%%%%%%%%%%%%%%
\textit{Introduction.---}
%%%%%%%%%%%%%%%%%%%%%%
The thermal Hall effect (THE), the induction of a transverse heat current by longitudinal temperature gradient and an external magnetic field, serves as a fundamental probe of charge-neutral excitations in quantum materials, offering insights into complex spin~\cite{katsura2010theory,onose2010observation,matsumoto2011rotational,matsumoto2014thermal,mook2014magnon,hirschberger2015large,hirschberger2015thermal,owerre2016first,owerre2017topological,kasahara2018majorana,laurell2018magnon,park2019topological,zhang2019thermal,zhang2021anomalous,akazawa2022topological,zhang2024thermal} and lattice~\cite{strohm2005phenomenological,sheng2006theory,inyushkin2007phonon,kagan2008anomalous,wang2009phonon,zhang2010topological,agarwalla2011phonon,qin2012berry,mori2014origin,ideue2017giant,hentrich2018unusual,saito2019berry,grissonnanche2019giant,boulanger2020thermal,li2020phonon,sim2021sizable,uehara2022phonon,sharma2024phonon,chen2024planar,oh2025phonon,oh2025spin,behnia2025phonon} dynamics.
In Mott insulators, where electrons are frozen out, a transverse heat current arises primarily from collective bosonic excitations such as magnons and phonons.
However, unlike in metallic systems governed by the Wiedemann-Franz law, it remains challenging to separate the contributions of different heat carriers in experiments.
Thus, it is crucial to understand how the charge-neutral excitations experience an effective Lorentz force under broken time-reversal symmetry.
In particular, the phonon thermal Hall effect (PTHE) in Mott insulators has recently attracted significant attention since the observed phonon contribution to the THE was comparable to the spin contribution~\cite{chen2022large,li2023phonon,kim2024thermal,jin2025discovery,xiang2026phonon,shragai2026phonon}.
The PTHE arises when the nuclear motion is deflected by an effective magnetic (gauge) field.
Conventionally, the effective field is described by a phenomenological Raman spin-lattice interaction, $\al_R \bb{M} \cdot (\bb{u} \times \bb{P})$, where $\bb{M}$ is the bulk magnetization.
Because this mechanism typically requires strong spin-orbit coupling (SOC), it struggles to account for the PTHE in weak-SOC Mott insulators.
Recent studies have shown that an emergent gauge field can be generated by the scalar spin chirality of the local spin clusters, even in SOC-free systems~\cite{oh2025phonon,oh2025spin}.
However, the model applies only to the spin-fluctuating regime, which features solely short-ranged spin correlations.
It remains an open challenge to develop a formalism for general lattices, especially those with extensive spin ordering.
In this work, we present a microscopic theory of the PTHE in general chiral Mott insulators.
First, we derive the analytic form of the emergent gauge field in half-filled Mott insulators for arbitrary lattices, which acts as an effective Raman spin-lattice interaction.
We show that the emergent gauge field is directly proportional to the scalar spin chirality by employing a systematic inverse Schrieffer-Wolff (SW) perturbation theory in the strong Hund coupling limit of the double-exchange model.
Next, to evaluate the macroscopic transport, we introduce a gauge-invariant multiband linear response formalism and demonstrate the intrinsic PTHE on the kagome lattice.
Crucially, we reveal a temperature-dependent crossover in the transport behavior under isotopic substitution, which primarily acts on phonons.
By leveraging this isotope effect, we establish a scaling ansatz that isolates the PTHE signal from the background at intermediate-to-high temperatures.
Our study provides the first microscopic derivation of the effective Raman interaction in chiral Mott insulators, and establishes a comprehensive experimental roadmap for identifying the microscopic origin of the THE in these systems.

%%%%%%%%%%%%%%%%%%%%%%

\textit{Phonon dynamics and the emergent gauge field.---}
%%%%%%%%%%%%%%%%%%%%%%
We begin from the total Hamiltonian of a solid $\Htot = \kion + \hHel(\rion)$, where the kinetic energy of the nuclei is $\kion = \sum_{i} (\bP_i - Z_i e \bb{A}_i^{ext})^2/2M_i$, and $\hHel(\rion)$ contains the electronic kinetic energy and all Coulomb interaction terms.
Here, $\bP_i$ is the momentum operator of the nuclei, $Z_i$ is the atomic number, $M_i$ is the atomic mass, $\bb{A}_i^{ext}$ is the external vector potential evaluated at the $i$-th nucleus, and $\rion = (\br_1, \dots, \br_{\nion})$ is the collective position of all the nuclei.
Within the Born-Oppenheimer approximation~\cite{born1927quantentheorie,mead1979determination,mead1992geometric,zhang2010topological,qin2012berry,saito2019berry,oh2025phonon,oh2025spin}, the electronic wave function adapts instantaneously to $\rion$, satisfying $\hHel(\rion) \ket{\Psi_{el} (\rion)} = U_{el}(\rion) \ket{\Psi_{el} (\rion)}$.
By integrating out $\ket{\Psi_{el}(\rion)}$ from the total Hamiltonian, we obtain an effective Hamiltonian governing the lattice dynamics,
%%%%%%%%%%
\ba
\hH_{BO} = \sum_{i} \frac{(\bP_i - Z_i e \bb{A}_i^{ext} - \hbar \bb{A}^{em}_i)^2}{2M_i} + \Vb (\rion),
\ea
%%%%%%%%%%
where $\aem_{i\mu} = i \brk{\Psi_{el}}{\frac{\der \Psi_{el}}{\der r_i^\mu}}~(\mu = x, y, \dots)$ is the emergent Berry connection parametrized by $\rion$, and $\Vb (\rion)$ is the Born-Oppenheimer potential energy surface.
To quantify the PTHE, we consider small displacements $\bu_i$ from equilibrium.
We work in natural units ($\hbar = c = 1$) and rescale the variables as $\bpi_i/\sqrt{M_i} \rightarrow \bpi_i$ and $\sqrt{M_i} \bu_i \rightarrow \bu_i$, where the kinematic momentum is $\bpi_i = \bP_i - \bb{A}_i$ and $\bb{A}_i = Z_i e \bb{A}_i^{ext} + \hbar \bb{A}_i^{em}$.
Applying the harmonic approximation to $\Vb (\rion)$ and partitioning the site index as $i = (\bR, \al)$, where $\bR$ denotes the unit cell and  $\al = 1, \dots, \ns$ labels the sublattice, the phonon Hamiltonian in momentum space takes the quadratic form
%%%%%%%%%%
\ba
\Hp = \hf \sum_{\bk} \bx_{-\bk} h(\bk) \bx_{\bk}, \quad
h(\bk) =
\bpm
\id & \\
& D(\bk)
\epm
.
\label{eq:hphonon}
\ea
%%%%%%%%%%
Here, $\bx_\bk = (\bpi_{\bk 1}, \dots, \bpi_{\bk \ns}, \bu_{\bk 1}, \dots, \bu_{\bk \ns})$ is a $2d\ns$-component operator for $d$ spatial dimensions, and $D(\bk)$ is the dynamical matrix.
The mass-rescaled gauge field manifests as an unconventional commutation relation $[\hat{\pi}_{\bR \al \mu}, \hat{\pi}_{\bR' \be \nu}] = i F(\bR - \bR')_{\al \mu, \be \nu}$.
Here, the gauge field $F(\bR - \bR')_{\al \mu, \be \nu} = \der_{u_{\bR \al}^{\mu}} A_{\bR' \be \nu} - \der_{u_{\bR' \be}^{\nu}} A_{\bR \al \mu}$ serves as an effective Raman spin-lattice coupling.
The momentum space operators obey~\cite{qin2012berry,saito2019berry}
%%%%%%%%%%
\ba
[\bx_{\bk}, \bx_{\bk'}] = 
i \delta_{\bk, -\bk'}
\bpm
F(\bk) & - \id \\
\id & 0
\epm
= i \delta_{\bk, -\bk'} \cf (\bk),
\label{eq:comm}
\ea
%%%%%%%%%%
where $F(\bk)$ is the Fourier transform of $F(\bR)$.
We detail the formalism in the Supplemental Materials (SM).
The resulting Heisenberg equation of motion $\der_t \bx_\bk = \cf(\bk) h(\bk) \bx_\bk$ has $d \ns$ (counter-) propagating normal modes with (negative) positive frequencies $\bz_{n\bk}(t) = \bz_{n \bk} e^{-i \om_{n\bk}t}~(n = \pm 1, \dots, \pm d \ns)$.
This requires solving the non-Hermitian eigenvalue equation $i \cf(\bk) h(\bk) \bz_{n \bk} = \om_{n \bk} \bz_{n \bk}$.
Despite the non-Hermiticity, the positive semidefiniteness of $h(\bk)$ ensures that $\om_{n\bk}$ is real, and that the eigenvectors obey a generalized orthonormality condition $\bz_{m\bk}^\dg h(\bk) \bz_{n\bk} = \delta_{mn}$.
These modes provide the basis for evaluating the phonon Berry curvature and the associated thermal Hall conductivity $\kmn~(\mu \neq \nu)$.
[See SM for details.]
%

%%%%%%%%%%%%%%%%%%%%%%
\textit{Microscopic derivation from the double exchange model.---}
%%%%%%%%%%%%%%%%%%%%%%
To derive an analytic expression of the emergent gauge field $\fem$, we utilize the double exchange model at half-filling in the strong Hund coupling regime to describe the Mott insulator.
The electronic Hamiltonian is given by~\cite{ye1999berry,hamamoto2015quantized}
%%%%%%%%%%
\ba
\hHel(\rion) = -J \sum_{i} \bn_i \cdot \bS_i + \sum_{ij\sg} t_{ij} \cdag_{i \sg} \chat_{j\sg},
\label{eq:doubleH}
\ea
%%%%%%%%%%
where $\hat{S}_i^\mu = \hf \sum_{s s'} \sg^{\mu}_{s s'} \cdag_{i s} \chat_{i s'}$ is the onsite spin operator, $\bn_i$ is a unit vector of the local magnetization, $J>0$ is the Hund coupling, and $t_{ij}(\rion)$ is the position-dependent hopping amplitude with $t_{ii} = 0$.
For a lattice of $\nion$ sites with one orbital, the electronic ground state at half-filling is formed by filling the $\nion$ lowest energy states.
We introduce a $2 \nion \times \nion$ matrix $\vc$, whose columns represent the occupied state eigenvectors.
The ground state projector is given by $P = \vc \vc^\dg$.
Using composite indices $a = (i,\mu)$ and $b = (j, \nu)$, the emergent Berry connection is compactly expressed as $\aem_a = iTr[\vc^{\dg} \der_a \vc]$.
Furthermore, the emergent gauge field $\fab = \der_a \aem_b - \der_b \aem_a$ is given by
%%%%%%%%%%
\ba
\fem_{ab} = - 2 \imtr{P \der_a P \der_b P}.
\label{eq:fab}
\ea
%%%%%%%%%%
Consequently, extracting an analytic form of the gauge field reduces to solving for the ground state projector $P$. 
To systematically calculate $P$ in the strong Hund coupling regime $J \gg |t_{ij}|$, we adapt the SW transformation~\cite{schrieffer1966relation,wurtz2020variational}.
Taking $\bn_i$ as the local $z$-axis, we separate the Hamiltonian into an unperturbed local part $\Ho = -\frac{J}{2} \sg_3 \otimes \id$ and the hopping perturbation
%%%%%%%%%%
\ba
V =
\bpm
\vin & W \\
\wdag & \vout
\epm
,
\ea
%%%%%%%%%%
which is proportional to the hopping scale $t$.
Defining spinors $\ket{i +}$ ($\ket{i -}$) parallel (antiparallel) to the local magnetization axis $\bn_i$, the matrix elements are given by $[\vin]_{ij} = t_{ij} \brk{i+}{j+}, W_{ij} = t_{ij} \brk{i+}{j-}$, and $[\vout]_{ij} = t_{ij} \brk{i-}{j-}$.
The SW method uses a unitary transformation $e^{-S} (\Ho + V)e^S$ that block-diagonalizes the Hamiltonian.
Consequently, the projector in the original basis evolves as $P = e^S P_0 e^{-S}$, where $P_0$ is the ground state projector of $\Ho$.
We obtain a perturbation series for the projector up to third order by expanding $S$ in powers of $t/J$: $P \approx P_0 + P_1 + P_2 + P_3$.
Because lower-order terms vanish in the trace formula in \eq{eq:fab}, we evaluate this expansion up to third order.
We detail the model and the explicit form of $P$ in the SM.

Next, we obtain the analytic form of $\fab$ by inserting the projector $P$ into \eq{eq:fab}, yielding the expansion
%%%%%%%%%%
\ba
\fem_{ab} = [\fem_{ab}]^{(2)} + [\fem_{ab}]^{(3)} + [\fem_{ab}]^{(4)} + O((t/J)^{5}).
\ea
%%%%%%%%%%
Here, we assume that near the equilibrium position, the hopping amplitudes $t_{ij}(\rion)$ depend solely on the interatomic displacement $\br_i-\br_j$.
We denote their spatial derivatives as $\der_a t_{ij} = \Delta^a_{ij}$ and assume that they linearly scale with the hopping amplitudes evaluated at equilibrium $t_{ij}^{eq}$~\cite{harrison1989electronic}.
Under this assumption, the leading second-order correction identically vanishes, $[\fab]^{(2)} = 0$.
The third-order correction evaluates to
%%%%%%%%%%
\ba
[\fab]^{(3)} &= \frac{2}{J^3} \sum_{ijk} Im[\Delta_{ij}^a \Delta_{jk}^b t_{ki}]
\nn
& \times (\bn_i \cdot \bn_j + \bn_j \cdot \bn_k -2 \bn_k \cdot \bn_i).
\label{eq:third}
\ea
%%%%%%%%%%
Because $\Delta_{ij}^a \propto t_{ij}^{eq}$, this term is proportional to the imaginary part of the three-site hopping loop.
Consequently, it vanishes for purely real hopping amplitudes.
While an external magnetic field can induce complex hopping via Peierls substitution, realistic laboratory fields ($\sim 10$ T) generate a negligible flux ($\Phi/\Phi_0\sim 1.7 \times 10^{-4}$) across a microscopic triangle (interatomic distance $\sim 4$ \r{A}). Here, $\Phi_0$ is the magnetic flux quantum.
Thus, this term is negligible in experimental regimes.
Only in some systems, where intrinsically strong complex hoppings arise, may this term become the dominant source of the emergent gauge field.
Prominent examples include the loop-current phases of cuprates~\cite{honda1993effects,varma1997non,brehmer1999effects,simon2002detection,simon2003symmetry,delannoy2005neel,varma2006theory,agterberg2015emergent,chatterjee2017intertwining,bounoua2020loop,bounoua2022hidden} or the Haldane model~\cite{haldane1988model,oh2025thermal}.
In the absence of such intrinsic complex hopping, the characteristic scale $t/J \sim 10^{-1}$ drives the system into an unusual regime where the fourth-order perturbation dominates the third-order term.
The fourth-order term arises from virtual processes in which electrons couple to lattice displacements in four-site hopping loops:
%%%%%%%%%%
\ba
[\fab]^{(4)}
=& \frac{20}{J^4} \sum_{ijkl}
Re[\Delta_{ij}^a \Delta_{jk}^b t_{kl} t_{li}]
\nn
& \times 
(\chi_{ijk}+\chi_{jkl}+\chi_{lij}-3\chi_{kli})
\nn
+& \frac{40}{J^4} \sum_{ij}
Re[\dtdt{a}{i}{b}{j} - \dtdt{a}{i}{j}{b}]
\nn
& \times
(\chi_{aib}+ \chi_{ibj} -\chi_{bja}-\chi_{jai}),
\label{eq:f4}
\ea
%%%%%%%%%%
where $\chi_{ijk} = \frac{1}{8} \bn_i \cdot (\bn_j \times \bn_k)$ is the scalar spin chirality of the spins $ijk$.
We detail the perturbation theory in the SM.
Notably, the magnitude of this emergent gauge field vastly exceeds that of external magnetic fields in experiments.
To estimate the ratio $R = |\hbar [F^{em}]^{(4)}| / |e F^{ext}|$, we assume a Harrison-type scaling rule~\cite{harrison1989electronic}, where $t_{ij}$ depends on the bond distance $|\br_i - \br_j|$.
Introducing a composite index $a = (i_a, \mu_a)$, the derivative of the hopping amplitude is expressed as $\Delta_{ij}^a = \gm t^{eq}_{ij} \overline{\delta}_{ij}^a$, where $\gm$ is a material specific constant, and $\overline{\delta}_{ij}^a$ is the $\mu_a$-direction cosine of the bond $\overline{ij}$ with $i_a$ as the origin.
This leads to the order of magnitude estimate $|[F^{em}]^{(4)}| \sim 20 (t/J)^4 \gm^2 \chi$.
In typical transition-metal oxides, the effective $d-d$ hopping mediated by intermediate $p$ orbitals scales as $t \sim t_{pd}^2 / (\vep_p-\vep_d)$.
By Harrison's rule, $t_{pd} \propto r_{pd}^{-4}$, yielding $t \propto r_{pd}^{-8}$.
For a metal-oxygen bond length $r_{pd} \approx 2$ \r{A}, this gives the scaling factor $\gm = |\der_r t/t| \approx 4$ \r{A}$^{-1}$.
We insert this alongside $t/J = 0.2$ and a scalar spin chirality $\chi \sim 0.1$, which is readily achievable via Dzyaloshinskii-Moriya (DM) interactions with a 10 T magnetic field~\cite{grohol2005spin,laurell2018magnon,oh2025phonon}.
This yields a massive factor $R\sim 300$.
Thus, we hereafter define the total gauge field acting on the phonons strictly by $[F^{em}]^{(4)}$.
This analytic result demonstrates that the gauge field is fully characterized by the local lattice geometry, the normalized hopping scale $t/J$, and the scalar spin chirality of the underlying magnetic texture.

%%%%%%%%%%%%%%%%%%%%%%

\textit{Thermal transport and phonon symmetries.---}
%%%%%%%%%%%%%%%%%%%%%%
To quantify the macroscopic transport, the phonon thermal Hall conductivity $\kmn~(\mu\neq\nu)$ is evaluated by integrating the phonon Berry curvature over the Brillouin zone~\cite{qin2012berry,saito2019berry,sign}
%%%%%%%%%%
\ba
\kmn = -\frac{k_B^2 T}{V \hbar} \sum_{\bk} \sum_{n > 0}  \big[
c_2 (g(\hbar \omega_{n\bk})) - \frac{\pi^2}{3}
\big] \Omega_{n}(\bk)_{\mu \nu},
\label{eq:kxy}
\ea
%%%%%%%%%%
where $V$ is the system volume, $c_2(x) = \int_0^x dt \big(\ln \frac{1+t}{t} \big)^2$, $g(\vep)$ is the Bose-Einstein distribution, and the sum runs over the particle bands ($n > 0$).
Unlike electronic systems, the phonon Berry curvature $\Omega_{n}(\bk)_{\mu \nu} = \dmu \ca_n(\bk)_\nu - \dnu \ca_n(\bk)_\mu$ is derived from a generalized connection that incorporates the matrix $h(\bk)$ to account for bosonic orthonormality,
%%%%%%%%%%
\ba
\ca_n(\bk)_\mu = -Im [\bz_{n\bk}^{\dg} h(\bk) \dmu \bz_{n \bk}].
\ea
%%%%%%%%%%
When performing numerics, however, the derivatives of the eigenvectors are unstable due to arbitrary gauge choices and band degeneracies at each $\bk$ point.
Thus, we introduce a gauge-invariant formula for the multiband Abelian phonon Berry curvature
%%%%%%%%%%
\ba
\Omega_{[n]}(\bk)_{\mu \nu} = -\imtr{
\cp_{[n]\bk} [\dmu \cp_{[n]\bk}, \dnu \cp_{[n]\bk}]
}.
\ea
%%%%%%%%%%
Here, $[n]$ labels the degenerate subspaces at $\bk$, and $\cp_{[n]\bk}$ is a generalized non-Hermitian projector.
One can evaluate the macroscopic thermal transport for systems with complex band degeneracies by replacing the summation $n \to [n]$ in \eq{eq:kxy}.
We detail the generalized projector formalism in the SM.
%

%%%%%%%%%%%%%%%%%%%%%%
\begin{figure}[t]
	\centering\includegraphics[width=0.49\textwidth]{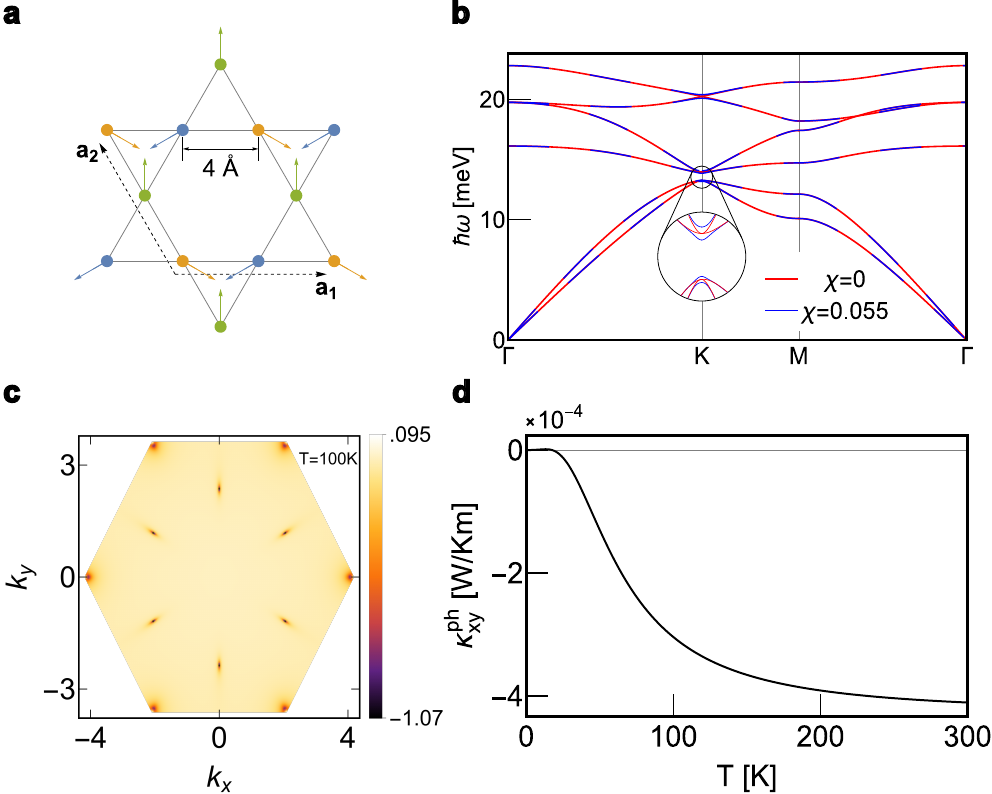}
	\caption{{\bf Intrinsic phonon thermal Hall effect in the kagome Mott insulator.}
		(a) Kagome lattice geometry with a canted antiferromagnetic order. The $10^{\circ}$ out-of-plane canting yields a uniform scalar spin chirality $\chi = 0.055$. (b) Phonon band structure. Solid red curves denote the unperturbed dispersion $(\chi = 0)$; dashed blue curves incorporate the emergent gauge field $(t/J = 0.3, \gm = 4$ \r{A}$^{-1})$. Calculations assume an interatomic spacing $4$ \r{A}, nuclear mass $10^{-25}~kg$, and longitudinal/transverse spring constants $20$ N/m and $10$ N/m, respectively. (c) Momentum-resolved thermal Hall conductivity integrand evaluated at $T = 100$ K. (d) Macroscopic thermal Hall conductivity $\kxy$ as a function of temperature assuming an interlayer spacing of $4$ \r{A}.}
	\label{fig1}
\end{figure}
%%%%%%%%%%%%%%%%%%%%%%

Crucially, the survival of $\kmn$ is dictated by global lattice symmetries.
However, phonon symmetries must be defined carefully, since $\fem$ does not appear directly in the phonon Hamiltonian $\Hp$ [\eq{eq:hphonon}].
Thus, we define a phonon symmetry $g$ through the invariance of the Heisenberg equation of motion.
When $\fem \neq 0$, this requires the invariance of both the electronic Hamiltonian $\hHel(\rion^{eq})$ at equilibrium nuclear configuration and the bare phonon Hamiltonian $\Hp|_{\fem = 0}$.
We consider general spin space group symmetries $g = (\mc{T})^{\phg}\{\cug | \Og | \bd_g \}$~\cite{brinkman1966theory,xiao2024spin,chen2024enumeration,jiang2024enumeration}, where $\mc{T}$ is the time reversal operation, $\phg = 0~(1)$ for unitary (antiunitary) symmetries, and $\cug \in SU(2)$ specifies the spin rotation.
$\Og$ and $\bd_g$ are the orthogonal matrix and the translation vector representing the pure spatial action of $g$.
Then, we find that the thermal conductivity tensor $\kappa^{ph}$ [see \eq{eq:kxy}] satisfies
%%%%%%%%%%
\ba
\kappa^{ph} = (-1)^{\phg} \Og \kappa^{ph} \Og^T.
\label{eq:kmnrule}
\ea
%%%%%%%%%%
In two dimensions (2D), this is equivalent to $\kxy = (-1)^{\phg} \det(\Og) \kxy$, meaning that $\kxy = 0$ if unitary improper or antiunitary proper rotations exist.
In 3D, \eq{eq:kmnrule} implies that the dual vector of the conductivity tensor $[\bka]_{\rho} = \hf \sum_{\mu\nu} \vep_{\mu\nu\rho} \kmn$ satisfies $\bka = (-1)^{\phg} \det(\Og) \Og \bka$.
Note that $\bka$ transforms identically to magnetic fields.
We detail the phonon symmetries in the SM.
Due to these symmetry constraints, highly symmetric systems such as the triangular or square lattices cannot exhibit a PTHE caused by the emergent gauge field, since such systems exhibit unitary mirror operations with $\cug = \id$.
Breathing distortions recover a finite PTHE by breaking the mirror symmetry.
On the other hand, the kagome lattice inherently lacks such mirror symmetries, providing an ideal platform to demonstrate the intrinsic PTHE.
Accordingly, we assume a canted antiferromagnetic order in a 2D kagome lattice insulator [see \fig{fig1}(a)], where the non-coplanar spin ordering can be achieved by DM interactions~\cite{grohol2005spin,laurell2018magnon} with external magnetic fields.
We present the explicit form of the dynamical matrix and $[\fem(\bk)]^{(4)}$ in the SM.
This configuration generates a uniform scalar spin chirality throughout
the entire system, inducing an emergent gauge field for
the phonons.
As shown in \fig{fig1}(b), the emergent gauge field alters the phonon dispersion.
The symmetry-protected degeneracies at the $K$ point and the accidental degeneracies on the $\Gamma-M$ line are lifted when a finite scalar spin chirality is introduced.
However, because the emergent gauge field vanishes at the $\Gamma$ point, the degeneracies there remain unaffected.
We demonstrate the physical consequence of these gap openings in the momentum-resolved thermal Hall integrand of \eq{eq:kxy} [see \fig{fig1}(c)].
The thermodynamically weighted phonon Berry curvature forms sharply localized hotspots centered at the gap openings at the $K$ point and on the $\Gamma-M$ line.
Conversely, at the $\Gamma$ point, the vanishing emergent field dictates that the local Berry curvature vanishes.
We note that the multiband projector formalism accurately captures this aspect, avoiding artificial gap-openings that generate unphysical Berry curvatures.
We integrate this curvature over the Brillouin zone to obtain the intrinsic phonon thermal Hall conductivity $\kxy$ [\fig{fig1}(d)].
At low temperatures ($T \lesssim 18$ K), the response is small and positive, dominated by the phonon Berry curvature of the lowest acoustic modes near $\Gamma$.
However, as the temperature increases, the Berry curvature hotspots at $K$ and the $\Gamma-M$ line become thermally populated.
Consequently, $\kxy$ undergoes a sign change and rapidly climbs to a large magnitude.
The robust signal persists throughout the antiferromagnetically ordered phase and is comparable in magnitude to the magnon thermal Hall conductivity in the high temperature regime~\cite{mook2014magnon}.
This highlights that phonons, as well as magnons, play a significant role in thermal transport.

%%%%%%%%%%%%%%%%%%%%%%

\begin{figure}[t]
	\centering\includegraphics[width=0.49\textwidth]{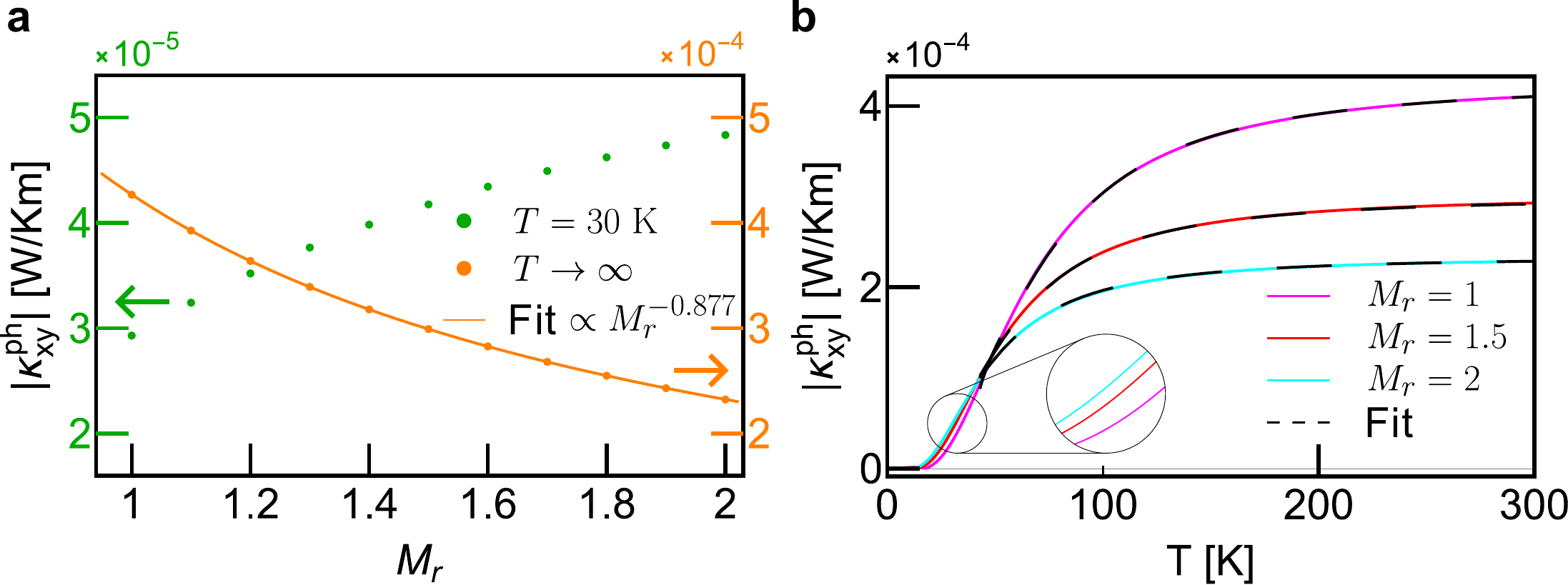}
	\caption{{\bf Isotope effect and mass scaling.}
		(a) Absolute value of thermal Hall conductivity as a function of the relative isotopic mass $\mr$. At low temperatures ($T=30$ K, green dots, left axis), heavier masses enhance the transport, whereas in the high-temperature limit (orange dots, right axis), the macroscopic signal is strictly suppressed. The solid orange line indicates a power-law fit $\kxy \propto \mr^{-0.877}$.
		(b) Temperature dependence of $|\kxy|$ for $\mr = 1$ (magenta), $1.5$ (red), and $2$ (cyan). The dotted lines are fits to the ansatz [\eq{eq:asym}] over the range $43 \text{ K} \leq T \leq 300 \text{ K}$.}
	\label{fig2}
\end{figure}

\textit{Isotope effect and scaling law.---}
%%%%%%%%%%%%%%%%%%%%%%
In chiral Mott insulators, the total transverse thermal transport $\ktot = \kmn + \kmag~(\mu \neq \nu)$ is a sum of the phonon contribution and other signals, primarily from magnons~\cite{laurell2018magnon}.
To isolate the pure phonon response, we turn to isotopic substitution.
This probe alters the lattice dynamics without affecting mass-independent backgrounds.
Isotopic substitution ($M_0 \rightarrow \mr M_0$, with relative mass $\mr > 1$) modifies the phonon properties in two competing ways.
First, the kinematic softening of the dynamical matrix lowers the phonon frequencies ($\om \sim \mr^{-1/2}$).
Second, the emergent gauge field is suppressed by the heavier mass, as it is rescaled by the mass of the kinematic momentum ($F^{em} \propto \mr^{-1}$) [see \eq{eq:comm}], thereby reducing the phonon Berry curvature. 
This dichotomy causes an opposite isotopic mass dependence of the phonon thermal Hall conductivity $|\kmn|$ depending on the temperature regime.
We demonstrate this in \fig{fig2}(a), where we plot $|\kxy|$ obtained at low ($T = 30$ K) and high ($T \to \infty$) temperatures as a function of $\mr$.
At low temperatures, where the small thermal occupation bottlenecks the transport, the softened phonon dispersion increases the thermal population of the Berry curvature hotspots, thereby enhancing $|\kxy|$.
Conversely, at high temperatures where these modes are sufficiently populated, the dominant effect is the suppression of the phonon Berry curvature, causing a reduction in $|\kxy|$.
%

%%%%%%%%%%%%%%%%%%%%%%
%%%%%%%%%%%%%%%%%%%%%%

%
In the $T \rightarrow \infty$ limit, $|\kxy|$ precisely scales as $\kxy \propto \mr^{-0.877}$ [orange line in \fig{fig2}(a)].
To quantify the scaling behavior, we expand \eq{eq:kxy} in the high-temperature limit ($k_B T \gg \hbar \om$), which yields
%%%%%%%%%%
\ba
\kmn = \frac{k_B}{V} \sum_{[n]\bk} \om_{[n]\bk} \Omega_{[n]}(\bk)_{\mu \nu}
\Big[1-\frac{1}{36} \big(\frac{\hbar \om_{[n]\bk}}{k_B T}\big)^2 + \dots
\Big].
\label{eq:taylor}
\ea
%%%%%%%%%%
This expansion consists of even powers of $\hbar \om/k_B T$, and is valid for $k_B T > \max(\hbar \om_{[n]\bk})/2\pi$, corresponding to $T \sim 43$ K for the kagome lattice model.
The $T \rightarrow \infty$ mass scaling behavior and \eq{eq:taylor} suggest that the mass dependencies can effectively be factored into a frequency softening $\om \sim \mr^{-1/2}$ and an effective Berry curvature suppression $\Omega \sim \mr^{-b}$ with $b > 0$.
This observation motivates the ansatz for the mass and temperature scaling,
%%%%%%%%%%
\ba
\kmn = \sum_{i=0}^{\infty} a_i \mr^{-(\hf + b + i)} T^{-2i}.
\label{eq:asym}
\ea
%%%%%%%%%%
As shown in \fig{fig2}(b), \eq{eq:asym} provides an accurate fit down to $T \sim 43$ K, which coincides with the radius of convergence of the expansion.
Additionally, \eq{eq:asym} robustly captures the continuous mass scaling across a dense sweep of $\mr \in [1,2]$ at fixed temperatures.
We detail the fit to the kagome lattice in the SM.

The success of this ansatz establishes a practical experimental protocol for isolating $\kmn$.
By measuring $\ktot(T)$ for two isotopically substituted samples in an intermediate-temperature window below the N\'eel temperature, one can subtract the signals to eliminate the background signal $\kmag$, which is mostly independent of the nuclear mass.
Then, the pure PTHE can be extracted by fitting this difference to \eq{eq:asym}.
This method provides a definitive tool for isolating the heat carriers in chiral Mott insulators.
%

%%%%%%%%%%%%%%%%%%%%%%
\textit{Discussion.---}
%%%%%%%%%%%%%%%%%%%%%%
In summary, we have rigorously established that the scalar spin chirality governs the microscopic spin-lattice coupling in half-filled Mott insulators.
Beyond the quantitative microscopic prediction of the intrinsic PTHE, the isotopic scaling law [\eq{eq:asym}] delivers a practical diagnostic toolkit for experimentally isolating the phononic heat carriers.
Although the scaling describes the intermediate-to-high temperature regime, a phononic response can still be isolated from the magnon contribution in the low temperature regime.
When a strong magnetic field is applied, the $SU(2)$ symmetry is broken and a Goldstone mass gap is introduced in the magnon spectrum.
Consequently, the magnon thermal Hall conductivity is suppressed via the thermal depletion of low-energy modes~\cite{onose2010observation}, while the phonon signal mostly remains.
Crucially, our microscopic derivation reveals qualitative differences from conventional phenomenological Raman interactions $\al_R \bb{M} \cdot (\bb{u} \times \bb{P})$.
First, the conventional and emergent field-induced Raman interactions exhibit opposite high-field behaviors.
Because the conventional Raman interaction couples to the bulk magnetization, the resulting PTHE saturates at extreme magnetic fields.
In contrast, when a massive external field forces the spins into a collinear polarized state, the scalar spin chirality vanishes.
Consequently, the emergent gauge field vanishes, and the associated PTHE collapses.
Second, the isotope effects for the two Raman interactions are different.
While the emergent field is suppressed by heavier isotopic masses ($\propto \mr^{-1}$), the mass scaling of the conventional Raman interaction is governed by the mass dependence of the material-specific phenomenological coupling constant $\al_R$.
These distinct scaling behaviors could serve as a discriminator between the mechanisms.
Despite these differences, however, the two mechanisms share an identical form~\cite{sign,matsumoto2014thermal,zhang2016berry,park2019topological} of the transport integral [\eq{eq:kxy}].
Hence, the isotopic scaling ansatz [\eq{eq:asym}] remains a ubiquitous tool to separate the phonon signal from other contributions beyond Mott insulators.
Next, we suggest the material candidates.
For immediate 2D experimental realization, we propose iron jarosites~\cite{grohol2005spin,laurell2018magnon} due to their kagome geometry, large canting angles, and high N\'eel temperatures.
Extending this framework to 3D, we anticipate a massive intrinsic PTHE in pyrochlore antiferromagnets, where the non-coplanar geometry naturally guarantees a macroscopic emergent gauge field~\cite{kim2020strain}.
We also note that $[\fab]^{(3)}$ in \eq{eq:third} can be realized in the loop-current physics historically explored in cuprate Mott insulators~\cite{honda1993effects,varma1997non,brehmer1999effects,simon2002detection,simon2003symmetry,delannoy2005neel,varma2006theory,agterberg2015emergent,chatterjee2017intertwining,bounoua2020loop,bounoua2022hidden}.

Finally, our microscopic formulation quantifies and generalizes the insights of recent studies~\cite{oh2025phonon,oh2025spin}, which identified the scalar spin chirality as the driver of the extrinsic PTHE via skew-scattering in frustrated magnets.
Because the analytic formulation of the emergent field derived here does not require translational symmetry and is dimension independent, it provides a universal platform for evaluating both intrinsic and extrinsic PTHE across arbitrary lattice geometries.
Consequently, our work opens a compelling theoretical direction for thermal transport studies in strongly correlated phases.
%
%%%%%%%%%%%%%%%

%\let\oldaddcontentsline\addcontentsline
%\renewcommand{\addcontentsline}[3]{}
\begin{acknowledgments}
J.K. and T.O. were supported by the Pioneer Project for Future-Oriented Convergence Technology (Challenge Type; RS-2024-00416976) and Basic Science Research Program (RS-2025-16065011) through the National Research Foundation of Korea (NRF) funded by the Ministry of Science and ICT of Korean government. T.O. was also supported by  Basic Science Research Program (RS-2026-25470523) through the National Research Foundation of Korea (NRF) funded by the Ministry of Science and ICT of Korean government, and G-LAMP (RS-2025-25441317) through the National Research Foundation of Korea (NRF) funded by the Ministry of Education of the Korean government.
\end{acknowledgments}
%%%%%%%%%%%%%%%%%%%%%%

%%%%%%%REFERENCES%%%%%%%%
%\bibliographystyle{apsrev}
\bibliography{Refs.bib}

@footnote{sign,
	note={Note that we have corrected a minor sign error in the general derivation of $\kxy$ present in Refs.~\cite{qin2012berry,saito2019berry}. Because the affected topological term vanishes in the systems studied in those works, their physical conclusions remain entirely unaffected. Reconciling the sign here, however, unifies the phonon Hall conductivity formulas across distinct theoretical frameworks~\cite{qin2012berry,matsumoto2014thermal,zhang2016berry} in the literature. For the explicit algebraic reconciliation, see the Supplemental Material.}
}

@article{haldane1988model,
	title={Model for a quantum Hall effect without Landau levels: Condensed-matter realization of the" parity anomaly"},
	author={Haldane, F Duncan M},
	journal={Physical review letters},
	volume={61},
	number={18},
	pages={2015},
	year={1988},
	publisher={APS},
	doi={
	10.1103/PhysRevLett.61.2015
	}
}

@article{honda1993effects,
	title={Effects of cyclic four-spin exchange on the magnetic properties of the CuO 2 plane},
	author={Honda, Y and Kuramoto, Y and Watanabe, T},
	journal={Physical Review B},
	volume={47},
	number={17},
	pages={11329},
	year={1993},
	publisher={APS},
	doi={
	10.1103/PhysRevB.47.11329
	}
}

@article{varma1997non,
	title={Non-Fermi-liquid states and pairing instability of a general model of copper oxide metals},
	author={Varma, CM},
	journal={Physical Review B},
	volume={55},
	number={21},
	pages={14554},
	year={1997},
	publisher={APS},
	doi={
	10.1103/PhysRevB.55.14554
	}
}

@article{brehmer1999effects,
	title={Effects of biquadratic exchange on the spectrum of elementary excitations in spin ladders},
	author={Brehmer, S and Mikeska, H-J and M{\"u}ller, M and Nagaosa, N and Uchida, S},
	journal={Physical Review B},
	volume={60},
	number={1},
	pages={329},
	year={1999},
	publisher={APS},
	doi={
	10.1103/PhysRevB.60.329
	}
}

@article{simon2002detection,
	title={Detection and implications of a time-reversal breaking state in underdoped cuprates},
	author={Simon, ME and Varma, CM},
	journal={Physical review letters},
	volume={89},
	number={24},
	pages={247003},
	year={2002},
	publisher={APS},
	doi={
	10.1103/PhysRevLett.89.247003
	}
}

@article{simon2003symmetry,
	title={Symmetry considerations for the detection of second-harmonic generation in cuprates in the pseudogap phase},
	author={Simon, ME and Varma, CM},
	journal={Physical Review B},
	volume={67},
	number={5},
	pages={054511},
	year={2003},
	publisher={APS},
	doi={
	10.1103/PhysRevB.67.054511
	}
}

@article{delannoy2005neel,
	title={N{\'e}el order, ring exchange, and charge fluctuations in the half-filled Hubbard model},
	author={Delannoy, J-YP and Gingras, MJP and Holdsworth, PCW and Tremblay, A-MS},
	journal={Physical Review B—Condensed Matter and Materials Physics},
	volume={72},
	number={11},
	pages={115114},
	year={2005},
	publisher={APS},
	doi={
	10.1103/PhysRevB.72.115114
	}
}

@article{varma2006theory,
	title={Theory of the pseudogap state of the cuprates},
	author={Varma, CM},
	journal={Physical Review B—Condensed Matter and Materials Physics},
	volume={73},
	number={15},
	pages={155113},
	year={2006},
	publisher={APS},
	doi={
	10.1103/PhysRevB.73.155113
	}
}

@article{agterberg2015emergent,
	title={Emergent loop current order from pair density wave superconductivity},
	author={Agterberg, Daniel F and Melchert, Drew S and Kashyap, Manoj K},
	journal={Physical Review B},
	volume={91},
	number={5},
	pages={054502},
	year={2015},
	publisher={APS},
	doi={
	10.1103/PhysRevB.91.054502
	}
}

@article{chatterjee2017intertwining,
	title={Intertwining topological order and broken symmetry in a theory of fluctuating spin-density waves},
	author={Chatterjee, Shubhayu and Sachdev, Subir and Scheurer, Mathias S},
	journal={Physical review letters},
	volume={119},
	number={22},
	pages={227002},
	year={2017},
	publisher={APS},
	doi={
	10.1103/PhysRevLett.119.227002
	}
}

@article{bounoua2020loop,
	title={Loop currents in two-leg ladder cuprates},
	author={Bounoua, Dalila and Mangin-Thro, Lucile and Jeong, Jaehong and Saint-Martin, Romuald and Pinsard-Gaudart, Loreynne and Sidis, Yvan and Bourges, Philippe},
	journal={Communications Physics},
	volume={3},
	number={1},
	pages={123},
	year={2020},
	publisher={Nature Publishing Group UK London},
	doi={
	10.1038/s42005-020-0388-1
	}
}

@article{bounoua2022hidden,
	title={Hidden magnetic texture in the pseudogap phase of high-Tc YBa2Cu3O6. 6},
	author={Bounoua, Dalila and Sidis, Yvan and Loew, Toshinao and Bourdarot, Fr{\'e}d{\'e}ric and Boehm, Martin and Steffens, Paul and Mangin-Thro, Lucile and Bal{\'e}dent, Victor and Bourges, Philippe},
	journal={Communications Physics},
	volume={5},
	number={1},
	pages={268},
	year={2022},
	publisher={Nature Publishing Group UK London},
	doi={
	10.1038/s42005-022-01048-1
	}
}

@article{born1927quantentheorie,
	title={Zur quantentheorie der molekeln},
	author={Born, Max and Oppenheimer, Robert},
	journal={Annalen der physik},
	volume={389},
	number={20},
	pages={457--484},
	year={1927},
	publisher={Wiley Online Library},
	doi={
	10.1002/andp.19273892002
	}
}

@article{mead1979determination,
	title={On the determination of Born--Oppenheimer nuclear motion wave functions including complications due to conical intersections and identical nuclei},
	author={Mead, C Alden and Truhlar, Donald G},
	journal={The Journal of Chemical Physics},
	volume={70},
	number={5},
	pages={2284--2296},
	year={1979},
	publisher={American Institute of Physics},
	doi={
	10.1063/1.437734
	}
}

@article{mead1992geometric,
	title={The geometric phase in molecular systems},
	author={Mead, C Alden},
	journal={Reviews of modern physics},
	volume={64},
	number={1},
	pages={51},
	year={1992},
	publisher={APS},
	doi={
	10.1103/RevModPhys.64.51
	}
}

@article{schrieffer1966relation,
	title={Relation between the anderson and kondo hamiltonians},
	author={Schrieffer, John R and Wolff, Peter A},
	journal={Physical Review},
	volume={149},
	number={2},
	pages={491},
	year={1966},
	publisher={APS},
	doi={
	10.1103/PhysRev.149.491
	}
}

@article{wurtz2020variational,
	title={Variational Schrieffer-Wolff transformations for quantum many-body dynamics},
	author={Wurtz, Jonathan and Claeys, Pieter W and Polkovnikov, Anatoli},
	journal={Physical Review B},
	volume={101},
	number={1},
	pages={014302},
	year={2020},
	publisher={APS},
	doi={
	10.1103/PhysRevB.101.014302
	}
}

@book{harrison1989electronic,
	title={Electronic structure and the properties of solids: the physics of the chemical bond},
	author={Harrison, Walter A},
	year={1989},
	publisher={Courier Corporation}
}

@article{ye1999berry,
	title={Berry phase theory of the anomalous Hall effect: Application to colossal magnetoresistance manganites},
	author={Ye, Jinwu and Kim, Yong Baek and Millis, AJ and Shraiman, BI and Majumdar, P and Te{\v{s}}anovi{\'c}, Z},
	journal={Physical review letters},
	volume={83},
	number={18},
	pages={3737},
	year={1999},
	publisher={APS},
	doi={
	10.1103/PhysRevLett.83.3737
	}
}

@article{hamamoto2015quantized,
	title={Quantized topological Hall effect in skyrmion crystal},
	author={Hamamoto, Keita and Ezawa, Motohiko and Nagaosa, Naoto},
	journal={Physical Review B},
	volume={92},
	number={11},
	pages={115417},
	year={2015},
	publisher={APS},
	doi={
	10.1103/PhysRevB.92.115417
	}
}

@article{lee1992gauge,
	title={Gauge theory of the normal state of high-T c superconductors},
	author={Lee, Patrick A and Nagaosa, Naoto},
	journal={Physical Review B},
	volume={46},
	number={9},
	pages={5621},
	year={1992},
	publisher={APS},
	doi={
	10.1103/PhysRevB.46.5621
	}
}

@article{strohm2005phenomenological,
	title={Phenomenological evidence for the phonon Hall effect},
	author={Strohm, C and Rikken, GLJA and Wyder, P},
	journal={Physical review letters},
	volume={95},
	number={15},
	pages={155901},
	year={2005},
	publisher={APS},
	doi={
	10.1103/PhysRevLett.95.155901
	}
}

@article{sheng2006theory,
	title={Theory of the phonon Hall effect in paramagnetic dielectrics},
	author={Sheng, L and Sheng, DN and Ting, CS},
	journal={Physical review letters},
	volume={96},
	number={15},
	pages={155901},
	year={2006},
	publisher={APS},
	doi={
	10.1103/PhysRevLett.96.155901
	}
}

@article{inyushkin2007phonon,
	title={On the phonon Hall effect in a paramagnetic dielectric},
	author={Inyushkin, Alexander Vasil’evich and Taldenkov, Alexandr Nikolaevich},
	journal={Jetp Letters},
	volume={86},
	number={6},
	pages={379--382},
	year={2007},
	publisher={Springer},
	doi={
	10.1134/S0021364007180075
	}
}

@article{kagan2008anomalous,
	title={Anomalous Hall effect for the phonon heat conductivity in paramagnetic dielectrics},
	author={Kagan, Yu and Maksimov, LA},
	journal={Physical review letters},
	volume={100},
	number={14},
	pages={145902},
	year={2008},
	publisher={APS},
	doi={
	10.1103/PhysRevLett.100.145902
	}
}

@article{wang2009phonon,
	title={Phonon Hall thermal conductivity from the Green-Kubo formula},
	author={Wang, Jian-Sheng and Zhang, Lifa},
	journal={Physical Review B—Condensed Matter and Materials Physics},
	volume={80},
	number={1},
	pages={012301},
	year={2009},
	publisher={APS},
	doi={
	10.1103/PhysRevB.80.012301
	}
}

@article{zhang2010topological,
	title={Topological nature of the phonon Hall effect},
	author={Zhang, Lifa and Ren, Jie and Wang, Jian-Sheng and Li, Baowen},
	journal={Physical review letters},
	volume={105},
	number={22},
	pages={225901},
	year={2010},
	publisher={APS},
	doi={
	10.1103/PhysRevLett.105.225901
	}
}

@article{agarwalla2011phonon,
	title={Phonon Hall effect in ionic crystals in the presence of static magnetic field},
	author={Agarwalla, Bijay K and Zhang, Lifa and Wang, Jian-Sheng and Li, Baowen},
	journal={The European Physical Journal B},
	volume={81},
	number={2},
	pages={197--202},
	year={2011},
	publisher={Springer},
	doi={
	10.1140/epjb/e2011-11002-x
	}
}

@article{qin2012berry,
	title={Berry curvature and the phonon Hall effect},
	author={Qin, Tao and Zhou, Jianhui and Shi, Junren},
	journal={Physical Review B—Condensed Matter and Materials Physics},
	volume={86},
	number={10},
	pages={104305},
	year={2012},
	publisher={APS},
	doi={
	10.1103/PhysRevB.86.104305
	}
}

@article{mori2014origin,
	title={Origin of the phonon Hall effect in rare-earth garnets},
	author={Mori, Michiyasu and Spencer-Smith, Alexander and Sushkov, Oleg P and Maekawa, Sadamichi},
	journal={Physical review letters},
	volume={113},
	number={26},
	pages={265901},
	year={2014},
	publisher={APS},
	doi={
	10.1103/PhysRevLett.113.265901
	}
}

@article{ideue2017giant,
	title={Giant thermal Hall effect in multiferroics},
	author={Ideue, Toshiya and Kurumaji, Takashi and Ishiwata, Shintaro and Tokura, Yoshinori},
	journal={Nature materials},
	volume={16},
	number={8},
	pages={797--802},
	year={2017},
	publisher={Nature Publishing Group UK London},
	doi={
	10.1038/nmat4905
	}
}

@article{hentrich2018unusual,
	title={Unusual phonon heat transport in $\alpha$-RuCl 3: Strong spin-phonon scattering and field-induced spin gap},
	author={Hentrich, Richard and Wolter, Anja UB and Zotos, Xenophon and Brenig, Wolfram and Nowak, Domenic and Isaeva, Anna and Doert, Thomas and Banerjee, Arnab and Lampen-Kelley, Paula and Mandrus, David G and others},
	journal={Physical review letters},
	volume={120},
	number={11},
	pages={117204},
	year={2018},
	publisher={APS},
	doi={
	10.1103/PhysRevLett.120.117204
	}
}

@article{saito2019berry,
	title={Berry phase of phonons and thermal Hall effect in nonmagnetic insulators},
	author={Saito, Takuma and Misaki, Kou and Ishizuka, Hiroaki and Nagaosa, Naoto},
	journal={Physical Review Letters},
	volume={123},
	number={25},
	pages={255901},
	year={2019},
	publisher={APS},
	doi={
	10.1103/PhysRevLett.123.255901
	}
}

@article{grissonnanche2019giant,
	title={Giant thermal Hall conductivity in the pseudogap phase of cuprate superconductors},
	author={Grissonnanche, Ga{\"e}l and Legros, Ana{\"e}lle and Badoux, Sven and Lefran{\c{c}}ois, Etienne and Zatko, Victor and Lizaire, Maude and Lalibert{\'e}, Francis and Gourgout, Adrien and Zhou, J-S and Pyon, Sunseng and others},
	journal={Nature},
	volume={571},
	number={7765},
	pages={376--380},
	year={2019},
	publisher={Nature Publishing Group UK London},
	doi={
	10.1038/s41586-019-1375-0
	}
}

@article{boulanger2020thermal,
	title={Thermal Hall conductivity in the cuprate Mott insulators Nd2CuO4 and Sr2CuO2Cl2},
	author={Boulanger, Marie-Eve and Grissonnanche, Ga{\"e}l and Badoux, Sven and Allaire, Andr{\'e}anne and Lefran{\c{c}}ois, {\'E}tienne and Legros, Ana{\"e}lle and Gourgout, Adrien and Dion, Maxime and Wang, CH and Chen, XH and others},
	journal={Nature communications},
	volume={11},
	number={1},
	pages={5325},
	year={2020},
	publisher={Nature Publishing Group UK London},
	doi={
	10.1038/s41467-020-18881-z
	}
}

@article{li2020phonon,
	title={Phonon thermal Hall effect in strontium titanate},
	author={Li, Xiaokang and Fauqu{\'e}, Beno{\^\i}t and Zhu, Zengwei and Behnia, Kamran},
	journal={Physical review letters},
	volume={124},
	number={10},
	pages={105901},
	year={2020},
	publisher={APS},
	doi={
	10.1103/PhysRevLett.124.105901
	}
}

@article{jin2025discovery,
  title={Discovery of universal phonon thermal Hall effect in crystals},
  author={Jin, XB and Zhang, Xu and Wan, WB and Wang, HR and Jiao, YH and Li, SY},
  journal={Physical Review Letters},
  volume={135},
  number={19},
  pages={196302},
  year={2025},
  publisher={APS},
  doi={
  10.1103/r572-5dfm
  }
}

@article{xiang2026phonon,
  title={Phonon thermal Hall effect: the roles of disorder, annealing, and metallic contacts},
  author={Xiang, Qiaochao and Li, Xiaokang and Guo, Xiaodong and Behnia, Kamran and Zhu, Zengwei},
  journal={npj Quantum Materials},
  year={2026},
  publisher={Nature Publishing Group},
  doi={
  10.1038/s41535-026-00876-6
  }
}

@article{kim2020strain,
  title={Strain engineering of the magnetic multipole moments and anomalous Hall effect in pyrochlore iridate thin films},
  author={Kim, Woo Jin and Oh, Taekoo and Song, Jeongkeun and Ko, Eun Kyo and Li, Yangyang and Mun, Junsik and Kim, Bongju and Son, Jaeseok and Yang, Zhuo and Kohama, Yoshimitsu and others},
  journal={Science Advances},
  volume={6},
  number={29},
  pages={eabb1539},
  year={2020},
  publisher={American Association for the Advancement of Science},
  doi={
  10.1126/sciadv.abb1539
  }
}

@article{oh2025thermal,
  title={Thermal Hall effect induced by phonon skew-scattering via orbital magnetization},
  author={Oh, Taekoo},
  journal={arXiv preprint arXiv:2507.22436},
  year={2025},
  doi={
  10.48550/arXiv.2507.22436
  }
}

@article{chen2022large,
  title={Large phonon thermal Hall conductivity in the antiferromagnetic insulator Cu3TeO6},
  author={Chen, Lu and Boulanger, Marie-Eve and Wang, Zhi-Cheng and Tafti, Fazel and Taillefer, Louis},
  journal={Proceedings of the National Academy of Sciences},
  volume={119},
  number={34},
  pages={e2208016119},
  year={2022},
  publisher={National Academy of Sciences},
  doi={
  10.1073/pnas.2208016119
  }
}

@article{li2023phonon,
  title={The phonon thermal Hall angle in black phosphorus},
  author={Li, Xiaokang and Machida, Yo and Subedi, Alaska and Zhu, Zengwei and Li, Liang and Behnia, Kamran},
  journal={Nature Communications},
  volume={14},
  number={1},
  pages={1027},
  year={2023},
  publisher={Nature Publishing Group UK London},
  doi={
  10.1038/s41467-023-36750-3
  }
}

@article{behnia2025phonon,
  title={Phonon thermal Hall as a lattice Aharonov-Bohm effect},
  author={Behnia, Kamran},
  journal={SciPost Physics Core},
  volume={8},
  number={3},
  pages={061},
  year={2025},
  doi={
  10.21468/SciPostPhysCore.8.3.061}
}

@article{shragai2026phonon,
  title={Phonon Hall viscosity and the intrinsic thermal Hall effect of $\alpha$-RuCl3},
  author={Shragai, Avi and Horsley, Ezekiel and Kim, Subin and Kim, Young-June and Ramshaw, BJ},
  journal={Nature},
  pages={1--7},
  year={2026},
  publisher={Nature Publishing Group UK London},
  doi={
  10.1038/s41586-026-10420-y
  }
}

@article{kim2024thermal,
  title={Thermal Hall effects due to topological spin fluctuations in YMnO3},
  author={Kim, Ha-Leem and Saito, Takuma and Yang, Heejun and Ishizuka, Hiroaki and Coak, Matthew John and Lee, Jun Han and Sim, Hasung and Oh, Yoon Seok and Nagaosa, Naoto and Park, Je-Geun},
  journal={Nature Communications},
  volume={15},
  number={1},
  pages={243},
  year={2024},
  publisher={Nature Publishing Group UK London},
  doi={
  10.1038/s41467-023-44448-9
  }
}

@article{sim2021sizable,
	title={Sizable suppression of thermal Hall effect upon isotopic substitution in SrTiO 3},
	author={Sim, Sangwoo and Yang, Heejun and Kim, Ha-Leem and Coak, Matthew J and Itoh, Mitsuru and Noda, Yukio and Park, Je-Geun},
	journal={Physical Review Letters},
	volume={126},
	number={1},
	pages={015901},
	year={2021},
	publisher={APS},
	doi={
	10.1103/PhysRevLett.126.015901
	}
}

@article{uehara2022phonon,
	title={Phonon thermal Hall effect in a metallic spin ice},
	author={Uehara, Taiki and Ohtsuki, Takumi and Udagawa, Masafumi and Nakatsuji, Satoru and Machida, Yo},
	journal={Nature Communications},
	volume={13},
	number={1},
	pages={4604},
	year={2022},
	publisher={Nature Publishing Group UK London},
	doi={
	10.1038/s41467-022-32375-0
	}
}

@article{sharma2024phonon,
	title={Phonon thermal Hall effect in charge-compensated topological insulators},
	author={Sharma, Rohit and Bagchi, Mahasweta and Wang, Yongjian and Ando, Yoichi and Lorenz, Thomas},
	journal={Physical Review B},
	volume={109},
	number={10},
	pages={104304},
	year={2024},
	publisher={APS},
	doi={
	10.1103/PhysRevB.109.104304
	}
}

@article{chen2024planar,
	title={Planar thermal Hall effect from phonons in a Kitaev candidate material},
	author={Chen, Lu and Lefran{\c{c}}ois, {\'E}tienne and Vallipuram, Ashvini and Barth{\'e}lemy, Quentin and Ataei, Amirreza and Yao, Weiliang and Li, Yuan and Taillefer, Louis},
	journal={Nature communications},
	volume={15},
	number={1},
	pages={3513},
	year={2024},
	publisher={Nature Publishing Group UK London},
	doi={
	10.1038/s41467-024-47858-5
	}
}

@article{oh2025phonon,
	title={Phonon thermal Hall effect in Mott insulators via skew scattering by the scalar spin chirality},
	author={Oh, Taekoo and Nagaosa, Naoto},
	journal={Physical Review X},
	volume={15},
	number={1},
	pages={011036},
	year={2025},
	publisher={APS},
	doi={
	10.1103/PhysRevX.15.011036
	}
}

@article{oh2025spin,
	title={Spin-phonon coupling and thermal Hall effect in the Kitaev model},
	author={Oh, Taekoo and Nagaosa, Naoto},
	journal={Physical Review B},
	volume={112},
	number={8},
	pages={L081104},
	year={2025},
	publisher={APS},
	doi={
	10.1103/ljlq-1kgl
	}
}

@article{zhang2016berry,
	title={Berry curvature and various thermal Hall effects},
	author={Zhang, Lifa},
	journal={New Journal of Physics},
	volume={18},
	number={10},
	pages={103039},
	year={2016},
	publisher={IOP Publishing},
	doi={
	10.1088/1367-2630/18/10/103039
	}
}

@article{grohol2005spin,
	title={Spin chirality on a two-dimensional frustrated lattice},
	author={Grohol, Daniel and Matan, Kittiwit and Cho, Jin-Hyung and Lee, Seung-Hun and Lynn, Jeffrey W and Nocera, Daniel G and Lee, Young S},
	journal={Nature Materials},
	volume={4},
	number={4},
	pages={323--328},
	year={2005},
	publisher={Nature Publishing Group UK London},
	doi={
	10.1038/nmat1353
	}
}

@article{kasahara2018majorana,
	title={Majorana quantization and half-integer thermal quantum Hall effect in a Kitaev spin liquid},
	author={Kasahara, Yuichi and Ohnishi, Tsuneya and Mizukami, Yuta and Tanaka, Osamu and Ma, Sixiao and Sugii, Kaori and Kurita, Nobuyuki and Tanaka, Hidekazu and Nasu, Joji and Motome, Yukitoshi and others},
	journal={Nature},
	volume={559},
	number={7713},
	pages={227--231},
	year={2018},
	publisher={Nature Publishing Group UK London},
	doi={
	10.1038/s41586-018-0274-0
	}
}

@article{katsura2010theory,
	title={Theory of the thermal Hall effect in quantum magnets},
	author={Katsura, Hosho and Nagaosa, Naoto and Lee, Patrick A},
	journal={Physical review letters},
	volume={104},
	number={6},
	pages={066403},
	year={2010},
	publisher={APS},
	doi={
	10.1103/PhysRevLett.104.066403
	}
}

@article{onose2010observation,
	title={Observation of the magnon Hall effect},
	author={Onose, Y and Ideue, T and Katsura, H and Shiomi, Y and Nagaosa, N and Tokura, Y},
	journal={Science},
	volume={329},
	number={5989},
	pages={297--299},
	year={2010},
	publisher={American Association for the Advancement of Science},
	doi={
	10.1126/science.1188260
	}
}

@article{matsumoto2011rotational,
	title={Rotational motion of magnons and the thermal Hall effect},
	author={Matsumoto, Ryo and Murakami, Shuichi},
	journal={Physical Review B—Condensed Matter and Materials Physics},
	volume={84},
	number={18},
	pages={184406},
	year={2011},
	publisher={APS},
	doi={
	10.1103/PhysRevB.84.184406
	}
}

@article{matsumoto2014thermal,
	title={Thermal Hall effect of magnons in magnets with dipolar interaction},
	author={Matsumoto, Ryo and Shindou, Ryuichi and Murakami, Shuichi},
	journal={Physical Review B},
	volume={89},
	number={5},
	pages={054420},
	year={2014},
	publisher={APS},
	doi={
	10.1103/PhysRevB.89.054420
	}
}

@article{mook2014magnon,
	title={Magnon Hall effect and topology in kagome lattices: A theoretical investigation},
	author={Mook, Alexander and Henk, J{\"u}rgen and Mertig, Ingrid},
	journal={Physical Review B},
	volume={89},
	number={13},
	pages={134409},
	year={2014},
	publisher={APS},
	doi={
	10.1103/PhysRevB.89.134409
	}
}

@article{hirschberger2015large,
	title={Large thermal Hall conductivity of neutral spin excitations in a frustrated quantum magnet},
	author={Hirschberger, Max and Krizan, Jason W and Cava, RJ and Ong, NP},
	journal={Science},
	volume={348},
	number={6230},
	pages={106--109},
	year={2015},
	publisher={American Association for the Advancement of Science},
	doi={
	10.1126/science.1257340
	}
}

@article{hirschberger2015thermal,
	title={Thermal Hall effect of spin excitations in a kagome magnet},
	author={Hirschberger, Max and Chisnell, Robin and Lee, Young S and Ong, Nai Phuan},
	journal={Physical review letters},
	volume={115},
	number={10},
	pages={106603},
	year={2015},
	publisher={APS},
	doi={
	10.1103/PhysRevLett.115.106603
	}
}

@article{owerre2016first,
	title={A first theoretical realization of honeycomb topological magnon insulator},
	author={Owerre, SA},
	journal={Journal of Physics: Condensed Matter},
	volume={28},
	number={38},
	pages={386001},
	year={2016},
	publisher={IOP Publishing},
	doi={
	10.1088/0953-8984/28/38/386001
	}
}

@article{owerre2017topological,
	title={Topological thermal Hall effect in frustrated kagome antiferromagnets},
	author={Owerre, SA},
	journal={Physical Review B},
	volume={95},
	number={1},
	pages={014422},
	year={2017},
	publisher={APS},
	doi={
	10.1103/PhysRevB.95.014422
	}
}

@article{laurell2018magnon,
	title={Magnon thermal Hall effect in kagome antiferromagnets with Dzyaloshinskii-Moriya interactions},
	author={Laurell, Pontus and Fiete, Gregory A},
	journal={Physical Review B},
	volume={98},
	number={9},
	pages={094419},
	year={2018},
	publisher={APS},
	doi={
	10.1103/PhysRevB.98.094419
	}
}

@article{park2019topological,
	title={Topological magnetoelastic excitations in noncollinear antiferromagnets},
	author={Park, Sungjoon and Yang, Bohm-Jung},
	journal={Physical Review B},
	volume={99},
	number={17},
	pages={174435},
	year={2019},
	publisher={APS},
	doi={
	10.1103/PhysRevB.99.174435
	}
}

@article{zhang2019thermal,
	title={Thermal Hall effect induced by magnon-phonon interactions},
	author={Zhang, Xiaoou and Zhang, Yinhan and Okamoto, Satoshi and Xiao, Di},
	journal={Physical review letters},
	volume={123},
	number={16},
	pages={167202},
	year={2019},
	publisher={APS},
	doi={
	10.1103/PhysRevLett.123.167202
	}
}

@article{zhang2021anomalous,
	title={Anomalous thermal Hall effect in an insulating van der Waals magnet},
	author={Zhang, Heda and Xu, Chunqiang and Carnahan, Caitlin and Sretenovic, Milos and Suri, Nishchay and Xiao, Di and Ke, Xianglin},
	journal={Physical Review Letters},
	volume={127},
	number={24},
	pages={247202},
	year={2021},
	publisher={APS},
	doi={
	10.1103/PhysRevLett.127.247202
	}
}

@article{akazawa2022topological,
	title={Topological thermal Hall effect of magnons in magnetic skyrmion lattice},
	author={Akazawa, Masatoshi and Lee, Hyun-Yong and Takeda, Hikaru and Fujima, Yuri and Tokunaga, Yusuke and Arima, Taka-hisa and Han, Jung Hoon and Yamashita, Minoru},
	journal={Physical Review Research},
	volume={4},
	number={4},
	pages={043085},
	year={2022},
	publisher={APS},
	doi={
	10.1103/PhysRevResearch.4.043085
	}
}

@article{zhang2024thermal,
	title={Thermal Hall effects in quantum magnets},
	author={Zhang, Xiao-Tian and Gao, Yong Hao and Chen, Gang},
	journal={Physics Reports},
	volume={1070},
	pages={1--59},
	year={2024},
	publisher={Elsevier},
	doi={
	10.1016/j.physrep.2024.03.004
	}
}

@article{herzog2022superfluid,
	title={Superfluid weight bounds from symmetry and quantum geometry in flat bands},
	author={Herzog-Arbeitman, Jonah and Peri, Valerio and Schindler, Frank and Huber, Sebastian D and Bernevig, B Andrei},
	journal={Physical review letters},
	volume={128},
	number={8},
	pages={087002},
	year={2022},
	publisher={APS},
	doi={
	10.1103/PhysRevLett.128.087002
	}
}

@article{brinkman1966theory,
	title={Theory of spin-space groups},
	author={Brinkman, WF and Elliott, Roger James},
	journal={Proceedings of the Royal Society of London. Series A. Mathematical and Physical Sciences},
	volume={294},
	number={1438},
	pages={343--358},
	year={1966},
	publisher={The Royal Society London},
	doi={
	10.1098/rspa.1966.0211
	}
}

@article{xiao2024spin,
	title={Spin space groups: Full classification and applications},
	author={Xiao, Zhenyu and Zhao, Jianzhou and Li, Yanqi and Shindou, Ryuichi and Song, Zhi-Da},
	journal={Physical Review X},
	volume={14},
	number={3},
	pages={031037},
	year={2024},
	publisher={APS},
	doi={
	10.1103/PhysRevX.14.031037
	}
}

@article{chen2024enumeration,
	title={Enumeration and representation theory of spin space groups},
	author={Chen, Xiaobing and Ren, Jun and Zhu, Yanzhou and Yu, Yutong and Zhang, Ao and Liu, Pengfei and Li, Jiayu and Liu, Yuntian and Li, Caiheng and Liu, Qihang},
	journal={Physical Review X},
	volume={14},
	number={3},
	pages={031038},
	year={2024},
	publisher={APS},
	doi={
	10.1103/PhysRevX.14.031038
	}
}

@article{jiang2024enumeration,
	title={Enumeration of spin-space groups: Toward a complete description of symmetries of magnetic orders},
	author={Jiang, Yi and Song, Ziyin and Zhu, Tiannian and Fang, Zhong and Weng, Hongming and Liu, Zheng-Xin and Yang, Jian and Fang, Chen},
	journal={Physical Review X},
	volume={14},
	number={3},
	pages={031039},
	year={2024},
	publisher={APS},
	doi={
	10.1103/PhysRevX.14.031039
	}
}

@article{hwang2026stable,
	title={Stable Wave-Function Zeros Indicate Exciton Topology},
	author={Hwang, Yoonseok and Davenport, Henry and Schindler, Frank},
	journal={arXiv preprint arXiv:2604.21643},
	year={2026},
	doi={
	10.48550/arXiv.2604.21643
	}
}

%%%%%%%%APPENDIX%%%%%%%%
\let\addcontentsline\oldaddcontentsline
%%%%%%%%%%%%%%%%%%%%%%

%%%%%%%%%%%%%%%%%%%%%%
\clearpage
\onecolumngrid
\begin{center}
\textbf{\large Supplemental Material for ``\ourtitle"}
\end{center}
%%%%%%%%%%%%%%%%%%%%%%
\setcounter{section}{0}
\setcounter{figure}{0}
\setcounter{equation}{0}
\setcounter{table}{0}
\renewcommand{\thefigure}{S\arabic{figure}}
\renewcommand{\theequation}{S\arabic{equation}}
\renewcommand{\thesection}{S\arabic{section}}
\renewcommand{\thetable}{S\arabic{table}}
\tableofcontents
%%%%%%%%%%%%%%%%%%%%%%
\hfill \\
\onecolumngrid
%%%%%%%%%%%%%%%%%%%%%%

%%%%%%%%%%%%%%%%%%%%%%
%In this Supplementary Material, we provide detailed discussions on the results presented in the main text.
%%%%%%%%%%%%%%%%%%%%%%

%%%%%%%%%%%%%%%%%%%%%%
\section{Born-Oppenheimer approximation}
\label{app:BO}
%%%%%%%%%%%%%%%%%%%%%%

In this Appendix, we provide the derivation of the effective Born-Oppenheimer (BO) Hamiltonian presented in the main text.
Let $\rion = (\br_1, \dots, \br_{\nion})$ denote the collection of all nuclear positions in $d$ spatial dimensions.
A general Hamiltonian describing the coupled system of interacting electrons and the nuclei is given by
%%%%%%%%%%
\ba
\hH = \sum_{i=1}^{\nion} \frac{(\bP_i - Z_i e \bb{A}^{ext}_i)^2}{2M_i} + \hHel(\rion) = \kion + \hHel(\rion),
\label{seq:htot}
\ea
%%%%%%%%%%
where $\bP_i^{\mu} = \frac{\hbar}{i} \frac{\der}{\der r_i^{\mu}}~(\mu = x, y, \dots)$ is the canonical nuclear momentum operator, $M_i$ and $Z_i e$ are the mass and effective charge of the $i$-th nucleus $(Z_i \in \mbb{Z})$, respectively, and $\bb{A}^{ext}_i = \bb{A}^{ext}(\br_i)$ is the external electromagnetic vector potential evaluated at $\br_i$.
The electronic Hamiltonian, $\hHel(\rion)$, is parametrically dependent on the nuclear coordinates $\rion$.
It incorporates the single-particle electronic kinetic energy, electron-electron interactions, and the Coulomb interactions between all the electrons and the nuclei.
Note that we do not require the single-particle electronic energy to be a strictly quadratic function of the electronic momentum, thereby allowing quite general forms including relativistic corrections.
The Born-Oppenheimer approximation assumes that the substantial mass difference between electrons and nuclei allows their dynamics to be adiabatically separated.
Thus, we assume an ansatz for the total system wave function
%%%%%%%%%%
\ba
\ket{\Psi_{BO}(\rion)} = \Psi_{nu}(\rion) \ket{\Psi_{el}(\rion)}.
\ea
%%%%%%%%%%
Here, $\ket{\Psi_{el}(\rion)}$ is defined as the instantaneous ground state of the electronic Hamiltonian at a fixed nuclear configuration, satisfying the eigenvalue equation
%%%%%%%%%%
\ba
\hHel(\rion) \ket{\Psi_{el}(\rion)} = U_{el}(\rion) \ket{\Psi_{el}(\rion)}.
\ea
%%%%%%%%%%
The Born-Oppenheimer wave function is then obtained by minimizing the variational energy $\bra{\Psi_{BO}(\rion)} \hH \ket{\Psi_{BO}(\rion)}$.

The effective Hamiltonian governing the nuclear dynamics is obtained by integrating out the electronic degrees of freedom as $\hH_{BO} \Psi_{nu}(\rion) = \bra{\Psi_{el}(\rion)} \hH \ket{\Psi_{BO}(\rion)}$, where the nuclear momentum operates on the total wave function.
Upon direct evaluation, one finds that the derivatives acting on $\ket{\Psi_{el}(\rion)}$ shift the nuclear momentum, resulting in
%%%%%%%%%%
\ba
\hH_{BO} = \sum_{i=1}^{\nion} \frac{(\bP_i - \bb{A}_i)^2}{2M_i} + \Vb (\rion)
, \quad
\bb{A}_i =Z_i e \bb{A}_i^{ext} + \hbar \bb{A}^{em}_i
\label{seq:bo}
\ea
%%%%%%%%%%
Here, the emergent gauge (Berry) potential is a real-valued connection given by
%%%%%%%%%%
\ba
A^{em}_{i\mu} \equiv i \brk{\Psi_{el}(\rion)}{\frac{\der \Psi_{el}(\rion)}{\der r_i^{\mu}}}.
\label{seq:gaugeconn}
\ea
%%%%%%%%%%
The corresponding Born-Oppenheimer potential energy, $\Vb (\rion)$, includes both the electronic ground state energy and an emergent geometric scalar potential arising from the second-order derivatives of the electronic wave function
%%%%%%%%%%
\ba
\Vb (\rion) = U_{el}(\rion) + \sum_{i=1}^{\nion} \sum_{\mu=1}^{d} \frac{\hbar^2}{2M_i} \Big( \brk{\frac{\der \Psi_{el}(\rion)}{\der r_i^{\mu}}}{\frac{\der \Psi_{el}(\rion)}{\der r_i^{\mu}}} + \brk{\Psi_{el}(\rion)}{\frac{\der \Psi_{el}(\rion)}{\der r_i^{\mu}}}^2 \Big).
\ea
%%%%%%%%%%
%
In the following sections, we adopt natural units by setting $\hbar = c = 1$.
We will explicitly restore these constants only when required for quantitative, material-specific calculations.

%%%%%%%%%%%%%%%%%%%%%%
\section{Analytic form of the emergent gauge field}
\label{app:gauge}
%%%%%%%%%%%%%%%%%%%%%%

%
A central object that governs the phonon dynamics is the emergent gauge field $\fem$.
In this Appendix, we present the detailed derivation of the analytic form presented in the main text.
In \Sec{app:gaugedef}, we introduce the double exchange model and express the emergent gauge field in terms of the electronic ground state projector.
Next, in \Sec{app:projpert}, we introduce a perturbative expansion for the projectors.
In \Sec{app:gaugecalc}, we apply this projector perturbation theory to prove the formula for $\fem$ utilized in the main text.

%%%%%%%%%%%%%%%%%%%%%%
\subsection{Definition and projector formula}
\label{app:gaugedef}
%%%%%%%%%%%%%%%%%%%%%%
%
We begin by defining the electronic environment.
We consider a double exchange model at half-filling on a general lattice of $\nion$ sites in the strong Hund's coupling regime.
The electronic Hamiltonian is given by
%%%%%%%%%%
\ba
\hHel(\rion) = -J \sum_{i} \bn_i \cdot \bS_i + \sum_{ij\sg} t_{ij}(\rion) \cdag_{i \sg} \chat_{j\sg},
\quad
\shat_i^\mu = \hf \sum_{s s'} \sg_{s s'}^\mu \cdag_{i s} \chat_{i s'},
\label{seq:double}
\ea
%%%%%%%%%%
where $\chat_{i\sg}$ is the electron annihilation operator at site $i$ with spin $\sg$, $\bn_i$ is a unit vector representing the local magnetization axis, and $J > 0$ is the Hund coupling.
We set $t_{ii} = 0$ and allow for complex hopping amplitudes $t_{ij}$.
This model is completely general regarding spatial dimension and lattice geometry.
Note that $t_{ij}(\rion)$ depends parametrically on the nuclear positions $\rion$; this label will henceforth be omitted for notational simplicity.
Let us denote the single-particle matrix representation of $\hHel$ as $\Hel$.
This $2\nion \times 2\nion$ matrix is characterized by the eigenvalue equation $\Hel \bv_n = E_n \bv_n$, with ordered eigenvalues $E_1 \leq \dots \leq E_{2\nion}$.
At half-filling, the many-body electronic ground state is formed by occupying the lowest $\nion$ single-particle states:
%%%%%%%%%%
\ba
\ket{\Psi_{el}} = \prod_{n=1}^{\nion} \pdag_n \vac,
\quad
\Big(
\pdag_n = \sum_{i \sg} [\bv_n]_{i\sg} \cdag_{i \sg}
\Big),
\ea
%%%%%%%%%%
where $\vac$ is the fermion vacuum.
Using the emergent gauge potential defined in \eq{seq:gaugeconn}, we define the corresponding emergent gauge field as its geometric curvature.
Introducing a composite index $a = (i, \mu)$ to accommodate both site and spatial indices ($\dera = \der / \der r_i^\mu$), the emergent gauge field reads
%%%%%%%%%%
\ba
\fem_{ab} \equiv \dera \aem_b - \derb \aem_a
= i \Big(
\brk{\dera \Psi_{el}}{\derb \Psi_{el}}
- \brk{\derb \Psi_{el}}{\dera \Psi_{el}}
\Big).
\label{seq:gaugecurv}
\ea
%%%%%%%%%%
Note that the emergent gauge field possesses both intra- and inter-site contributions, contrary to the external gauge field which only has intra-site components.
Furthermore, from $\aem_a \in \mbb{R}$ and $\fem_{ab} = - \fem_{ba}$, we readily find that $i \fem$ is a Hermitian matrix, a property that is highly advantageous when diagonalizing the effective phonon Hamiltonian in \Sec{app:diag}.
Substituting the Slater determinant ground state into \eq{seq:gaugecurv}, it is straightforward to verify that the emergent gauge potential and curvature reduce to traces over the occupied single-particle states:
%%%%%%%%%%
\ba
\aem_a &= i \sum_{n=1}^{\nion} \bv^{\dg}_n \dera \bv_n = i~Tr[\Vdag \dera \vc],
\nn
\fem_{ab} &= i~Tr[\dera \Vdag \derb \vc - \derb \Vdag \dera \vc] = -2 \imtr{\dera \Vdag \derb \vc},
\label{seq:gaugev}
\ea
%%%%%%%%%%
where $\vc = (\bv_1, \dots, \bv_{\nion})$ is the $2 \nion \times \nion$ matrix formed by the single-particle eigenvectors of the occupied states.
To facilitate a perturbative expansion in the strong coupling regime ($|t_{ij}|/J \ll 1$), we prove the following projector formula for the emergent gauge field as presented in the main text:
%%%%%%%%%%
\ba
\fab= -2\imtr{P \dera P \derb P},
\label{seq:fproj}
\ea
%%%%%%%%%%
where $P = \vc \Vdag$ is the ground state projector.
We begin by inserting the identity $\id = P + (\id - P)$ into \eq{seq:gaugev}:
%%%%%%%%%%
\ba
\fem_{ab}= -2\imtr{\dera \Vdag P \derb \vc} - 2 \imtr{\dera \Vdag (\id - P) \derb \vc}.
\ea
%%%%%%%%%%
We can show that the first term vanishes identically.
By substituting $P = \vc \Vdag$, applying the cyclic property of the trace, and using the identity $\Vdag \vc = \id_{\nion\times\nion}$ (which implies $\dera (\Vdag \vc) = 0$), we find
%%%%%%%%%%
\ba
\imtr{\dera \Vdag P \derb \vc}
&= \imtr{\dera \Vdag \vc \Vdag \derb \vc} \nn
&= \imtr{\Vdag \dera \vc \derb \Vdag \vc} \nn
&= \imtr{\derb \Vdag P \dera \vc} \nn
&= -\imtr{\dera \Vdag P \derb \vc} = 0.
\ea
%%%%%%%%%%
In the last line, we used $\imtr{X}=-\imtr{X^\dg}$.
For the remaining term, we utilize the identities $\Vdag (\id - P) = 0$ and $(\id - P) \vc = 0$ to express the derivatives entirely in terms of the projector:
%%%%%%%%%%
\ba
\fab
&= - 2 \imtr{(\Vdag \vc) \dera \Vdag (\id - P) \derb \vc} \nn
&= - 2 \imtr{\vc \dera \Vdag (\id - P) \derb \vc \Vdag} \nn
&= - 2 \imtr{(\dera P - \dera \vc \Vdag)(\id - P) (\derb P - \vc \derb \Vdag)} \nn
&= - 2 \imtr{\dera P (\id - P) \derb P}.
\ea
%%%%%%%%%%
Finally, by utilizing the idempotence of $P$ ($P^2 = P$), we note that
%%%%%%%%%%
\ba
\dera P P + P \dera P = \dera P
\Leftrightarrow
\dera P (\id - P) = P \dera P.
\ea
%%%%%%%%%%
Substituting this relation into the trace yields \eq{seq:fproj}. 
We note in passing that while we have derived this identity explicitly for the half-filled double exchange model, \eq{seq:fproj} is a general property of the Berry curvature for any noninteracting fermionic system described by a single Slater determinant ground state~\cite{herzog2022superfluid}.
Consequently, extracting the analytic form of $\fem$ is formally reduced to obtaining an analytic expression for the ground state projector $P$.

%%%%%%%%%%%%%%%%%%%%%%
\subsection{Projector perturbation theory}
\label{app:projpert}
%%%%%%%%%%%%%%%%%%%%%%
In this Appendix, we derive a perturbative expansion for the ground state projector by utilizing the Schrieffer-Wolff (SW) transformation.
While we ultimately apply this formalism to the double exchange model, we emphasize that the following approach is completely general.
It can be utilized to track the evolution of any unperturbed eigen-subspace within an arbitrary Hamiltonian, provided that the target subspace remains energetically isolated from the rest of the spectrum upon the introduction of a perturbation.
Consider a general Hamiltonian parametrized by a dimensionless parameter $\ld \in [0, 1]$,
%%%%%%%%%%
\ba
\Hld = \Ho + \ld V = \Ho + \ld (\vd + \vo),
\ea
%%%%%%%%%%
where the matrix has been decomposed into its unperturbed part $\Ho$ and a perturbation $V$.
We further separate the perturbation into block-diagonal ($\vd$) and block-off-diagonal ($\vo$) components with respect to the unperturbed basis.
In this basis, the matrices take the form
%%%%%%%%%%
\ba
\Ho =
\bpm
\ein & 0 \\
0 & \eout
\epm
, \quad
\vd =
\bpm
\vin & 0 \\
0 & \vout
\epm
, \quad
\vo =
\bpm
0 & W \\
\wdag & 0
\epm
.
\ea
%%%%%%%%%%
Here, $\ein$ and $\eout$ are diagonal matrices representing the unperturbed energies of the target subspace and the remainder of the Hilbert space, respectively.
We assume the gap between these two sectors is sufficiently large such that their perturbed eigenvalue spectra remain disjoint for all $\ld \in [0, 1]$.
The goal of the SW transformation is to find an skew-Hermitian generator $\Sld = - \Sld^\dg$ that unitarily rotates the Hamiltonian into a block-diagonal form, thereby decoupling the two subspaces:
%%%%%%%%%%
\ba
\Htd = e^{-\Sld} \Hld e^{\Sld}
=
\bpm
\td{\mc{H}}^{eff}_{\ld} & 0 \\
0 & *
\epm
,
\quad
\Sld =
\bpm
0 & \Tld \\
-\Tld^\dg & 0
\epm
.
\label{seq:SW}
\ea
%%%%%%%%%%
Here, $\Sld$ is a smooth function of $\ld$ satisfying $S(0) = 0$.
Since this transformation is continuous with respect to the tuning parameter $\ld$, the spectrum of the upper-left block $\td{\mc{H}}^{eff}_{\ld}$ is adiabatically connected to the unperturbed target spectrum $\ein$.
This explicitly guarantees that $\td{\mc{H}}^{eff}_{\ld}$ captures the true physical evolution of the target subspace, preventing the two distinct energetic sectors from mixing as the perturbation is introduced.

Let the eigenvectors of the block-diagonalized Hamiltonian in the target subspace be denoted by $\td{\bv}_n(\ld)$, satisfying $\Htd \td{\bv}_n(\ld) = E_n(\ld) \td{\bv}_n(\ld)$.
Because $\Htd$ is perfectly block-diagonal, the projector onto this subspace in the rotated frame is simply the unperturbed projector
%%%%%%%%%%
\ba
\td{P}(\ld) = \sum_n \td{\bv}_n(\ld) \td{\bv}_n^\dg(\ld) =
\bpm
\id & 0 \\
0 & 0
\epm
\equiv P_0.
\ea
%%%%%%%%%%
The true physical eigenvectors of the original Hamiltonian are given by applying the inverse rotation, $\bv_n(\ld) = e^{\Sld} \td{\bv}_n(\ld)$.
Consequently, the true ground state projector evolves according to an inverse SW transformation,
%%%%%%%%%%
\ba
P(\ld) = e^{\Sld} P_0 e^{-\Sld}.
\label{seq:projpert}
\ea
%%%%%%%%%%
Thus, determining the generator $\Sld$ immediately yields the perturbative expansion for the projector.
To find $\Sld$, we apply the Baker-Campbell-Hausdorff (BCH) lemma to expand $\Htd$
%%%%%%%%%%
\ba
\Htd = \Hld + [\Hld, \Sld] + \frac{1}{2!} [[\Hld, \Sld], \Sld] + \frac{1}{3!} [[[\Hld, \Sld], \Sld], \Sld] + \dots.
\ea
%%%%%%%%%%
We require $\Htd$ to be block-diagonal, meaning all off-diagonal terms in this expansion must vanish at every order of $\ld$.
Expanding $\Sld = \ld S_1 + \ld^2 S_2 + \ld^3 S_3 + \dots$, substituting $\Hld = \Ho + \ld\vd + \ld\vo$, and collecting the off-diagonal terms order by order yields the fundamental conditions for the SW generators
%%%%%%%%%%
\ba
& [\Ho, S_1] + \vo = 0, \nn
& [\Ho, S_2] + [\vd, S_1] = 0, \nn
& [\Ho, S_3] + [\vd, S_2] + \hf [[\vo, S_1], S_1] + \frac{1}{6} [[[\Ho, S_1],S_1],S_1] = 0.
\label{seq:swCond}
\ea
%%%%%%%%%%
These conditions define $\Sld$ up to third order in the perturbation.
By utilizing the analogous BCH expansion for the projector in \eq{seq:projpert},
%%%%%%%%%%
\ba
P(\ld) = P_0 + [\Sld, P_0] + \hf [[\Sld, P_0], \Sld] + \frac{1}{6} [[[\Sld, P_0], \Sld], \Sld] + \dots,
\ea
%%%%%%%%%%
and parameterizing the block-off-diagonal generator as $\Tld = \ld T_1 + \ld^2 T_2 + \ld^3 T_3 + \dots$ [see \eq{seq:SW}], we obtain the explicit perturbation series $P(\ld) = P_0 + \ld P_1 + \ld^2 P_2 + \ld^3 P_3 + \mc{O}(\ld^4)$.
Evaluating the commutators yields the matrix elements in block form
%%%%%%%%%%
\ba
P_0 =
\bpm
\id & 0 \\
0 & 0
\epm
, \quad
P_1 = -
\bpm
0 & T_1 \\
T_1^\dg & 0
\epm
, \quad
P_2 =
\bpm
-T_1 T_1^\dg & -T_2 \\
-T_2^\dg & T_1^\dg T_1
\epm
, \quad
P_3 = 
\bpm
-T_1 T_2^\dg - T_2 T_1^\dg & \frac{2}{3} T_1 T_1^\dg T_1 - T_3 \\
\frac{2}{3} T_1^\dg T_1 T_1^\dg - T_3^\dg & T_1^\dg T_2 + T_2^\dg T_1
\epm
.
\label{seq:projterms}
\ea
%%%%%%%%%%

%
We now apply this general formalism to the double exchange model.
We identify the unperturbed Hamiltonian as the local Hund's coupling, $\hat{\mc{H}}_0 = -J \sum_i \bn_i \cdot \bS_i$, and the perturbation as the inter-site hopping, $\ld \hat{V} = \ld \sum_{ij\sg} t_{ij} \cdag_{i\sg} \chat_{j \sg}$.
Since a global energy shift does not alter the eigenstates or the projectors, we subtract the single-particle ground state energy $-J/2$ to set the unperturbed energy of the half-filled target subspace to zero.
Thus, the unperturbed Hamiltonian matrix takes the simplified form $\Ho = \bpm 0 & 0 \\ 0 & J \id \epm$.
Solving the equations in \eq{seq:swCond} with this $\Ho$ yields the generators
%%%%%%%%%%
\ba
T_1 = \frac{1}{J} W, \quad
T_2 = \frac{1}{J^2}(\vin W - W \vout), \quad
T_3 = \frac{1}{J^3} \Big( \vin^2 W + W \vout^2 - 2 \vin W \vout - \frac{4}{3} W \wdag W \Big).
\ea
%%%%%%%%%%
Finally, inserting these $T_n$ matrices into \eq{seq:projterms}, we obtain the analytic perturbative expansion for the ground state projectors in the strong-coupling limit
%%%%%%%%%%
\ba
P_0 &=
\bpm
\id & 0 \\
0 & 0
\epm
, \quad
P_1 = -\frac{1}{J}
\bpm
0 & W \\
\wdag & 0
\epm
, \quad
P_2 = \frac{1}{J^2}
\bpm
-W \wdag & W \vout - \vin W \\
\vout \wdag - \wdag \vin & \wdag W
\epm
, \nn
P_3 &= \frac{1}{J^3}
\bpm
2W \vout \wdag - W \wdag \vin - \vin W \wdag & 2 W \wdag W - \vin^2 W - W \vout^2 + 2 \vin W \vout \\
2 \wdag W \wdag - \wdag \vin^2 - \vout^2 \wdag + 2 \vout \wdag \vin & 2 \wdag \vin W - \wdag W \vout - \vout \wdag W
\epm
.
\label{seq:projres}
\ea
%%%%%%%%%%
In the next section, we will utilize these explicit projector representations to evaluate the analytic form of the emergent gauge field order by order.

%%%%%%%%%%%%%%%%%%%%%%
\subsection{Calculation of the emergent gauge field}
\label{app:gaugecalc}
%%%%%%%%%%%%%%%%%%%%%%

To obtain the perturbative expansion for the emergent gauge field, we insert the projector series $P(\ld) = P_0 + \ld P_1 + \ld^2 P_2 + \ld^3 P_3 + \mc{O}(\ld^4)$ into the geometric formula $\fab = -2\imtr{P \dera P \derb P}$ derived in \eq{seq:fproj}.
Collecting the terms in powers of $\ld$, we find
%%%%%%%%%%
\ba
\fab
=& -2 Im\Big[Tr\big[
(P_0 + \ld P_1 + \ld^2 P_2)
(\ld \dera P_1 + \ld^2 \dera P_2 + \ld^3 \dera P_3)
(\ld \derb P_1 + \ld^2 \derb P_2 + \ld^3 \derb P_3)
\big]\Big]
\nn
=& -2 Im\Big[Tr\big[
\ld^2 P_0 \dera P_1 \derb P_1
+ \ld^3(
P_0 \dera P_1 \derb P_2 + P_0 \dera P_2 \derb P_1 + P_1 \dera P_1 \derb P_1
)
\nn
&+ \ld^4(
P_0 \dera P_1 \derb P_3 + P_0 \dera P_2 \derb P_2 + P_0 \dera P_3 \derb P_1 + P_1 \dera P_1 \derb P_2 + P_1 \dera P_2 \derb P_1 + P_2 \dera P_1 \derb P_1
)
\big]\Big]
\nn
=& \ld^2 [\fab]^{(2)} + \ld^3 [\fab]^{(3)} + \ld^4 [\fab]^{(4)}.
\label{seq:fpert}
\ea
%%%%%%%%%%
Note that the perturbation series naturally starts at second order ($\dera P_0 = 0$).
This is because the spatial dependence of the electronic Hamiltonian is encapsulated entirely within the hopping perturbation; the unperturbed projector $P_0$ is determined strictly by the local magnetization axes $\bn_i$, rendering it independent of the nuclear coordinates $\rion$.
Inserting the explicit projector terms from \eq{seq:projres} into this expression yields the formal perturbation series.
By utilizing the idempotence of $P_0$, the cyclicity of the trace, and the fundamental properties of block matrices—specifically that the product of a block-diagonal and a block-off-diagonal matrix is block-off-diagonal (meaning its trace vanishes), while the product of two block-off-diagonal matrices is block-diagonal—we can systematically extract the non-zero contributions for the emergent field:
%%%%%%%%%%
\ba
[\fab]^{(2)} = -\frac{2}{J^2} \imtr{\dera W \derb \wdag},
\label{seq:f2}
\ea
%%%%%%%%%%
%
%%%%%%%%%%
\ba
[\fab]^{(3)} = \frac{1}{J^3} \imtr{
	&(2 \dera \vout \derb \wdag W + 2 \dera W \derb \vout \wdag - 4 \dera W \derb \wdag \vin)
	\nn
	&+ (-2\derb \vin \dera W \wdag -2 \derb \wdag \dera \vin W +4 \derb \wdag \dera W \vout)
},
\label{seq:f3}
\ea
%%%%%%%%%%
%
%%%%%%%%%%
\ba
[\fab]^{(4)} = \frac{1}{J^4} \imtr{
	(& -2 \dera \vout \derb \vout \wdag W - 2 \dera \vout \derb \wdag W \vout - 2 \dera W \derb \vout \vout \wdag
	\nn
	& + 6 \dera \vout \derb \wdag \vin W + 6 \dera W \derb \vout \wdag \vin + 6 \dera W \derb \wdag W \wdag - 6 \dera W \derb \wdag \vin^2
	)
	\nn
	+ (&
	-2 \derb \vin \dera \vin W \wdag -2 \derb \vin \dera W \wdag \vin -2 \derb \wdag \dera \vin \vin W
	\nn
	& +6 \derb \vin \dera W \vout \wdag + 6 \derb \wdag  \dera \vin W \vout + 6 \derb \wdag \dera W \wdag W -6 \derb \wdag \dera W \vout^2
	)
	\nn
	+ (&
	2 \dera\vin W \derb\vout \wdag+2 \dera\vout \wdag \derb\vin W +12 \dera W \vout \derb\wdag \vin
	\nn
	&-4 \dera\vin W \derb\wdag \vin -4 \dera\vout \vout \derb\wdag W -4 \dera W \vout \derb\vout \wdag -4 \dera W \wdag \derb\vin \vin
	\nn
	&+2 \dera W \wdag \derb W \wdag + 2 \dera\wdag W \derb\wdag W
	)
}.
\label{seq:f4}
\ea
%%%%%%%%%%
We can slightly simplify this result by noting that the last line of \eq{seq:f4} identically vanishes.
This follows directly from the property $\imtr{X}=-\imtr{X^\dg}$; by taking the Hermitian conjugate and applying the cyclicity of the trace, we find
%%%%%%%%%%
\ba
\imtr{\dera \wdag W \derb \wdag W}
= - \imtr{\wdag \derb W \wdag \dera W}
= -\imtr{\dera W \wdag \derb W \wdag}.
\ea
%%%%%%%%%%

%
Equations \eqref{seq:f2}--\eqref{seq:f4} represent the formal perturbative expressions of the emergent gauge field for any half-filled double exchange model in the strong Hund's coupling limit.
To evaluate these traces using the physical hopping amplitudes and local magnetization axes $\bn_i$, we construct the single-particle basis of the unperturbed Hamiltonian $\hat{\mc{H}}_0 = -J \sum_i \bn_i \cdot \bS_i$ as
%%%%%%%%%%
\ba
\hat{\mc{H}}_0 \ket{i} \otimes \ket{i \pm} = \mp \frac{J}{2} \ket{i} \otimes \ket{i \pm},
\ea
%%%%%%%%%%
where $\ket i$ is the spatial wave function localized at site $i$, and $\ket{i +}$ (or $\ket{i -}$) are two-component spinors aligned parallel (or antiparallel) to the local magnetization $\bn_i$.
For analytical convenience, we fix the spin gauge freedom with the convention
%%%%%%%%%%
\ba
\ket{i+} = \hat U^z(\vph_i) \hat U^y (\vth_i) \ket{\ua},
\quad
\ket{i-} = \hat U^z(\vph_i) \hat U^y (\vth_i) \ket{\da},
\quad (\hat U^\mu (\theta) = e^{-i \theta \shat^\mu}),
\label{seq:spingauge}
\ea
%%%%%%%%%%
where $(\vth_i, \vph_i)$ are the spherical angles of $\bn_i$, and $\hat S^z (\ket{\ua}, \ket \da) = (\hf \ket \ua, -\hf \ket \da)$.
Upon introducing this basis, the elements of the block matrices from \eq{seq:projterms} are explicitly specified as
%%%%%%%%%%
\ba
[\vin]_{ij} = t_{ij} \brk{i+}{j+}, \quad
[\vout]_{ij} = t_{ij} \brk{i-}{j-}, \quad
[W]_{ij} = t_{ij} \brk{i+}{j-}, \quad
[\wdag]_{ij} = t_{ij} \brk{i-}{j+}.
\label{seq:entries}
\ea
%%%%%%%%%%

Since the emergent gauge field $\fem$ is constructed from spatial derivatives of the projector, we must explicitly specify how the underlying hopping matrices respond to nuclear displacements.
We assume that near the equilibrium lattice configuration, the hopping amplitudes depend on the relative displacement between the orbitals: $t_{ij}(\rion) = t_{ij}(\br_i - \br_j) = t_{ji}^*(\br_j-\br_i)$.
To prevent index collisions with the internal dummy variables $(i,j,k,l)$ used in the subsequent trace expansions, we refine our external composite index notation to $a = (i_a, \mu_a)$, where $i_a$ explicitly denotes the fixed site and $\mu_a$ the spatial direction.
The derivative of the hopping amplitude evaluated at the equilibrium position is then
%%%%%%%%%%
\ba
\Delta_{ij}^{a} \equiv \dera t_{ij}(\rion^{eq}) = (\delta_{i_a, i} - \delta_{i_a, j}) \der_{\mu_a} t_{ij} (\br_i^{eq} - \br_j^{eq}),
\label{seq:der}
\ea
%%%%%%%%%%
where we have $\Delta_{ij}^a = (\Delta_{ji}^a)^*$ by definition.
Using this formalism, the remainder of this section is dedicated to proving the final analytic identities presented in the main text
%%%%%%%%%%
\ba
[\fab]^{(2)} =& 0,
\nn
[\fab]^{(3)} =& \frac{2}{J^3} \sum_{ijk} Im[\Delta_{ij}^a \Delta_{jk}^b t_{ki}]  (\bn_i \cdot \bn_j + \bn_j \cdot \bn_k -2 \bn_k \cdot \bn_i),
\nn
[\fab]^{(4)} =& \frac{20}{J^4}
\Big[
\sum_{ijkl} Re[\Delta_{ij}^a \Delta_{jk}^b t_{kl} t_{li}]  (\chi_{ijk}+\chi_{jkl}+\chi_{lij}-3\chi_{kli})
\nn
&+ 2 \sum_{ij} Re[\Delta_{ai}^a t_{ib} \Delta_{bj}^b t_{ja} - \Delta_{ai}^a t_{ij} \Delta_{jb}^b t_{ba}] 
(\chi_{aib}+\chi_{ibj}-\chi_{bja}-\chi_{jai})
\Big],
\label{seq:ftotal}
\ea
%%%%%%%%%%
where $\chi_{ijk} = \expv{\bS_i}_0 \cdot (\expv{\bS_j}_0 \times \expv{\bS_k}_0) = \frac{1}{8} \bn_i \cdot (\bn_j \times \bn_k)$ is the \textit{scalar spin chirality} of the spins at sites $i, j,$ and $k$~\cite{lee1992gauge}. 
Note that when the composite indices $a$ or $b$ appear in the subscript of the scalar spin chirality (e.g., $\chi_{aib}$), they refer strictly to the site components $i_a$ or $i_b$ of those indices.
Here, $\expv{\hat O}_0$ denotes the expectation value of $\hat O$ evaluated in the ground state of the unperturbed Hamiltonian $\hat{\mc{H}}_0$.

\paragraph*{Spinor identities.---}
Before directly evaluating the traces, we present a set of fundamental spinor identities necessary for simplifying successive products of spinor wave functions. 
First, based on our explicit gauge choice in \eq{seq:spingauge}, the inner products obey the conjugation relations
%%%%%%%%%%
\ba
\brk{i+}{j+}^* = \brk{i-}{j-},
\quad
\brk{i+}{j-}^* = -\brk{i-}{j+}.
\label{seq:conj}
\ea
%%%%%%%%%%
This can be verified by direct evaluation:
%%%%%%%%%%
\ba
& \brk {i+}{j+} = 
\ijp \cos \hthi \cos \hthj + \ijm \sin \hthi \sin \hthj,
\nn
&
\brk{i-}{j-} = \ijp \sin \hthi \sin \hthj + \ijm \cos \hthi \cos \hthj,
\nn
& \brk {i+}{j-} = \ijm \sin \hthi \cos \hthj - \ijp \cos \hthi \sin \hthj,
\nn
& \brk{i-}{j+} = -\ijp \sin \hthi \cos \hthj + \ijm \cos \hthi \sin \hthj.
\ea
%%%%%%%%%%

%
The next class of identities relates closed loops of spinor inner products to combinations of $\bn_i$ and $\chi_{ijk}$ that, in contrast to \eq{seq:conj}, are manifestly gauge-independent~\cite{lee1992gauge}:
%%%%%%%%%%
\ba
& \brk{i+}{j+}\brk{j+}{i+} = \frac{1 + \bn_i \cdot \bn_j}{2},
\nn
& \brk{i+}{j+}\brk{j+}{k+}\brk{k+}{i+} =  \frac{1}{4}(1+\bn_i \cdot \bn_j + \bn_j \cdot \bn_k + \bn_k \cdot \bn_i) + 2i \chi_{ijk},
\nn
& \brkf{i+}{j+}{k+}{l+} = \mathcal{R} + i(\chi_{ijk} + \chi_{jkl} + \chi_{kli} + \chi_{lij}),
\label{seq:spinor}
\ea
%%%%%%%%%%
where $\mathcal{R} \in \mbb{R}$ is a purely real quantity irrelevant to our subsequent calculations.
To rigorously prove these relations, consider the two-dimensional matrix representation $H_i = -\bn_i \cdot \boldsymbol{\sg}$.
The eigenstates of the Hamiltonian operator $\hat{H}_i$ are defined by $\hat H_i \ket{i+} = -\ket{i+}$ and $\hat H_i \ket{i-} = \ket{i-}$.
We can express $\hat{H}_i$ in terms of the ground state projector $\hat{P}_i = \kbr{i+}{i+}$ as $\hat{H}_i = -\hat{P}_i + (\id - \hat{P}_i)$, which upon rearranging the corresponding matrices yields
%%%%%%%%%%
\ba
P_i = \frac{\id - H_i}{2} = \frac{\id + \bn_i \cdot \boldsymbol{\sg}}{2}.
\ea
%%%%%%%%%%
This allows us to recast the closed spinor loops entirely as traces over products of local projectors
%%%%%%%%%%
\ba
Tr[P_i P_j] &= \brk{i+}{j+}\brk{j+}{i+},
\nn
Tr[P_i P_j P_k] &= \brk{i+}{j+}\brk{j+}{k+}\brk{k+}{i+},
\nn
Tr[P_i P_j P_k P_l] &= \brkf{i+}{j+}{k+}{l+}.
\ea
%%%%%%%%%%
Applying the standard Pauli matrix identity $(\bn_i \cdot \boldsymbol{\sg}) (\bn_j \cdot \boldsymbol{\sg}) = (\bn_i \cdot \bn_j) \id + i (\bn_i \times \bn_j) \cdot \boldsymbol{\sg}$ directly proves the relations in \eq{seq:spinor}.
Furthermore, by substituting any $P_i$ with $\id - P_i$, one can show that flipping a state ($\ket{i+} \rightarrow \ket{i-}$) maps identically to inverting the local magnetization ($\bn_i \rightarrow -\bn_i$), which consequently flips the sign of any scalar spin chirality term containing that site ($\chi_{ijk} \rightarrow - \chi_{ijk}$).
Equipped with these tools, we now systematically evaluate the emergent gauge field order by order.

\textit{Second order contribution $[\fab]^{(2)}$.---}
We begin with the second-order term.
Substituting \eq{seq:entries} and \eq{seq:der} into \eq{seq:f2}, we find
%%%%%%%%%%
\ba
[\fab]^{(2)}
&= -\frac{2}{J^2} Im \sum_{ij} \Delta_{ij}^a \Delta_{ji}^b \brk{i+}{j-} \brk{j-}{i+}
\nn
&= -\frac{1}{J^2} Im \sum_{ij} (\Delta_{ij}^a \Delta_{ji}^b \brk{i+}{j-} \brk{j-}{i+}
+ \Delta_{ji}^a \Delta_{ij}^b \brk{j+}{i-} \brk{i-}{j+}).
\ea
%%%%%%%%%%
From $\Delta_{ij}^a = (\Delta_{ji}^a)^*$ and \eq{seq:conj}, we find $[\fab]^{(2)} = -\frac{2}{J^2} Im \sum_{ij} Re[\Delta_{ij}^a \Delta_{ji}^b \brk{i+}{j-} \brk{j-}{i+}] = 0$.
\textit{Third order contribution $[\fab]^{(3)}$.---} 
Utilizing \eq{seq:f3}, \eq{seq:entries}, and \eq{seq:der}, the third order term evaluates to
%%%%%%%%%%
\ba
[\fab]^{(3)} =
\frac{1}{J^3} Im \sum_{ijk}
\big(
\Delta_{ij}^a \Delta_{jk}^b t_{ki} [X_1]_{ijk}
- \Delta_{ij}^b \Delta_{jk}^a t_{ki} 
[X_2]_{ijk}
\big),
\label{seq:f31}
\ea
%%%%%%%%%%
where we have defined the purely spinor-dependent quantities
%%%%%%%%%%
\ba
& [X_1]_{ijk} = 
2 \brk{i-}{j-} \brk{j-}{k+} \brk{k+}{i-} + 2 \brk{i+}{j-} \brk{j-}{k-} \brk{k-}{i+} -4 \brk{i+}{j-} \brk{j-}{k+} \brk{k+}{i+},
\nn
& [X_2]_{ijk} = 
2 \brk{i+}{j+} \brk{j+}{k-} \brk{k-}{i+} +2 \brk{i-}{j+} \brk{j+}{k+} \brk{k+}{i-} -4 \brk{i-}{j+} \brk{j+}{k-} \brk{k-}{i-}.
\ea
%%%%%%%%%%
By interchanging the dummy indices $i \leftrightarrow k$ in the second term of \eq{seq:f31} and using the property $Im[X] = -Im[X^*]$, the summation condenses to
%%%%%%%%%%
\ba
[\fab]^{(3)} &=
\frac{1}{J^3} Im \sum_{ijk}
\big(
\Delta_{ij}^a \Delta_{jk}^b t_{ki}  [X_1]_{ijk}
- \Delta_{kj}^b \Delta_{ji}^a t_{ik}  [X_2]_{kji}
\big)
\nn
&=
\frac{1}{J^3} Im \sum_{ijk}
\Delta_{ij}^a \Delta_{jk}^b t_{ki}  ([X_1]_{ijk}+[X_2]_{kji}^*).
\ea
%%%%%%%%%%
We can relate these two spinor blocks by explicitly taking the complex conjugate of $[X_2]_{kji}$:
%%%%%%%%%%
\ba
[X_2]_{kji}^*
&= 
(2 \brk{k+}{j+}\brk{j+}{i-}\brk{i-}{k+} +2\brk{k-}{j+}\brk{j+}{i+}\brk{i+}{k-} -4\brk{k-}{j+}\brk{j+}{i-}\brk{i-}{k-})^*
\nn
&=
2\brk{i+}{j+}\brk{j+}{k-}\brk{k-}{i+} +2\brk{i-}{j+}\brk{j+}{k+}\brk{k+}{i-} -4\brk{i-}{j+}\brk{j+}{k-}\brk{k-}{i+}
.
\ea
%%%%%%%%%%
Applying the conjugation rules from \eq{seq:conj}, it immediately follows that $[X_2]_{kji}^*=[X_1]_{ijk}^*$.
Thus, extracting the imaginary part yields
%%%%%%%%%%
\ba
[\fab]^{(3)} =& \frac{2}{J^3} \sum_{ijk} Im[\Delta_{ij}^a \Delta_{jk}^b t_{ki}] Re[X_1]_{ijk}.
\ea
%%%%%%%%%%
To resolve $Re[X_1]_{ijk}$, we deploy the second identity in \eq{seq:spinor}, recognizing that flipping the spinors introduces the appropriate minus signs into the scalar products:
%%%%%%%%%%
\ba
Re[X_1]_{ijk}
=&
\frac{1}{4}
\big\{
2(1+\bn_i \cdot \bn_j - \bn_j \cdot \bn_k - \bn_k \cdot \bn_i)
+ 2(1-\bn_i \cdot \bn_j +\bn_j \cdot \bn_k - \bn_k \cdot \bn_i)
\nn
&-4 (1 - \bn_i \cdot \bn_j - \bn_j \cdot \bn_k + \bn_k \cdot \bn_i)
\big\}
\nn
=& \bn_i \cdot \bn_j + \bn_j \cdot \bn_k -2 \bn_k \cdot \bn_i.
\ea
%%%%%%%%%%
This completes the proof for the third order contribution.

\textit{Fourth order contribution $[\fab]^{(4)}$.---}
Finally, we evaluate the fourth-order perturbation. Dropping the vanishing last line of \eq{seq:f4} and substituting \eq{seq:entries} and \eq{seq:der}, we obtain the extensive summation
%%%%%%%%%%
\ba
[\fab]^{(4)} = \frac{1}{J^4} Im \sum_{ijkl} (
\Delta_{ij}^a \Delta_{jk}^b t_{kl} t_{li} [Y_1]_{ijkl}
+ \Delta_{ij}^b \Delta_{jk}^a t_{kl} t_{li} [Y_2]_{ijkl} + \Delta_{ij}^a t_{jk} \Delta_{kl}^b t_{li} Z_{ijkl}
).
\label{seq:f41}
\ea
%%%%%%%%%%
Here, the four-site spinor terms are defined as
%%%%%%%%%%
\ba
[Y_1]_{ijkl} =&
-2 \brkf{i-}{j-}{k-}{l+} -2 \brkf{i-}{j-}{k+}{l-}
\nn & -2\brkf{i+}{j-}{k-}{l-}
+ 6\brkf{i-}{j-}{k+}{l+}
\nn
&+ 6\brkf{i+}{j-}{k-}{l+} + 6 \brkf{i+}{j-}{k+}{l-}
\nn
&-6 \brkf{i+}{j-}{k+}{l+},
\nn
[Y_2]_{ijkl} =&
-2 \brkf{i+}{j+}{k+}{l-} -2 \brkf{i+}{j+}{k-}{l+}
\nn
&-2 \brkf{i-}{j+}{k+}{l+} + 6 \brkf{i+}{j+}{k-}{l-}
\nn
&+ 6\brkf{i-}{j+}{k+}{l-} + 6\brkf{i-}{j+}{k-}{l+}
\nn
& -6 \brkf{i-}{j+}{k-}{l-},
\nn
Z_{ijkl} =&
~2 \brkf{i+}{j+}{k-}{l-} +2 \brkf{i-}{j-}{k+}{l+}
\nn
& +12 \brkf{i+}{j-}{k-}{l+} -4 \brkf{i+}{j+}{k-}{l+}
\nn
& -4 \brkf{i-}{j-}{k-}{l+} -4 \brkf{i+}{j-}{k-}{l-}
\nn
& -4 \brkf{i+}{j-}{k+}{l+}.
\label{seq:yz}
\ea
%%%%%%%%%%
We follow a similar symmetry strategy to simplify the first two terms of \eq{seq:f41}.
By interchanging the indices $i \leftrightarrow k$ and applying $Im[X] = -Im[X^*]$, we can group the $Y_1$ and $Y_2$ terms:
%%%%%%%%%%
\ba
& \frac{1}{J^4} Im \sum_{ijkl}
(
\Delta_{ij}^a \Delta_{jk}^b t_{kl} t_{li} [Y_1]_{ijkl}
+ \Delta_{kj}^b \Delta_{ji}^a t_{il} t_{lk} [Y_2]_{kjil})
\nn
=& \frac{1}{J^4} Im \sum_{ijkl} 
\big\{
(\Delta_{ij}^a \Delta_{jk}^b t_{kl} t_{li}) [Y_1]_{ijkl}
+ (\Delta_{ij}^a \Delta_{jk}^b t_{kl} t_{li})^* [Y_2]_{kjil}
\big\} \nn
=& \frac{1}{J^4} Im \sum_{ijkl}
\Delta_{ij}^a \Delta_{jk}^b t_{kl} t_{li} ([Y_1]_{ijkl} - [Y_2]_{kjil}^*).
\label{seq:f42}
\ea
%%%%%%%%%%
Upon rearranging the internal products and explicitly evaluating the complex conjugate of $[Y_2]$, we find
%%%%%%%%%%
\ba
[Y_2]_{kjil}^*=
&
(
-2 \brkf{k+}{j+}{i+}{l-} -2\brkf{k+}{j+}{i-}{l+}\nn
& -2\brkf{k-}{j+}{i+}{l+} +6 \brkf{k+}{j+}{i-}{l-}\nn 
& +6 \brkf{k-}{j+}{i+}{l-} +6 \brkf{k-}{j+}{i-}{l+}\nn
& -6 \brkf{k-}{j+}{i-}{l-}
)^* \nn
=&
( -2 \brk{j+}{i+}\brk{k+}{j+}\brk{l-}{k+}\brk{i+}{l-} -2 \brk{j+}{i-}\brk{k+}{j+}\brk{l+}{k+}\brk{i-}{l+} \nn
& -2 \brk{j+}{i+}\brk{k-}{j+}\brk{l+}{k-}\brk{i+}{l+} +6 \brk{j+}{i-}\brk{k+}{j+}\brk{l-}{k+}\brk{i-}{l-} \nn
& +6 \brk{j+}{i+}\brk{k-}{j+}\brk{l-}{k-}\brk{i+}{l-} +6 \brk{j+}{i-}\brk{k-}{j+}\brk{l+}{k-}\brk{i-}{l+} \nn
& -6 \brk{j+}{i-}\brk{k-}{j+}\brk{l-}{k-}\brk{i-}{l-})^*
\nn
=&
-2 \brkf{i+}{j+}{k+}{l-} -2\brkf{i-}{j+}{k+}{l+} \nn
& -2 \brkf{i+}{j+}{k-}{l+} +6 \brkf{i-}{j+}{k+}{l-} \nn
& +6 \brkf{i+}{j+}{k-}{l-} +6 \brkf{i-}{j+}{k-}{l+} \nn
& -6 \brkf{i-}{j+}{k-}{l-}
.
\ea
%%%%%%%%%%
Applying \eq{seq:conj}, one finds that $[Y_2]_{kjil}^*=[Y_1]_{ijkl}^*$.
Thus, \eq{seq:f42} reduces to
%%%%%%%%%%
\ba
\frac{2}{J^4} \sum_{ijkl} Re[\Delta_{ij}^a \Delta_{jk}^b t_{kl} t_{li}] Im [Y_1]_{ijkl}.
\label{seq:f4y}
\ea
%%%%%%%%%%

To systematically reduce the $Z$-dependent term in \eq{seq:f41}, we leverage the spatial connectivity restrictions embedded in \eq{seq:der}.
Since the derivatives strictly pin specific indices, the summation simplifies to
%%%%%%%%%%
\ba
& \frac{1}{J^4} Im \sum_{ij} 
(
\dtdt{a}{i}{b}{j} Z_{aibj} + \dtdt{a}{i}{j}{b} Z_{aijb} + \dtdt{i}{a}{b}{j} Z_{iabj} + \dtdt{i}{a}{j}{b} Z_{iajb}
)
\nn
=& \frac{1}{J^4} Im \sum_{ij} 
\Big[
\big\{\dtdt{a}{i}{b}{j} Z_{aibj} + (\dtdt{a}{i}{b}{j})^* Z_{iajb} \big\}
+
\big\{\dtdt{a}{i}{j}{b} Z_{aijb} + (\dtdt{a}{i}{j}{b})^* Z_{iabj} \big\}
\Big].
\label{seq:f43}
\ea
%%%%%%%%%%
Upon permuting the indices, it can be shown that
%%%%%%%%%%
\ba
Z_{jilk} =& ~2 \brkf{j+}{i+}{l-}{k-} +2 \brkf{j-}{i-}{l+}{k+} \nn
&+12 \brkf{j+}{i-}{l-}{k+} -4 \brkf{j+}{i+}{l-}{k+} \nn
&-4 \brkf{j-}{i-}{l-}{k+} -4 \brkf{j+}{i-}{l-}{k-} \nn
&-4 \brkf{j+}{i-}{l+}{k+}
\nn
=& ~2 \brk{j+}{i+}\brk{k-}{j+}\brk{l-}{k-}\brk{i+}{l-} +2 \brk{j-}{i-}\brk{k+}{j-}\brk{l+}{k+}\brk{i-}{l+} \nn
&+12 \brk{j+}{i-}\brk{k+}{j+}\brk{l-}{k+}\brk{k-}{l-} -4 \brk{j+}{i+}\brk{k+}{j+}\brk{l-}{k+}\brk{i+}{l-} \nn
&-4 \brk{j-}{i-}\brk{k+}{j-}\brk{l-}{k+}\brk{i-}{l-} -4 \brk{j+}{i-}\brk{k-}{j+}\brk{l-}{k-}\brk{i-}{l-} \nn
&-4 \brk{j+}{i-}\brk{k+}{j+}\brk{l+}{k+}\brk{i-}{l+}
\nn
=&
(
2 \brkf{i+}{j+}{k-}{l-} +2 \brkf{i-}{j-}{k+}{l+} \nn
&+12 \brkf{i-}{j+}{k+}{l-} -4 \brkf{i+}{j+}{k+}{l-} \nn
& -4 \brkf{i-}{j-}{k+}{l-} -4 \brkf{i-}{j+}{k-}{l-} \nn
& -4 \brkf{i-}{j+}{k+}{l+}
)
.
\ea
%%%%%%%%%%
%
Applying \eq{seq:conj}, one finds $Z_{jilk}= Z_{ijkl}$.
Therefore, \eq{seq:f43} reduces to
%%%%%%%%%%
\ba
\frac{2}{J^4} \sum_{ij} 
(
Re[\dtdt{a}{i}{b}{j}] Im Z_{aibj}
+ Re[\dtdt{a}{i}{j}{b}] Im Z_{aijb}
).
\ea
%%%%%%%%%%

We now possess an intermediate expression for the fourth order perturbation
%%%%%%%%%%
\ba
[\fab]^{(4)} =&
\frac{2}{J^4} \sum_{ijkl} Re[\Delta_{ij}^a \Delta_{jk}^b t_{kl} t_{li}] Im [Y_1]_{ijkl}
+ \frac{2}{J^4} \sum_{ij} 
(
Re[\dtdt{a}{i}{b}{j}] Im Z_{aibj}
+ Re[\dtdt{a}{i}{j}{b}] Im Z_{aijb}
).
\label{seq:f4int}
\ea
%%%%%%%%%%
By mapping the four-site spinor identities (\eq{seq:spinor}) onto $Y_1$, the imaginary part reduces to a sum of scalar spin chiralities:
%%%%%%%%%%
\ba
Im[Y_1]_{ijkl}
=&
-2 \chif{-}{+}{+}{+} -2\chif{}{+}{+}{-} -2 \chif{}{-}{+}{+} \nn
&+ 6 \chif{}{-}{-}{+} +6 \chif{}{+}{-}{-} +6 \chif{-}{+}{-}{+} \nn
&- 6 \chif{-}{-}{+}{-}
\nn
=& 10\chif{}{+}{-3}{+}.
\label{seq:imy}
\ea
%%%%%%%%%%
Similarly, the first two terms in $Z_{ijkl}$ sum to a purely real value, resulting in
%%%%%%%%%%
\ba
Im[Z]_{ijkl} =&
12 \chif{}{+}{-}{-} -4 \chif{-}{-}{-}{+} -4 \chif{-}{+}{+}{+}
\nn
&-4 \chif{}{-}{+}{+} -4 \chif{-}{-}{+}{-}
\nn
=& 20 \chif{}{+}{-}{-}.
\label{seq:imz}
\ea
%%%%%%%%%%
Substituting these quantities back into \eq{seq:f4int} gives
%%%%%%%%%%
\ba
[\fab]^{(4)} &= \frac{20}{J^4}
\Big[
\sum_{ijkl} Re[\Delta_{ij}^a \Delta_{jk}^b t_{kl} t_{li}] (\chi_{ijk}+\chi_{jkl}+\chi_{lij}-3\chi_{kli})
\nn
&+ \sum_{ij} Re[\dtdt{a}{i}{b}{j}](\chi_{aib}+\chi_{ibj}-\chi_{bja}-\chi_{jai})
+ \sum_{ij} Re[\dtdt{a}{i}{j}{b}] (\chi_{aij}+\chi_{ijb}-\chi_{jba}-\chi_{bai})
\Big].
\ea
%%%%%%%%%%
Finally, by exploiting the total antisymmetry of the scalar spin chirality upon interchanging indices, we relate the spinor part of the final two terms:
%%%%%%%%%%
\ba
(\chi_{aij}+\chi_{ijb}-\chi_{jba}-\chi_{bai})
= -(\chi_{aib}+\chi_{ibj}-\chi_{bja}-\chi_{jai}),
\ea
%%%%%%%%%%
which definitively yields \eq{seq:ftotal} and completes the derivation of the analytic emergent gauge field.

%%%%%%%%%%%%%%%%%%%%%%
\section{Dynamical and geometric properties of the phonon system}
\label{app:phonon}
%%%%%%%%%%%%%%%%%%%%%%

In this Appendix, we provide a rigorous mathematical formulation of the phonon system in the presence of the emergent gauge field.
In \Sec{app:phonondef}, we establish the lattice notation, define the coupled phonon Hamiltonian, and explore its underlying momentum space algebraic structure.
Next, in \Sec{app:diag}, we detail the diagonalization procedure for this system, specifically addressing the non-Hermitian nature of the coupled equations of motion.
Finally, in \Sec{app:phononbc}, we construct the phonon Berry curvature using a generalized multiband projector formalism.
This approach naturally accommodates band degeneracies and provides an efficient framework for numerical evaluation.

%%%%%%%%%%%%%%%%%%%%%%
\subsection{General definition and algebraic structure}
\label{app:phonondef}
%%%%%%%%%%%%%%%%%%%%%%
To define the phonon Hamiltonian, we consider a $d$-dimensional ($d$-D) lattice with discrete translational symmetry, subject to periodic boundary conditions (PBC).
We index the unit cells by the lattice vectors $\bR = \sum_{i=1}^{d} n_i \bba_i~(n_i \in \mbb{Z}_{N_i})$, where $\bba_i$ are the primitive lattice vectors, and the integers $N_i$ define the periodic boundaries of the crystal.
The total number of unit cells is $\nc = \prod_{i=1}^d N_i$.
This structure allows us to uniquely specify any nuclear site using the composite index $(\bR, \al)$, where $\al = 1, \dots, \ns$ denotes the sublattice index within the unit cell.
The equilibrium position of a nucleus is given by $\bR + \bt_\al$, where $\bt_\al$ is the vector connecting the unit cell origin to the sublattice site.
Starting from the effective Born-Oppenheimer dynamics in \eq{seq:bo}, the phonon Hamiltonian $\Hp$ is obtained by expanding the potential energy $\Vb(\rion)$ up to quadratic order in the nuclear displacements $\bb{u}_{\bR \al}$ around their equilibrium positions, and subtracting the constant equilibrium energy $\Vb(\rion^{eq})$.
To simplify the resulting expressions, we absorb the nuclear masses $M_{\al}$ by rescaling the operators.
We redefine the displacement $\bu_{\bR \al} \to \bu_{\bR \al} / \sqrt{M_\al}$, the kinematic momentum $\bpi_{\bR \al} \to \sqrt{M_\al} \bpi_{\bR \al}$, the gauge connection $\bb{A}_{\bR \al} \to \sqrt{M_\al} \bb{A}_{\bR \al}$, and the gauge field $F_{\bR \al \mu, \bR' \be \nu} \to \sqrt{M_\al M_\be} F_{\bR \al \mu, \bR' \be \nu}$.
Upon such transformation, the harmonic phonon Hamiltonian takes the quadratic form
%%%%%%%%%%
\ba
\Hp = \hf \sum_{\bR \al} \bpi_{\bR \al}^2
+ \hf \sum_{\bR \bR' \al \be \mu \nu} \uhat_{\bR \al \mu} D(\bR - \bR')_{\al \mu, \be \nu}  \uhat_{\bR' \be \nu}.
\ea
%%%%%%%%%%
The matrix $D$ is the dynamical matrix, representing the mass-rescaled second-order spatial derivatives of the Born-Oppenheimer potential.
By definition, the dynamical matrix obeys the symmetries $D(\bR)_{\al \mu, \be \nu} = D(\bR)_{\al \nu, \be \mu} = D(-\bR)_{\be \mu, \al \nu}$.
The inclusion of the gauge potential fundamentally alters the underlying quantum algebra of the lattice operators.
Since the gauge potential depends on the nuclear coordinates, the kinematic momenta do not commute.
Specifically, the operators obey the following modified commutation relations
%%%%%%%%%%
\ba
[\uhat_{\bR \al \mu}, \uhat_{\bR' \be \nu}] = 0, \quad
[\pihat_{\bR \al \mu}, \uhat_{\bR' \be \nu}] = -i \delta_{\bR \bR'} \delta_{\al \be} \delta_{\mu \nu}, \quad
[\pihat_{\bR \al \mu}, \pihat_{\bR' \be \nu}] = i F_{\bR \al \mu,\bR' \be \nu}.
\ea
%%%%%%%%%%
By definition, the gauge field is antisymmetric upon simultaneously interchanging all the indices $F_{\bR \al \mu, \bR' \be \nu} = -F_{\bR' \be \nu, \bR \al \mu}$.
Since we consider spatially uniform external magnetic fields and a periodic single-particle electronic Hamiltonian obeying \eq{seq:der}, the total gauge field inherently respects discrete lattice translational symmetry at the equilibrium position [see \Sec{app:esym} for a proof].
Consequently, $F$ depends only on the relative distance between unit cells: $F_{\bR \al \mu, \bR' \be \nu} = F(\bR - \bR')_{\al \mu, \be \nu}$.
To leverage this discrete translational symmetry, we express the Hamiltonian in the momentum space basis.
We define the reciprocal lattice vectors $\bb{b}_j$ such that $\bba_i \cdot \bb{b}_j = 2\pi \delta_{ij}$.
The crystal momentum is defined as $\bk = \sum_{i=1}^d k_i \bb{b}_i = \sum_{\mu} k_\mu \hat{\bb{e}}_{\mu}$, with $k_i = n_i/N_i~(n_i \in \mbb{Z}_{N_i})$ restricted to the first unit cell. 
The discrete spatial Fourier transforms of the operators are defined as
%%%%%%%%%%
\ba
\bpi_{\bR \al}
&=
\frac{1}{\sqrt{\nc}} \sum_{\bk} e^{i \bk \cdot (\bR + \bt_{\al})}
\bpi_{\bk \al},
\quad
\bpi_{\bk \al}
=
\frac{1}{\sqrt{\nc}} \sum_{\bR} e^{-i \bk \cdot (\bR + \bt_{\al})}
\bpi_{\bR \al},
\nn
\bu_{\bR \al}
&=
\frac{1}{\sqrt{\nc}} \sum_{\bk} e^{i \bk \cdot (\bR + \bt_{\al})}
\bu_{\bk \al},
\quad
\bu_{\bk \al}
=
\frac{1}{\sqrt{\nc}} \sum_{\bR} e^{-i \bk \cdot (\bR + \bt_{\al})}
\bu_{\bR \al},
\label{seq:fourier}
\ea
%%%%%%%%%%
where the phase factors explicitly account for the intra-cell sublattice positions $\bt_\al$.
Substituting these Fourier expansions into the real-space equations yields
%%%%%%%%%%
\ba
\Hp = \hf \sum_{\bk} \bx_{-\bk} h(\bk) \bx_{\bk},
\quad
h(\bk) = 
\bpm
\id & 0 \\
0 & D(\bk)
\epm,
\label{seq:hphonon}
\ea
%%%%%%%%%%
where we have grouped the operators into a single multi-component vector $\bx_{\bk} = (\bpi_{\bk, 1}, \dots, \bpi_{\bk, \ns}, \bu_{\bk, 1}, \dots, \bu_{\bk, \ns})$, which inherently satisfies $\bx_{-\bk} = \bx_{\bk}^{\dg}$.
The momentum-space dynamical matrix is defined as $D(\bk)_{\al \mu, \be \nu} = \sum_{\bR} e^{-i \bk \cdot (\bR + \bt_{\al} - \bt_{\be})} D(\bR)_{\al \mu, \be \nu}$.
By applying the Fourier transformations to the real-space commutation relations and utilizing the translational invariance of $F$, we derive the corresponding momentum-space commutation relations.
These operators obey the compact matrix equation presented in the main text:
%%%%%%%%%%
\ba
[\xhat_{\bk p}, \xhat_{\bk' q}] 
= i \delta_{\bk, -\bk'} \cf (\bk)_{pq},
\quad
\cf(\bk) = 
\bpm
F(\bk) & - \id \\
\id & 0
\epm
,
\quad
F(\bk)_{\al \mu, \be \nu} = \sum_{\bR} e^{-i \bk \cdot (\bR + \bs{\tau}_{\al} - \bs{\tau}_{\be})} F(\bR)_{\al \mu, \be \nu}
\label{seq:comm}
\ea
%%%%%%%%%%
where we have introduced a composite index $p$ for $\bx_{\bk}$.
Note that one must multiply $\hbar /\sqrt{M_\al M_\be}$ to the gauge field expression in \eq{seq:ftotal} to recover the standard units when performing numerical calculations.
From the underlying real-space symmetries of $D$ and $F$, it directly follows that the momentum-space matrices satisfy:
%%%%%%%%%%
\ba
h(-\bk)^T = h(\bk),
\quad
\fc(-\bk)^T = -\fc(\bk).
\label{seq:transpose}
\ea
%%%%%%%%%%
Furthermore, we note that the full real-space dynamical matrix $D$---the massive $\nc \ns d \times \nc \ns d$ matrix with entries $D_{\bR \al \mu, \bR' \be \nu}$---is positive semidefinite due to the lattice stability at equilibrium.
Due to the translational invariance of the lattice, this large matrix can be block-diagonalized into distinct momentum sectors, rendering it unitarily equivalent to the direct sum $\oplus_{\bk} D(\bk)$.
Consequently, the spectrum of the full real-space matrix is simply the union of the spectra of $D(\bk)$ across all discrete momentum points.
Since the parent real-space matrix is positive semidefinite, it strictly follows that $D(\bk)$ must be positive semidefinite for every momentum $\bk$. 
These symmetry properties and structural features of $\cf(\bk)$ and $h(\bk)$ form the mathematical foundation required to diagonalize the phonon Hamiltonian, which we address in the following section.

%%%%%%%%%%%%%%%%%%%%%%
\subsection{Diagonalization of the phonon Hamiltonian}
\label{app:diag}
%%%%%%%%%%%%%%%%%%%%%%
We diagonalize the phonon Hamiltonian by decomposing the momentum-space operator vector $\bx_{\bk}$ into normal modes.
These normal modes are defined as the eigensolutions of the Heisenberg equation of motion (EOM), $\der_t \bx_\bk = i [\Hp, \bx_\bk]$.
Utilizing the Hamiltonian definition and the commutation relations derived in \eq{seq:comm}, we evaluate the EOM component by component
%%%%%%%%%%
\ba
\der_t \xhat_{\bk p}
&= \frac{i}{2} \sum_{\bk' p' q'} h(\bk')_{p'q'} [\xhat_{-\bk' p'} \xhat_{\bk' q'}, \xhat_{\bk p}]
\nn
&= -\hf \sum_{\bk' p' q'} h(\bk')_{p'q'}
(
\xhat_{-\bk' p'} \delta_{\bk',-\bk} \fc(\bk')_{q' p}
+
\delta_{\bk \bk'} \fc(-\bk')_{p' p} \xhat_{\bk' q'}
)
\nn
&= -\hf \sum_{p'q'}
(
\fc(-\bk)_{q' p} h(-\bk)_{p' q'} \xhat_{\bk p'}
+
\fc(-\bk)_{p' p} h(\bk)_{p' q'} \xhat_{\bk q'}
)
\nn
&= \hf \sum_{p' q'}
(
\fc(\bk)_{pq'} h(\bk)_{q' p'} \xhat_{\bk p'}
+
\fc(\bk)_{p p'} h(\bk)_{p' q'} \xhat_{\bk q'}
)
\nn
&= [\fc(\bk) h(\bk) \bx_\bk]_{p},
\label{seq:heom}
\ea
%%%%%%%%%%
where we used $h(-\bk)^T = h(\bk)$ and $\fc(-\bk)^T = -\fc(\bk)$ from \eq{seq:transpose}.
This reduces the EOM to the compact matrix form $\der_t \bx_\bk = \fc(\bk) h(\bk) \bx_\bk$.
Assuming a harmonic time dependence $\bz_{n\bk}(t) = \bz_{n\bk} e^{-i\om_{n\bk} t}$, the normal modes are obtained by solving the eigenvalue equation
%%%%%%%%%%
\ba
\om_{n\bk} \bz_{n\bk}
= i \fc(\bk) h(\bk) \bz_{n\bk},
\quad (n = \pm 1, \dots, \pm d \ns).
\label{seq:normal}
\ea
%%%%%%%%%%
A straightforward calculation shows that both $i\fc(\bk)$ and $h(\bk)$ are Hermitian matrices.
However, their product $i\fc(\bk) h(\bk)$ is generally non-Hermitian.
Thus, solving for the normal modes requires diagonalizing a non-Hermitian matrix, which is not guaranteed to be diagonalizable by a unitary transformation and could, in principle, possess complex eigenvalues.
\textit{Existence of real eigenvalues.---}
We first prove that despite the non-Hermiticity of $i\fc(\bk) h(\bk)$, there exist exactly $2d\ns$ distinct solutions to \eq{seq:normal} with purely real frequencies $\om_{n\bk} \in \mbb{R}$ at any given momentum $\bk$.
Temporarily suppressing the $(n, \bk)$ indices, note that the matrix $i \cf h$ has dimension $2d\ns \times 2d\ns$, meaning there are at most $2d\ns$ real eigenvalues.
By partitioning the eigenvector into momentum and displacement components $\bz = \bpm \bmu \\ \bep \epm$, the eigenvalue equation expands to
%%%%%%%%%%
\ba
\om
\bpm \bmu \\ \bep \epm
=
i
\bpm
F & -D \\
\id & 0
\epm
\bpm \bmu \\ \bep \epm
=
\bpm
i F \bmu - i D \bep \\
i \bmu
\epm
.
\ea
%%%%%%%%%%
This coupled system is algebraically equivalent to the conditions
%%%%%%%%%%
\ba
\om^2 \bep = (i \om F + D)\bep
, \quad
\bmu = -i \om \bep.
\label{seq:ep}
\ea
%%%%%%%%%%

For any purely real frequency $\om \in \mbb{R}$, the matrix $X(\om) = i\om F + D$ is explicitly Hermitian.
Therefore, it possesses a complete set of real eigenvalues $\ld_m(\om) \in \mbb{R}$ and orthonormal eigenvectors $\bep_m(\om)$, satisfying $X(\om) \bep_{m}(\om) = \ld_m(\om) \bep_m (\om)$ for $m = 1, \dots, d \ns$.
Finding the real-frequency normal modes of \eq{seq:ep} is thus equivalent to finding the roots of the implicit equation $\om^2 = \ld_m(\om)$.
At strictly zero frequency, $X(0) = D \succeq 0$, which implies $\ld_m(0) \geq 0$.
We categorize these modes into two cases.
If $\ld_m(0) = 0$, the equation $\om^2 = \ld_m(\om)$ possesses a double root at $\om = 0$.
These correspond to Goldstone modes (eigenmodes with strictly zero frequency at the specified momentum $\bk$), indicating that the emergent gauge field cannot gap out these fundamental zero-energy modes.
We denote the number of such Goldstone modes at $\bk$ as $N_G$.
Conversely, if $\ld_m(0) > 0$, then at $\om = 0$, we clearly have $\om^2 < \ld_m(\om)$.
However, in the limit of large $|\om|$, the matrix $X(\om)$ is dominated by the linear term $i \om F$.
Because $\lambda_m(\omega)$ scales at most linearly with $\omega$ at large frequencies, the quadratic $\omega^2$ term must eventually overtake it, meaning $\om^2 > \ld_m(\om)$ for sufficiently large $|\om|$.
By continuity, there must exist at least one crossing where $\om^2 = \ld_m(\om)$ for $\om > 0$, and another for $\om < 0$.

Counting these solutions at a fixed momentum $\bk$, there are $2 N_G$ solutions at $\om = 0$ (counting the double roots as separate eigenmodes), plus at least $d\ns - N_G$ strictly positive and strictly negative frequency solutions.
Summing these yields a minimum of $2 d \ns$ real-frequency eigenmodes.
Since the total matrix dimension is $2d\ns$, this proves that all eigenvalues of \eq{seq:normal} must be purely real.
We adopt the convention of indexing the propagating modes with positive frequencies ($\om \geq 0$) as $n = 1, \dots, d \ns$, and the counterpropagating modes with negative frequencies ($\om \leq 0$) as $n = -1, \dots, -d \ns$.
Applying complex conjugation to \eq{seq:normal} and utilizing the real-space conditions $F(\bR), D(\bR) \in \mbb{R}_{d \ns \times d \ns}$ yields
%%%%%%%%%%
\ba
\om_{n\bk} \bz_{n\bk}^*
&= -i \cf(\bk)^* h(\bk)^* \bz_{n\bk}^*
\nn
&= -i \cf(-\bk) h(-\bk) \bz_{n\bk}^*.
\ea
%%%%%%%%%%
Thus, we enforce the effective particle-hole symmetry constraints on the normal modes
%%%%%%%%%%
\ba
\om_{-n -\bk} = - \om_{n \bk},
\quad
\bz_{-n -\bk} = \bz_{n \bk}^*.
\ea
%%%%%%%%%%
This specific gauge choice for the particle-hole related eigenvectors is required to properly define the canonical phonon creation and annihilation operators later in this section.
\textit{Numerical diagonalization and normalization.---}
Having established that the eigenvalues are purely real, we now present a systematic method for numerically obtaining the normal modes $\bz_{n\bk}$ and enforcing their generalized normalization condition.
Before proceeding, we invoke a standard linear algebraic identity: for any two general square matrices $A$ and $B$, the products $AB$ and $BA$ share the exact same spectrum ($spec[AB] = spec[BA]$).
This is easily proven by noting that the block matrices $\mf{M}_1 = \left(\begin{smallmatrix} AB & 0 \\ B & 0 \end{smallmatrix}\right)$ and $\mf{M}_2 = \left(\begin{smallmatrix} 0 & 0 \\ B & BA \end{smallmatrix}\right)$ are related by the similarity transformation $\mf{M}_2 = \mc{S}^{-1} \mf{M}_1 \mc{S}$ via the $\mc{S} = \left(\begin{smallmatrix} \id & A \\ 0 & \id \end{smallmatrix}\right)$, whose inverse is given by $\left(\begin{smallmatrix} \id & -A \\ 0 & \id \end{smallmatrix}\right)$.
From $h(\bk) \succeq 0$, we can uniquely define its principal square root and rewrite the EOM matrix as $i\fc(\bk) h(\bk) = [i\fc(\bk) \sqrt{h(\bk)}] \sqrt{h(\bk)}$.
Using the spectral identity $spec[AB] = spec[BA]$, we deduce that this non-Hermitian matrix shares the same eigenvalues as the reversed product
%%%%%%%%%%
\ba
\heff(\bk) = i \sqrt{h(\bk)} \fc(\bk) \sqrt{h(\bk)}
= i
\bpm
F(\bk) & - \sqrt{D(\bk)} \\
\sqrt{D(\bk)} & 0
\epm.
\label{seq:heff}
\ea
%%%%%%%%%%
Crucially, the effective phonon Hamiltonian matrix $\heff(\bk)$ is explicitly Hermitian.
Thus, the real frequencies $\om_{n\bk}$ can be robustly obtained through standard unitary diagonalization
%%%%%%%%%%
\ba
\heff \bby_{n\bk} = \om_{n\bk} \bby_{n\bk},
\quad
\bby_{m\bk}^{\dg} \bby_{n\bk} = \delta_{mn}.
\ea
%%%%%%%%%%
To relate $\bby_{n\bk}$ back to the normal modes $\bz_{n\bk}$, we multiply the original EOM (\eq{seq:normal}) from the left by $\sqrt{h(\bk)}$:
%%%%%%%%%%
\ba
\heff(\bk) \big(\sqrt{h(\bk)} \bz_{n\bk}\big) = \om_{n\bk} \big(\sqrt{h(\bk)} \bz_{n\bk}\big).
\ea
%%%%%%%%%%
This structural similarity motivates defining the physical normal modes as
%%%%%%%%%%
\ba
\bz_{n\bk} \equiv \sqrt{h(\bk)^+} \bby_{n\bk},
\ea
%%%%%%%%%%
where $h(\bk)^+$ denotes the Moore-Penrose pseudoinverse.
The pseudoinverse safely handles the non-invertibility introduced by the presence of zero-frequency Goldstone modes in the null space of $h(\bk)$.
This definition naturally enforces the generalized normalization condition
%%%%%%%%%%
\ba
\bz_{m\bk}^\dg h(\bk) \bz_{n \bk} = \delta_{mn}.
\label{seq:porth}
\ea
%%%%%%%%%%
Strictly speaking, the explicit eigenvectors for the zero-frequency Goldstone modes at a given momentum $\bk$ cannot be constructed via this pseudoinverse relation.
However, because zero-frequency modes carry no energy, they do not contribute to the thermal Hall conductivity (as detailed in \Sec{app:response_series}). Consequently, their precise analytic form is safely ignored in practical numerical evaluations.
\textit{Phonon operators.---}
Using these properly normalized modes, we map the displacement vectors to canonical phonon creation and annihilation operators $\bhat_{n\bk}$, defined by the standard commutation relations
%%%%%%%%%%
\ba
[\bhat_{m\bk}, \bhat_{n \bk'}] = \delta_{m, -n}^z \delta_{\bk, -\bk'},
\quad
\bhat_{-n-\bk} = \bhat_{n\bk}^\dg,
\label{seq:pcomm}
\ea
%%%%%%%%%%
where we have introduced the signed Kronecker delta $\delta^z_{m,-n} = \text{sign}(m) \delta_{m, -n}$.
Utilizing the completeness relation $\sum_n [\bby_{n \bk}]_p [\bby_{n \bk}^*]_q = \delta_{pq}$ (which follows directly from the unitarity of the matrix $Y_\bk = (\bby_{-d\ns \bk}, \dots, \bby_{d \ns \bk})$), we construct the transformation between the physical coordinates and the canonical operators
%%%%%%%%%%
\ba
\bx_\bk = \sum_n \sqrt{|\om_{n\bk}|} \bz_{n\bk} \bhat_{n\bk},
\quad
\bhat_{n\bk} = \frac{1}{\sqrt{|\om_{n\bk}|}} \bz_{n\bk}^{\dg} h(\bk) \bx_\bk,
\label{seq:phonodef}
\ea
%%%%%%%%%%
where we implicitly exclude any $\omega_{n\bk} = 0$ Goldstone mode at a given momentum $\bk$.
Substituting this transformation back into the phonon Hamiltonian yields a completely decoupled, diagonal basis
%%%%%%%%%%
\ba
\Hp
&= \hf \sum_{\bk} \bx_{\bk}^\dg h(\bk) \bx_\bk
\nn
&= \hf \sum_{mn\bk} \sqrt{|\om_{m \bk} \om_{n \bk}|} \bz_{m\bk}^\dg h(\bk) \bz_{n\bk} \bhat_{m\bk}^\dg \bhat_{n\bk}
\nn
&=
\hf \sum_{n \bk} |\om_{n\bk}| \bhat_{n\bk}^\dg \bhat_{n\bk}
\nn
&= \hf \sum_{n >0} \sum_{\bk}
(
\om_{n\bk} \bhat_{n \bk}^\dg \bhat_{n\bk} - \om_{-n-\bk} \bhat_{-n-\bk}^\dg \bhat_{-n-\bk}
)
\nn
&= \hf \sum_{n >0} \sum_\bk \om_{n\bk} (\bhat_{n\bk}^\dg \bhat_{n\bk} + \bhat_{n\bk} \bhat_{n\bk}^\dg)
\nn
&= \sum_{n >0} \sum_\bk (\bhat_{n\bk}^\dg \bhat_{n\bk} + \hf),
\ea
%%%%%%%%%%
This final result confirms that the complex coupled dynamics induced by the emergent gauge field have been successfully reduced to a collection of independent quantum harmonic oscillators.

%%%%%%%%%%%%%%%%%%%%%%
\subsection{Projector formalism for multiband phonon Berry curvature}
\label{app:phononbc}
%%%%%%%%%%%%%%%%%%%%%%
In this Appendix, we present a projector formula for the Abelian phonon Berry curvature, which is highly efficient for numerical calculations and naturally handles band degeneracies.
The phonon Berry curvature differs from its electronic counterpart primarily due to the non-unitary nature of the eigenvectors. 
To properly account for bosonic orthonormality, we introduce the metric $h(\bk)$.
Away from degenerate points, the phonon Berry connection for a single isolated band is defined as~\cite{qin2012berry,saito2019berry}
%%%%%%%%%%
\ba
\ca_{n}(\bk)_\mu = -Im[\bxi_{n\bk}^\dg \dmu \bz_{n\bk}],
\quad \text{where} \quad
\bxi_{n\bk} \equiv h(\bk) \bz_{n\bk},
\ea
%%%%%%%%%%
where $\dmu \bz_{n \bk} \equiv \lim_{\vep \rightarrow 0} (\bz_{n, \bk + \vep \ehat_{\mu}} - \bz_{n \bk})/\vep$. 
Unlike electronic eigenstates, the inner product $\bxi_{n\bk}^\dg \dmu \bz_{n\bk}$ is not automatically purely imaginary; thus, we must explicitly extract the imaginary part to define a real-valued connection. 
The corresponding single band Berry curvature is given by
%%%%%%%%%%
\ba
\Omega_n(\bk)_{\mu\nu} = \dmu \ca_n(\bk)_\nu - \dnu \ca_n(\bk)_\mu
=
-Im[ \dmu \bxi_{n\bk}^\dg \dnu \bz_{n\bk} -\dnu \bxi_{n\bk}^\dg \dmu \bz_{n\bk}].
\ea
%%%%%%%%%%

%
In generic phonon systems, the single band Berry curvature $\Omega_n(\bk)$ exhibits singular divergences at band crossings.
These divergences pose a significant challenge when evaluating physical observables, such as the phonon thermal Hall conductivity $\kmn \propto \sum_{\bk, n > 0} \big[c_2(g(\hbar\omega_{n\bk})) - \frac{\pi^2}{3}\big] \Omega_{n}(\bk)_{\mu \nu}$ [see \eq{seq:kmn}].
However, when multiple bands are (nearly) degenerate at a given momentum, their energy-dependent weighting factors converge to a common value.
The transport integral locally reduces to this common weight multiplied by the sum of the individual Berry curvatures of the constituent bands.
Crucially, even if the individual curvatures diverge, their sum across the degenerate subspace remains entirely smooth and well-defined. 
We formalize this by introducing a multiband Abelian generalization.
We define a multiplet matrix $\Zeta_{\nk}$ formed by grouping the eigenvectors of these (nearly) degenerate bands.
The multiband Abelian Berry connection and curvature are then defined as
%%%%%%%%%%
\ba
\ca_{[n]}(\bk)_\mu = 
-\imtr{\Xi_{\nk}^\dg \dmu \Zeta_{\nk}},
\quad
\Omega_{\ndeg }(\bk)_{\mu\nu} =
-\imtr{
	\dmu \Xi_{\nk}^\dg \dnu \Zeta_{\nk} - \dnu \Xi_{\nk}^\dg \dmu \Zeta_{\nk}
},
\label{seq:pcurv}
\ea
%%%%%%%%%%
where $\Xi_{\nk} = h(\bk) \Zeta_{\nk}$.
To implement these formulas robustly in numerical calculations—where derivatives like $\dmu \Zeta_{\nk}$ are evaluated using a finite-difference stencil (e.g., at $\bk$ and $\bk \pm \vep \ehat_{\mu}$)—we must ensure that the column dimension of $\Zeta_{\nk}$ is consistent across the local neighborhood. 
To elaborate, the strict degeneracy structure often fluctuates between adjacent grid points.
For example, a four-band system might exhibit degeneracies grouped as $\{1\}, \{2,3\}, \{4\}$ exactly at $\bk$, but split into $\{1,2\}, \{3\}, \{4\}$ at $\bk + \vep \ehat_{\mu}$.
If we were to define $\Zeta_{\nk}$ using the exact degeneracies at each independent grid point, the matrix dimension would inconsistently change across the stencil, rendering the finite-difference derivative ill-defined.
To resolve this, we uniformly define the multiplet matrix $\Zeta_{\nk}$ by grouping all bands that cross or come within a specified small energy tolerance of one another \textit{anywhere} within the local finite-difference neighborhood.
In the previous example, we would uniformly group the bands across the entire stencil as $\Zeta_{[1]\bk} = (\bz_{1\bk}, \bz_{2\bk}, \bz_{3\bk})$ and $\Zeta_{[2]\bk} = (\bz_{4\bk})$. 
Provided the energy tolerance and the momentum grid spacing are sufficiently small, this multiband curvature naturally allows us to generalize transport formulas for nondegenerate bands to arbitrary phonon systems with complex band crossings.
To compute this multiband curvature efficiently, we define non-Hermitian generalized phonon projector for the multiplet $[n]$:
%%%%%%%%%%
\ba
\cp_{\nk} = \Zeta_{\nk} \Xi_{\nk}^\dg.
\label{seq:pproj}
\ea
%%%%%%%%%%
This quantity satisfies all the fundamental properties of a projector.
Specifically, it is idempotent and acts as the identity within its subspace
%%%%%%%%%%
\ba
\Xi_{[m]\bk}^\dg \cp_{\nk} = \delta_{[m],[n]} \Xi_{[m]\bk}^\dg,
\quad
\cp_{\nk} \Zeta_{[m] \bk} = \delta_{[m],[n]} \Zeta_{[m]\bk}.
\label{seq:pid}
\ea
%%%%%%%%%%
Furthermore, $\cp_{\nk}$ is manifestly invariant under any unitary band mixing $\Zeta_{\nk} \to \Zeta_{\nk} U_{\nk}$ within the degenerate subspace, and its trace accurately returns the dimension of the degenerate subspace.
We now show that the Abelian phonon Berry curvature in \eq{seq:pcurv} can be elegantly recast entirely in terms of these generalized projectors and their derivatives.
Inserting the identity $\id = \cp_{\nk} + (\id - \cp_{\nk})$ into \eq{seq:pcurv} and temporarily suppressing the $([n], \bk)$ indices, we separate the curvature into two traces:
%%%%%%%%%%
\ba
\Omega_{\mu\nu} = 
-\imtr{
	\dmu \Xi^\dg \cp \dnu \Zeta - \dnu \Xi^\dg \cp \dmu \Zeta
}
-
\imtr{
	\dmu \Xi^\dg (\id - \cp) \dnu \Zeta
	- \dnu \Xi^\dg (\id - \cp) \dmu \Zeta
}
.
\ea
%%%%%%%%%%
By applying the relation $\Xi^\dg \Zeta = \id$ (which implies $\dmu \Xi^\dg \Zeta = - \Xi^\dg \dmu \Zeta$) and utilizing the cyclicity of the trace, the first term identically vanishes:
%%%%%%%%%%
\ba
-\imtr{\dmu \Xi^\dg (\Zeta \Xi^\dg) \dnu \Zeta - \dnu \Xi^\dg (\Zeta \Xi^\dg) \dmu \Zeta }
&= -\imtr{ \dmu \Xi^\dg \Zeta \Xi^\dg \dnu \Zeta - \Xi^\dg \dnu \Zeta \dmu \Xi^\dg \Zeta }
\nn
&= -\imtr{ \dmu \Xi^\dg \Zeta \Xi^\dg \dnu \Zeta - \dmu \Xi^\dg \Zeta \Xi^\dg \dnu \Zeta }
\nn
&= 0.
\ea
%%%%%%%%%%
To evaluate the remaining term, we insert the identity matrix $\id = \Xi^\dg \Zeta$ into the trace and again use cyclicity to pull $\Xi^\dagger$ to the right
%%%%%%%%%%
\ba
\Omega_{\mu\nu} &=
-\imtr{
	(\Xi^\dg \Zeta) \dmu \Xi^\dg (\id - \cp) \dnu \Zeta - (\mu \leftrightarrow \nu)
}
\nn
&=
-\imtr{
	\Zeta \dmu \Xi^\dg (\id - \cp) \dnu \Zeta \Xi^\dg - (\mu \leftrightarrow \nu)
}.
\ea
%%%%%%%%%%
Substituting $\dmu \cp = \dmu \Zeta \Xi^\dg + \Zeta \dmu \Xi^\dg$ into the trace yields
%%%%%%%%%%
\ba
\Omega_{\mu\nu} 
&=
-\imtr{
	(\dmu \cp - \dmu \Zeta \Xi^\dg) (\id - \cp) (\dnu \cp - \Zeta \dnu \Xi^\dg) - (\mu \leftrightarrow \nu)
}.
\ea
%%%%%%%%%%
Utilizing $\Xi^\dg (\id - \cp) = \Xi^\dg - \Xi^\dg \cp = 0$ and $(\id - \cp) \Zeta = \Zeta - \cp \Zeta = 0$ [see \eq{seq:pid}], the curvature reduces to
%%%%%%%%%%
\ba
\Omega_{\mu\nu} 
&=
-\imtr{
	\dmu \cp (\id - \cp) \dnu \cp - (\mu \leftrightarrow \nu)
}.
\ea
%%%%%%%%%%
Finally, taking the derivative of the idempotence relation $\cp^2 = \cp$ yields $\dmu \cp \cp + \cp \dmu \cp = \dmu \cp$, which rearranges to $\dmu \cp (\id - \cp) = \cp \dmu \cp$.
Substituting this identity into our expression, we arrive at the projector formula for the multiband phonon Berry curvature
%%%%%%%%%%
\ba
\Omega_{\mu\nu} =
-\imtr{
	\cp \dmu \cp \dnu \cp - \cp \dnu \cp \dmu \cp
}
= -\imtr{\cp [\dmu \cp, \dnu \cp]}.
\label{seq:phononbc}
\ea
%%%%%%%%%%

%%%%%%%%%%%%%%%%%%%%%%
\section{Symmetries of the electron and phonon systems}
\label{app:sym}
%%%%%%%%%%%%%%%%%%%%%%

In this Appendix, we formally define the relevant symmetry groups of the system and examine how they constrain the phonon dynamics coupled to the emergent gauge field.
In \Sec{app:esym}, we establish the electronic symmetries and derive the resulting transformation rules for the emergent gauge field $\fem$.
Next, in \Sec{app:psym}, we deduce the symmetry properties of the coupled phonon system.
Establishing these geometric transformation rules is a necessary prerequisite for determining the symmetry constraints on the macroscopic phonon thermal Hall conductivity.
Closely following the notation of Ref.~\cite{hwang2026stable}, we denote a general spin space group symmetry element~\cite{brinkman1966theory,xiao2024spin,chen2024enumeration,jiang2024enumeration} as $g = (\trs)^{\phg}\{\cug | \Og | \bd_g \}$, where $\trs$ is the time-reversal operator, $\cug$ is the $SU(2)$ action of $g$ [see \eq{seq:esym}], and $\phg = 0$ ($\phg = 1$) for unitary (antiunitary) symmetries.
As we consider general spin space groups, $\Og$ does not restrict the form of $\cug$.
The symmetry $g$ acts on real and reciprocal space coordinates as
%%%%%%%%%%
\ba
g \br = \Og \br + \bd_g,
\quad
g \bk = (-1)^{\phg} \Og \bk,
\quad
(\Og \in O(d)).
\ea
%%%%%%%%%%
Its action on a general complex number (as opposed to an operator) is simply complex conjugation for antiunitary operations
%%%%%%%%%%
\ba
g x g^{-1} = \gbar{x} =
\begin{cases}
	x & (\phg = 0) \\
	x^* & (\phg = 1)
\end{cases}
.
\ea
%%%%%%%%%%

%%%%%%%%%%%%%%%%%%%%%%
\subsection{Electronic symmetries}
\label{app:esym}
%%%%%%%%%%%%%%%%%%%%%%
A critical subtlety in our formalism is that we define the electronic space group $\gel$ based strictly on the equilibrium nuclear configuration.
We say $g \in \gel$ if it leaves the electronic Hamiltonian [see \eq{seq:double}] invariant at the \textit{equilibrium} configuration.
Consequently, at finite displacements, $g$ enforces a covariant symmetry transformation between the symmetry-related displaced configurations
%%%%%%%%%%
\ba
g \hHel(\bb 0) g^{-1} = \hHel(\bb 0),
\quad
g \hHel(\uion) g^{-1} = \hHel(g \uion),
\ea
%%%%%%%%%%
where $\uion = (\bb{u}_1, \dots, \bb{u}_{\nion})$ denotes the nuclear displacements.
We denote the action of $g$ on the localized electron creation operators as
%%%%%%%%%%
\ba
g \cdag_{i \sg} g^{-1} = \sum_{j \sg'} [\Ug]_{j \sg', i \sg} \cdag_{j \sg'}
= \sum_{\sg'} \cdag_{i' \sg'} e^{i \theta_{g}(i)}[\mc{U}_g]_{\sg' \sg},
\label{seq:esym}
\ea
%%%%%%%%%%
where $e^{i \theta_g(i)} \in U(1)$ captures an on-site phase, and $\cug$ is the $SU(2)$ action of $g$.
Note that the primed site index $i'$ denotes the spatial coordinate mapped by the symmetry ($g: i \mapsto i'$).
\eq{seq:esym} implies that the single-particle matrix representation of the electronic Hamiltonian transforms covariantly as
%%%%%%%%%%
\ba
\Hel(g \uion) = \Ug \gbar{\Hel(\uion)} \Ug^{\dg}.
\label{seq:esymrep}
\ea
%%%%%%%%%%

%
To determine how this constrains the real-space emergent gauge field $\fem_{i\mu, j \nu} = -2 \imtr{P \der_{i\mu} P \der_{j\nu} P}$ [see \eq{seq:fproj}], we note that \eq{seq:esymrep} forces the electronic ground state projector to obey the identical transformation rule
%%%%%%%%%%
\ba
P(g \uion) = \Ug \gbar{P(\uion)} \Ug^{\dg}.
\ea
%%%%%%%%%%
By explicitly evaluating the derivative of the projector with respect to a displacement at the mapped site $i'$, and letting $\hat{\bs{\ep}}_{i\mu}$ denote the basis vectors of the $d \nion$-dimensional displacement space, we find
%%%%%%%%%%
\ba
\der_{i'\mu'} P(\bb 0)
& \equiv \lim_{\vep \to 0} \frac{P(\vep \hat{\bs{\ep}}_{i' \mu'}) - P(\bb 0)}{\vep}
\nn
&= \lim_{\vep \to 0} \Ug \frac{\sum_{\mu} \gbar{P(\vep [\Og]_{\mu' \mu}\hat{\bs{\ep}}_{i \mu})} - \gbar{P(\bb 0)}}{\vep} \Ug^\dg
\nn
&= \sum_{\mu} \Ug \der_{i\mu} \gbar{P(\bb 0)} \Ug^\dg [\Og]_{\mu' \mu}.
\ea
%%%%%%%%%%
Substituting this into the definition of the gauge field and utilizing the cyclicity of the trace, the unitary matrices strictly cancel, revealing that the equilibrium emergent gauge field rotates as a tensor
%%%%%%%%%%
\ba
\fem_{i'\mu', j'\nu'} = (-1)^{\phg} \sum_{\mu \nu} [\Og]_{\mu' \mu} \fem_{i\mu, j\nu} [\Og^T]_{\nu \nu'}.
\label{seq:fsym}
\ea
%%%%%%%%%%
An immediate consequence is that for a crystal with discrete translational symmetry, the gauge field depends only on the relative distance between unit cells.
Separating the site index into unit cell and sublattice components $i = (\bR, \al)$, we can write $\fem_{\bR_1 \al \mu, \bR_2 \be \nu} = \fem(\bR_1 - \bR_2)_{\al \mu, \be \nu}$.
Furthermore, we can apply a spatial Fourier transform to determine the transformation of the momentum-space gauge field $\fem(\bk)_{\al \mu, \be \nu} = \sum_{\bR} e^{-i \bk \cdot (\bR + \bt_\al - \bt_\be)} \fem(\bR)_{\al \mu, \be \nu}$. 
Tracking the mapped indices $g: \al \mapsto \al'$ and relative distances $\bR' + \bt_{\al'} - \bt_{\be'} = \Og(\bR + \bt_\al - \bt_\be)$ yields
%%%%%%%%%%
\ba
\fem(g \bk)_{\al' \mu', \be' \nu'}
&= \sum_{\bR'} e^{-i (g\bk) \cdot(\bR' + \bt_{\al'} - \bt_{\be'})} \fem(\bR')_{\al' \mu', \be' \nu'}
\nn
&= (-1)^{\phg} \sum_{\bR \mu \nu} [\Og]_{\mu' \mu} e^{-i (-1)^{\phg} \bk \cdot (\bR + \bt_\al - \bt_\be)} \fem(\bR)_{\al \mu, \be \nu} [\Og^T]_{\nu \nu'}
\nn
&=
(-1)^{\phg} \sum_{\mu \nu} [\Og]_{\mu' \mu} \gbar{\fem(\bk)_{\al \mu, \be \nu}} [\Og^T]_{\nu \nu'}.
\ea
%%%%%%%%%%
By introducing the sublattice permutation matrix $[\Pg]_{\be \al} = \delta_{\be \al'}$ (not to be confused with the electronic projector $P$), this momentum-space constraint can be compactly expressed in matrix form as
%%%%%%%%%%
\ba
\fem(g\bk) = (-1)^{\phg} \Gg \gbar{\fem(\bk)} \Gg^T,
\quad \text{where} \quad \Gg = \Pg \otimes \Og.
\label{seq:ftransform}
\ea
%%%%%%%%%%

%%%%%%%%%%%%%%%%%%%%%%
\subsection{Phonon symmetries}
\label{app:psym}
%%%%%%%%%%%%%%%%%%%%%%
Defining symmetry in the phonon sector requires care due to the presence of the gauge field.
Because the kinematic momentum $\bpi$ completely absorbs the gauge potential, the standard kinetic energy term $\bpi^2/2$ is formally invariant regardless of the underlying gauge field structure [see \eq{seq:hphonon}].
Consequently, declaring a symmetry based on the invariance of the Hamiltonian $\Hp$ is insufficient.
Instead, we define the phonon space group $\gph$ based on the invariance of the Heisenberg equations of motion, $i \der_t \bx_{\bk} = i \cf(\bk) h(\bk) \bx_{\bk}$ [see \eq{seq:heom}]. 
In the following, we assume the total effective gauge field is given entirely by the electronic emergent gauge field, $F = \fem$.
Let us denote the action of $g$ on the lattice operators as
%%%%%%%%%%
\ba
g \pihat_{\bR \al \mu} g^{-1} = (-1)^{\phi_g} \sum_{\be \nu} \pihat_{\bR' \be \nu} [\Gg]_{\be \nu, \al \mu},
\quad
g \uhat_{\bR \al \mu} g^{-1} = \sum_{\be \nu} \uhat_{\bR' \be \nu} [\Gg]_{\be \nu, \al \mu},
\ea
%%%%%%%%%%
where $\Gg = \Pg \otimes \Og$ and $\bR' + \br_{\be} = \Og(\bR + \br_{\al}) + \bd_g$.
Transforming to momentum space, this becomes
%%%%%%%%%%
\ba
g \pihat_{\bk \al \mu} g^{-1} = (-1)^{\phi_g} \sum_{\be \nu} \pihat_{g \bk, \be \nu} e^{i g \bk \cdot \bd_g} [\Gg]_{\be \nu, \al \mu},
\quad
g \uhat_{\bk \al \mu} g^{-1} = \sum_{\be \nu} \uhat_{g \bk, \be \nu} e^{i g \bk \cdot \bd_g} [\Gg]_{\be \nu, \al \mu},
\ea
%%%%%%%%%%
which can be written compactly in terms of the composite operator vector $\bx_\bk$ as
%%%%%%%%%%
\ba
g \xhat_{\bk p} g^{-1} = \sum_q \xhat_{g \bk, q} e^{i g \bk \cdot \bd_g} [\gtd]_{qp},
\quad \text{where} \quad
\gtd = \bpm (-1)^{\phi_g} \Gg & 0 \\ 0 & \Gg \epm
.
\ea
%%%%%%%%%%
Noting that time derivatives transform as $g i \der_t g^{-1} = (-1)^{2 \phi_g} i \der_t = i \der_t$, the Heisenberg EOM is invariant under $g \in \gph$ if and only if the EOM matrix transforms as
%%%%%%%%%%
\ba
i \cf(g \bk) h(g \bk) = \gtd \big(\gbar{i \cf(\bk) h(\bk)}\big) \gtd^T.
\label{seq:psym}
\ea
%%%%%%%%%%
This single matrix equation strictly dictates two independent conditions for the block matrices
%%%%%%%%%%
\ba
\fem (g \bk) = (-1)^{\phi_g} \Gg \gbar{\fem (\bk)} \Gg^T
\quad \text{and} \quad
D(g \bk) = \Gg \gbar{D(\bk)} \Gg^T.
\label{seq:psym2}
\ea
%%%%%%%%%%
The first condition demands that either $g \in \gel$ (satisfying \eq{seq:ftransform}) or $\fem (\bk) = 0$.
The second condition dictates that $g$ must be a valid symmetry of the bare phonon system (i.e., the standard lattice dynamical matrix without the gauge field).
Thus, one can diagnose whether a specific operation is a valid symmetry of the fully coupled system simply by checking if it is a joint symmetry of the electronic space group and the bare lattice structure.
With these transformation rules established, we can deduce the symmetry properties of the phonon Berry curvature, $\Omega_{[n]}(\bk)_{\mu \nu} = -\imtr{\cp_{[n]\bk} [\dmu \cp_{[n]\bk}, \dnu \cp_{[n]\bk}]}$ [see \eq{seq:phononbc}].
From the EOM eigenvalue problem $i \cf(\bk) h(\bk) \bz_{n\bk}= \om_{n\bk} \bz_{n\bk}$ and the transformation rule in \eq{seq:psym}, it immediately follows that $\gtd \gbar{\bz}_{n\bk}$ is a valid normal mode with frequency $\om_{n\bk}$ at the transformed momentum $g \bk$.
Combined with \eq{seq:psym2}, the generalized non-Hermitian phonon projector $\cp_{[n]\bk}$ must therefore transform as $\cp_{[n] g\bk} = \gtd \gbar{\cp}_{[n]\bk} \gtd^T$.
Evaluating the momentum derivative of the projector under this mapping yields
%%%%%%%%%%
\ba
\der_{\mu} \cp_{[n],g \bk}
&= \lim_{\vep \to 0} \frac{\cp_{[n] g \bk + \vep \ehat_{\mu}} - \cp_{[n] g \bk}}{\vep}
\nn
&= \lim_{\vep \to 0} \frac{\gtd \big(\gbar{\cp}_{[n] \bk + \vep \Og^{-1} \ehat_{\mu}} - \gbar{\cp_{[n] \bk}} \big) \gtd^{T}}{\vep}
\nn
&= \sum_{\nu} [\Og]_{\mu \nu} \gtd \der_{\nu} \gbar{\cp}_{[n]\bk} \gtd^T.
\ea
%%%%%%%%%%
Inserting this into the trace formula for the curvature, the internal $\gtd$ matrices cancel, leaving the fundamental symmetry transformation rule for the phonon Berry curvature
%%%%%%%%%%
\ba
\Omega_{[n]}(g\bk) = (-1)^{\phi_g} \Og \Omega_{[n]} (\bk) \Og^T.
\label{seq:omtransform}
\ea
%%%%%%%%%%

%
The general transformation rule in \eq{seq:omtransform} manifests differently depending on the dimensionality of the system. 
In purely two-dimensional systems confined to the $xy$-plane, the Berry curvature possesses only a single independent component, $\Omega_{[n]}(\bk)_{xy}$. For spatial symmetries confined to this plane, the transformation rule strictly simplifies to a scalar equation
%%%%%%%%%%
\ba
\Omega_{[n]}(g\bk)_{xy} = (-1)^{\phi_g} \det(\Og) \Omega_{[n]}(\bk)_{xy}.
\ea
%%%%%%%%%%
This immediately provides powerful local constraints.
Under in-plane unitary and antiunitary rotations, $\det(\Og) = 1$ and the curvature obeys
%%%%%%%%%%
\ba
\Omega_{[n]}(C_{nz} \bk)_{xy} = \Omega_{[n]}(\bk)_{xy},
\quad
\Omega_{[n]}(\mc{T} C_{nz} \bk)_{xy} = -\Omega_{[n]}(\bk)_{xy}.
\label{seq:rot2d}
\ea
%%%%%%%%%%
Conversely, under unitary and antiunitary mirror reflections (including glide mirrors), $\det(O_g) = -1$, and the curvature is constrained as
%%%%%%%%%%
\ba
\Omega_{[n]}(m_{2D} \bk)_{xy} = -\Omega_{[n]}(\bk)_{xy},
\quad
\Omega_{[n]}(\mc{T} m_{2D} \bk)_{xy} = \Omega_{[n]}(\bk)_{xy}.
\label{seq:mirror2d}
\ea
%%%%%%%%%%

%

%
In three-dimensional systems, the curvature tensor is completely determined by three independent components.
It is therefore highly convenient to map the tensor to a dual vector, defined as $\cb_{[n]}^\rho (\bk) \equiv \hf \sum_{\mu \nu} \vep_{\mu \nu \rho} \Omega_{[n]}(\bk)_{\mu \nu}$.
In terms of this dual vector, \eq{seq:omtransform} is mathematically equivalent to
%%%%%%%%%%
\ba
\bs{\cb}_{[n]}(g \bk) = (-1)^{\phi_g} \det(\Og) \Og \bs{\cb}_{[n]}(\bk).
\label{seq:om3d}
\ea
%%%%%%%%%%
This elegantly demonstrates that the 3D Berry curvature transforms precisely as a time-reversal odd pseudovector. 
This pseudovector nature easily resolves standard 3D symmetry constraints.
For instance, under spatial inversion $\mc{I}$ ($\phi_g = 0$, $O_g = -\id$, $|O_g| = -1$), the dual vector maps as $\bs{\cb}_{[n]}(-\bk) = (1)(-1)(-\id)\bs{\cb}_{[n]}(\bk) = \bs{\cb}_{[n]}(\bk)$, demonstrating that inversion preserves the local Berry curvature.
\paragraph*{Emergent time-reversal symmetry at $\fem = 0$.---}
To conclude this section, we examine the limit where the emergent gauge field vanishes. 
Because the system hosts local magnetic moments, the true time-reversal operator $\mc{T}$ is intrinsically broken ($\mc{T} \notin \gel$).
However, if the underlying magnetic configuration is strictly collinear, or in the limit $t/J \to 0$, the electron wave is independent of the nuclear displacements, causing the emergent gauge field to strictly vanish ($\fem = 0$). 
Crucially, the bare dynamical matrix inherently respects time-reversal symmetry ($D(-\bk) = D(\bk)^*$).
Therefore, turning off the emergent gauge field restores an \textit{effective} antiunitary time-reversal symmetry to the lattice dynamics. 
Under this emergent symmetry operation ($\phi_g = 1$, $O_g = \id$), the Berry curvature transformation rule strictly enforces
%%%%%%%%%%
\ba
\Omega_{[n]}(-\bk) = -\Omega_{[n]}(\bk)
\quad \text{if} \quad
\fem = 0.
\ea
%%%%%%%%%%
The form of the intrinsic phonon thermal Hall conductivity readily shows that this antisymmetry constraint dictates that the intrinsic phonon thermal Hall effect must strictly vanish when $\fem = 0$ [see \eq{seq:kmn}].
This confirms the central physical premise of our formalism: the emergent gauge field is the fundamental required mechanism for generating a finite intrinsic phonon thermal Hall effect in these systems.

%%%%%%%%%%%%%%%%%%%%%%
\section{Phonon thermal Hall conductivity in linear response theory}
\label{app:response}
%%%%%%%%%%%%%%%%%%%%%%

In this Appendix, we derive the intrinsic phonon thermal Hall conductivity $\kmn$ and examine its fundamental algebraic properties. 
In \Sec{app:response_deriv}, we establish the exact linear response formula for a multiband system coupled to a gauge field, correcting a minor historical sign error to unify the result with standard canonical bosonic formalisms. 
In \Sec{app:response_series}, we derive a high-temperature Taylor expansion for the transport integral. 
Finally, in \Sec{app:response_sym}, we apply the geometric transformation rules derived in \Sec{app:psym} to establish the macroscopic symmetry constraints on the thermal Hall tensor.

%%%%%%%%%%%%%%%%%%%%%%
\subsection{Multiband linear response formula}
\label{app:response_deriv}
%%%%%%%%%%%%%%%%%%%%%%
We begin by deriving the explicit form of the thermal Hall conductivity introduced in the main text
%%%%%%%%%%
\ba
\kmn = -\frac{k_B^2 T}{V \hbar} \sum_{\bk} \sum_{[n] > 0}  \big[
c_2 (g(\hbar \omega_{[n]\bk})) - \frac{\pi^2}{3}
\big] \omn
\quad
(\mu \neq \nu),
\label{seq:kmn}
\ea
%%%%%%%%%%
where $V$ is the system volume, $g(\vep) = \frac{1}{e^{\be \vep} - 1}$ is the Bose-Einstein distribution function with $\be = \frac{1}{k_B T}$, and
%%%%%%%%%%
\ba
c_2(x) = \int_0^x dt~\Big(\ln \frac{1+t}{t}\Big)^2
= (1+x) \Big( \ln \frac{1+x}{x} \Big)^2 - (\ln x)^2 - 2 \text{Li}_2(-x),
\ea
%%%%%%%%%%
where $\text{Li}_2(x)$ is the dilogarithm (or second polylogarithm) function~\cite{matsumoto2014thermal}.
To obtain this expression, we build upon the foundational derivation by Qin \textit{et al.}~\cite{qin2012berry}, who obtained a linear response expression applicable to systems where the gauge field is absorbed directly into the kinematic momentum.
Correcting a minor sign error in the first term of Eq. (37) in their original manuscript, the macroscopic phonon Hall conductivity is given by
%%%%%%%%%%
\ba
\kmn = \frac{(\pi k_B)^2}{3h} Z_{\mu \nu} T - \frac{1}{T} \int_0^{\infty} d\vep~ \vep^2 \sg_{\mu \nu} (\vep) \frac{d g(\vep)}{d\vep}
,
\label{seq:kref}
\ea
%%%%%%%%%%
where the auxiliary variables are defined as sums over the particle bands ($n>0$)
%%%%%%%%%%
\ba
Z_{\mu\nu} = \frac{2\pi}{V} \sum_{\bk} \sum_{n > 0} \Omega_{n}(\bk)_{\mu \nu},
\quad
\sg_{\mu\nu}(\vep) = -\frac{1}{V \hbar} \sum_{\hbar \om_{n\bk} \leq \vep} \Omega_{n}(\bk)_{\mu \nu}.
\ea
%%%%%%%%%%
Substituting these definitions into \eq{seq:kref} yields
%%%%%%%%%%
\ba
\kmn = \frac{k_B^2 T}{V \hbar}
\Big(
\frac{\pi^2}{3} \sum_{\bk} \sum_{n > 0} \Omega_n(\bk)_{\mu\nu}
+
\int_0^\infty d \vep ~ (\be \vep)^2  \frac{d g(\vep)}{d\vep} \sum_{\hbar \om_{n\bk} \leq \vep} \Omega_n(\bk)_{\mu \nu}
\Big).
\ea
%%%%%%%%%%
To evaluate the energy integral, we track the integration boundaries for each individual eigenstate.
Introducing the dimensionless variable $x = \be \vep$, the second term can be rewritten as
%%%%%%%%%%
\ba
\sum_{\bk} \sum_{n > 0} \int_{\be \hbar \om_{n\bk}}^\infty dx~x^2 \frac{d}{dx} \Big( \frac{1}{e^x-1} \Big)
\Omega_n(\bk)_{\mu\nu}
= - \sum_{\bk} \sum_{n > 0}
\int_0^{g(\hbar \om_{n\bk})} dt~\Big(\ln \frac{1+t}{t}\Big)^2 \Omega_n(\bk)_{\mu \nu},
\ea
%%%%%%%%%%
where we have applied the integration substitution $t = \frac{1}{e^x-1}$, which implies $x = \ln \frac{1+t}{t}$.
Substituting this analytically evaluated integral back into the total conductivity expression yields exactly
%%%%%%%%%%
\ba
\kmn = -\frac{k_B^2 T}{V \hbar} \sum_{\bk} \sum_{n > 0}  \Big[
c_2 (g(\hbar \omega_{n\bk})) - \frac{\pi^2}{3}
\Big] \Omega_{n}(\bk)_{\mu \nu}.
\ea
%%%%%%%%%%
To make this formula robust against band degeneracies and numerical singularities, we elevate the single-band summation to the multiband Abelian generalization detailed in \Sec{app:phononbc}, implementing the substitution $n \to [n]$.
This formally yields the expression in \eq{seq:kmn}.
Crucially, this sign-corrected linear response derivation exactly reproduces the thermal Hall formulas derived independently via semiclassical wave-packet dynamics~\cite{zhang2016berry}, as well as formulas derived for canonical phonon Hamiltonians (such as those governing conventional Raman spin-lattice interactions~\cite{park2019topological}), demonstrating universal agreement across distinct theoretical frameworks.
%

%%%%%%%%%%%%%%%%%%%%%%
\subsection{High-temperature series expansion}
\label{app:response_series}
%%%%%%%%%%%%%%%%%%%%%%
To obtain a physically transparent expansion of \eq{seq:kmn} at high temperatures, we analyze the analytic structure of the dimensionless weighting factor
%%%%%%%%%%
\ba
\mc{W}(y) \equiv c_2 \Big(\frac{1}{e^y - 1}\Big) - \frac{\pi^2}{3},
\quad \text{where} \quad
y = \frac{\hbar \om_{[n]\bk}}{k_B T}.
\label{seq:cterm}
\ea
%%%%%%%%%%
In the high-temperature limit $T \to \infty$, or equivalently the low-energy limit $\om_{[n]\bk} \to 0^+$, the variable $y \to 0^+$.
In this limit, the integral evaluates to $c_2(\infty) - \pi^2/3 = 0$~\cite{matsumoto2014thermal}.
To determine the higher-order behavior, we differentiate $\mc{W}(y)$ to obtain
%%%%%%%%%%
\ba
\frac{d \mc{W}}{dy} =
- \Big(
\frac{y}{e^{y/2}-e^{-y/2}}
\Big)^2 = - \Big(\frac{(y/2)}{\sinh(y/2)}\Big)^2.
\ea
%%%%%%%%%%
As this derivative is strictly an even function of $y$, its integral $\mc{W}(y)$ must consist exclusively of odd powers of $y$ in its polynomial expansion around $y=0$.
Crucially, the analytic structure of this derivative precisely dictates the radius of convergence for the resulting series.
In the complex plane, $\frac{d\mc{W}}{dy}$ possesses a removable singularity at $y=0$ and true poles where $\sinh(y/2) = 0$ for $y \neq 0$, which occur at $y = \pm 2\pi i, \pm 4\pi i, \dots$.
According to complex analysis, the radius of convergence for a Taylor series is strictly bounded by the distance to the nearest non-removable singularity, which is given by $|2\pi i| = 2\pi$ in this case.
Therefore, the Taylor series for $\mc{W}(y)$ converges absolutely for $|y| < 2\pi$, and strictly diverges for $|y| > 2\pi$.
Translated to our physical parameters, this guarantees that the polynomial expansion is mathematically exact so long as $k_B T > \max(\hbar \om_{[n]\bk}) / 2\pi$.
Evaluating the successive derivatives analytically, we construct the convergent high-temperature series expansion for the intrinsic thermal Hall conductivity
%%%%%%%%%%
\ba
\kxy = \frac{k_B}{V} \sum_{[n]\bk} \om_{[n]\bk} \Omega_{[n]}(\bk)_{xy}
\Big[1-\frac{1}{36} \Big(\frac{\hbar \om_{[n]\bk}}{k_B T}\Big)^2 +
\frac{1}{1200} \Big(\frac{\hbar \om_{[n]\bk}}{k_B T}\Big)^4
- \frac{1}{42336} \Big(\frac{\hbar \om_{[n]\bk}}{k_B T}\Big)^6 +
\dots
\Big].
\ea
%%%%%%%%%%
As the prefactor $T$ in the original $\kappa_{xy}$ formula cancels with the $T^{-1}$ scaling of the leading $y^1$ term, the leading-order thermal Hall conductivity approaches a constant value at high temperatures. 
Note that this series expansion is a universal property of the geometric function $\mc{W}(y)$ and thus applies generically to the intrinsic thermal Hall conductivity of various bosonic excitations, including both phonons and magnons~\cite{matsumoto2014thermal}.
%

%%%%%%%%%%%%%%%%%%%%%%
\subsection{Symmetry constraints}
\label{app:response_sym}
%%%%%%%%%%%%%%%%%%%%%%
We conclude this section by explicitly establishing the spatial symmetry constraints governing the macroscopic tensor $\kmn$.
Substituting the momentum summation as $\bk \to g \bk$ and applying the symmetry constraints derived in \Sec{app:psym} directly into the transport integral in \eq{seq:kmn}, one immediately finds that the macroscopic conductivity tensor transforms covariantly
%%%%%%%%%%
\ba
\kmn = (-1)^{\phi_g} \Og \kmn \Og^T
\ea
%%%%%%%%%%
[see \eq{seq:omtransform}].
In purely two-dimensional systems, the orthogonal rotation matrices simplify this constraint.
As explicitly shown in \eq{seq:rot2d} and \eq{seq:mirror2d}, this relation forces $\kmn$ to vanish identically under any antiunitary rotation or unitary mirror operation (including glide mirrors).
All other standard operations preserve the Hall response.
In three-dimensional systems, the constraints are most easily diagnosed by mapping the anti-symmetric conductivity tensor to its dual pseudovector, defined as $[\bka]_{\rho} = \hf \sum_{\mu \nu} \vep_{\mu \nu \rho} \kmn$.
Following the Berry curvature derivation in \eq{seq:om3d}, the macroscopic dual vector strictly obeys
%%%%%%%%%%
\ba
\bka = (-1)^{\phi_g} \det(\Og) \Og \bka.
\ea
%%%%%%%%%%
This simple vectorial rule can be readily applied to any given 3D space group to determine which plane can support a finite thermal Hall heat current.

%%%%%%%%%%%%%%%%%%%%%%
\section{Explicit calculation in the kagome lattice}
\label{app:model}
%%%%%%%%%%%%%%%%%%%%%%
In this Appendix, we explicitly present the dynamical matrix, emergent gauge field, and isotopic scaling parameters for the kagome lattice model introduced in the main text.
Following the conventions established in \Sec{app:phonondef}, we assume a periodic lattice and decompose the lattice site index as $i = (\bR, \al)$, where $\bR$ labels the unit cell and $\al \in \{1,2,3\}$ labels the sublattice.
The primitive lattice vectors are $\bba_1 = (1, 0)$ and $\bba_2 = (-\hf, \frac{\sqrt{3}}{2})$ [see Fig. 1(a) of the main text].
The sublattice positions within the unit cell are defined as $\bt_1 = \bb 0$, $\bt_2 = \hf \bba_1$, and $\bt_3 = \hf(\bba_1 + \bba_2)$.
Using the reciprocal lattice vectors $\bb{b}_1, \bb{b}_2$ satisfying $\bba_i \cdot \bb{b}_j = 2\pi \delta_{ij}$, we parameterize the momentum in crystalline coordinates as $\bk = k_1 \bb{b}_1 + k_2 \bb{b}_2$.

\textit{Dynamical matrix.---}
We denote the longitudinal and transverse spring constants (divided by the atomic mass $M$) as $\kl$ and $\kt$, respectively.
In the composite basis $\al \otimes \mu$, the dynamical matrix takes a block-diagonal form [see \Sec{app:phonondef}]:
%%%%%%%%%%
\ba
D(\bk) =
\bpm
D_{11}(\bk) & D_{12}(\bk) & D_{13}(\bk) \\
D_{21}(\bk) & D_{22}(\bk) & D_{23}(\bk) \\
D_{31}(\bk) & D_{32}(\bk) & D_{33}(\bk)
\epm
,
\ea
%%%%%%%%%%
where the $2 \times 2$ sub-blocks satisfy the Hermiticity condition $D_{\al \be}(\bk) = D_{\be \al}(\bk)^\dg$.
Upon calculation, the diagonal blocks are given by
%%%%%%%%%%
\ba
D_{11} (\bk) &= \hf
\bpm
5 \kl + 3 \kt & \sqrt{3} (\kl - \kt)\\
\sqrt{3} (\kl - \kt) & 3 \kl + 5 \kt
\epm
,
\nn
D_{22}(\bk) &= \hf
\bpm
5 \kl + 3 \kt & -\sqrt{3}(\kl - \kt) \\
-\sqrt{3}(\kl - \kt) & 3 \kl + 5 \kt
\epm
,
\nn
D_{33}(\bk) &=
\bpm
\kl + 3 \kt & 0 \\
0 & 3 \kl + \kt
\epm
.
\ea
%%%%%%%%%%
The off-diagonal blocks connecting distinct sublattices evaluate to
%%%%%%%%%%
\ba
D_{12}(\bk) &= -2 \cos \pi k_1
\bpm
\kl & 0 \\
0 & \kt
\epm
,
\nn
D_{13}(\bk) &= - \frac{\cos \pi(k_1 + k_2)}{2}
\bpm
\kl + 3 \kt & \sqrt{3}(\kl - \kt) \\
\sqrt{3}(\kl - \kt) & 3 \kl + \kt
\epm
,
\nn
D_{23}(\bk) &= -\frac{\cos \pi k_2}{2}
\bpm
\kl + 3 \kt & \sqrt{3} (\kl - \kt) \\ 
\sqrt{3} (\kl - \kt) & 3 \kl + \kt
\epm
.
\ea
%%%%%%%%%%

%
\textit{Emergent gauge field.---}
To calculate the emergent gauge field, we assume the hopping amplitudes depend strictly on the interatomic distance, and we restrict our model to nearest-neighbor hopping amplitudes $t \in \mbb{R}$.
Using the composite index $a = (i_a, \mu_a)$, the spatial derivative of the hopping amplitude is [see \eq{seq:der}]:
%%%%%%%%%%
\ba
\Delta_{ij}^a \equiv \der_a t_{ij}(|\br_i - \br_j|)_{\rion=\rion^{eq}} = \gm t_{ij}^a \overline{\delta}_{ij}^a,
\quad
t_{ij}^a =
\begin{cases}
	t & \text{if} \quad i_a \in \{i,j\} \\
	0 & \text{else}
\end{cases}
,
\ea
%%%%%%%%%%
where $\gm \in \mbb{R}$ is a material-specific constant, and $\overline{\delta}_{ij}^a$ is the $\mu_a$-directional cosine of the line segment connecting sites $i$ and $j$ (with $i_a$ designated as the origin).
Following the fourth-order expansion derived in \eq{seq:ftotal}, the real-space emergent gauge field evaluates to
%%%%%%%%%%
\ba
[\fab]^{(4)}
=& \frac{20}{J^4} \sum_{ijkl}
Re[t_{ij}^a t_{jk}^b t_{kl} t_{li}] \gm^2 \overline{\delta}_{ij}^a \overline{\delta}_{jk}^b  
(\chi_{ijk}+\chi_{jkl}+\chi_{lij}-3\chi_{kli})
\nn
+& \frac{40}{J^4} \sum_{ij}
Re[t_{ai} t_{ib} t_{bj} t_{ja} - t_{ai}t_{ij}t_{jb} t_{ba}] 
\gm^2
\overline{\delta}_{ai}^a \overline{\delta}_{bj}^b 
(\chi_{aib}+ \chi_{ibj} -\chi_{bja}-\chi_{jai}).
\ea
%%%%%%%%%%

%
The momentum-space emergent gauge field is obtained via the Fourier transform $\fem(\bk)^{(4)}_{\al \mu, \be \nu} = \sum_{\bR} e^{-i \bk \cdot (\bR + \bt_{\al} - \bt_\be)} \fem(\bR)_{\al \mu, \be \nu}^{(4)}$ [see \eq{seq:comm}].
Dropping the superscript for brevity and expanding the field into the $\al \otimes \mu$ sublattice basis, we obtain:
%%%%%%%%%%
\ba
F(\bk) = 20 \chi_{123} \Big(\frac{t}{J}\Big)^{4}
\bpm
F_{11}(\bk) & F_{12}(\bk) & F_{13}(\bk) \\
F_{21}(\bk) & F_{22}(\bk) & F_{23}(\bk) \\
F_{31}(\bk) & F_{32}(\bk) & F_{33}(\bk)
\epm
,
\ea
%%%%%%%%%%
where each $2 \times 2$ block satisfies the anti-Hermiticity condition $F_{\al \be}(\bk) = -F_{\be \al}(\bk)^{\dg}$.
By utilizing graph-theoretic methods to sum the four-site hopping loops containing $(\bR, \al)$ and $(\bb 0, \be)$, we obtain
%%%%%%%%%%
\ba
F_{11}(\bk) &= F_{22}(\bk) = F_{33}(\bk) =
\bpm
0 & 0 \\
0 & 0
\epm
,
\nn
F_{12}(\bk) &= 4(\cos \pi k_1 - \cos \pi (k_1 +2 k_2))
\bpm
1 & -\sqrt{3} \\
\sqrt{3} & 3
\epm
, \nn
F_{13}(\bk) &= 8(\cos \pi(k_1 + k_2) - \cos \pi (k_1 - k_2))
\bpm
1 & -\sqrt{3} \\
0 & 0
\epm
, \nn
F_{23}(\bk) &= 8(\cos \pi (2k_1 + k_2) - \cos \pi k_2)
\bpm
1 & \sqrt{3} \\
0 & 0
\epm
.
\ea
%%%%%%%%%%

\textit{Isotopic scaling law.---}
We present the fitting parameters obtained for the intermediate-to-high temperature isotopic scaling law [see Fig. 2(b) in the main text]
%%%%%%%%%%
\ba
\kmn/[\kmn] = \sum_{i=0}^{\infty} a_i \mr^{-(\hf + b + i)} (T/[K])^{-2i}
\quad (\mu \neq \nu),
\ea
%%%%%%%%%%
where we divided $\kmn$ and $T$ by their standard units.
Evaluating up to the fourth order correction, we obtain
%%%%%%%%%%
\begin{center}
\begin{tabular}{ |c|c|c|c|c| }
\hline
$a_0$ & $a_1$ & $a_2$ & $a_3$ & $b$ \\
\hline
$-4.26268 \times 10^{-4}$ & $1.46013$ & $-2.58377 \times 10^{3}$ & $1.92423 \times 10^{6}$ & $0.370686$
\\
\hline
\end{tabular}
\end{center}
%%%%%%%%%%
The $p$-values for all fitting parameters are smaller than the numerical precision.

\end{document}